\documentclass[fleqn,usenatbib]{mnras}

\usepackage[T1]{fontenc}

\DeclareRobustCommand{\VAN}[3]{#2}
\let\VANthebibliography\thebibliography
\def\thebibliography{\DeclareRobustCommand{\VAN}[3]{##3}\VANthebibliography}

\usepackage{graphicx}	
\usepackage{newtxtext,newtxmath}
\usepackage{orcidlink}

\title[Stacking exoplanet spectra]{Stacking transmission spectra of {\it different} exoplanets}

\author[Kirk, J. \&  Owen, J. E.]{
James Kirk$^{\orcidlink{0000-0002-4207-6615},1}$\thanks{E-mail: j.kirk22@imperial.ac.uk} and James E. Owen$^{\orcidlink{0000-0002-4856-7837},1,2}$\\
$^{1}$Imperial Astrophysics, Imperial College London, Blackett Laboratory, Prince Consort Road, London SW7 2AZ, UK\\
$^{2}$Department of Earth, Planetary, and Space Sciences, University of California, Los Angeles, CA 90095, USA
}

\date{Accepted 2026 May 27. Received 2026 April 24; in original form 2025 October 31}

\pubyear{XXXX}

\begin{document}
\label{firstpage}
\pagerange{\pageref{firstpage}--\pageref{lastpage}}
\maketitle

\begin{abstract}
In many areas of astronomy, spectra of different objects are co-added or stacked to improve signal-to-noise and reveal population-level characteristics. As the number of exoplanets with measured transmission spectra grows, it becomes important to understand when stacking spectra from different exoplanets is appropriate and what stacked spectra physically represent. Stacking will be particularly valuable for long-period planets, where repeated observations of the same planet are time-consuming. Here, we show that stacked exoplanet transmission spectra can, under well-defined conditions, be represented by spectra generated from the geometric mean of each planet's abundance ratios. We test this by comparing stacked and geometric mean spectra across grids of forward models over JWST's NIRSpec/G395H wavelength range (2.8--5.2\,$\mu$m). For two dominant species (e.g., H$_2$O and CO$_2$), the geometric mean accurately reflects the stacked spectrum if abundance ratios are self-similar across planets. Introducing a third species (e.g., CH$_4$) makes temperature a critical factor, with stacking becoming inappropriate across the CO/CH$_4$ boundary, which is the primary chemical transition considered in this work. Surface gravity exerts only a minor influence when stacking within comparable planetary regimes. We further assess the number of stacked, \textit{distinct} sub-Neptunes with high-metallicity atmospheres and low-pressure, grey cloud decks required to rule out a flat spectrum at $>5\sigma$, as a function of both cloud deck pressure and per-planet spectral precision. These results provide guidance on when stacking is useful and on how to interpret stacked exoplanet spectra in the era of population studies of exoplanets.
\end{abstract}




\section{Introduction}

As the sample of exoplanets with measured transmission spectra continues to grow, population-level trends in exoplanetary spectra will become increasingly accessible. Previous efforts in this direction include exploring the relation between hazes and equilibrium temperature in sub-Neptunes \citep[e.g.,][]{Crossfield2017,Brande2024}; clouds, equilibrium temperature, and surface gravity in hot Jupiters \citep[e.g.,][]{Sing2016,Stevenson2016,Fu2017,Gao2020}; mass and metallicity \citep[e.g.,][]{Welbanks2019,Spake2021,Fu2025}; and atmospheric escape and irradiation \citep[e.g.,][]{Nortmann2018,Zhang2023,Orell-Miquel2024}. These studies have naturally treated each exoplanet individually, using its observables as a single point in a population-level relation. 

A complementary approach, employed in many other areas of astronomy, is to combine measurements across objects -- for instance, by co-adding or stacking different objects' spectra -- to directly reveal population-level characteristics that might be obscured by noise in individual observations. In extragalactic astronomy, for example, stacking galaxy spectra enables the recovery of average chemical abundances, star formation rates, or kinematic properties from data that would otherwise be too noisy to interpret individually \cite[e.g.,][]{Shapley2003,Steidel2016}. For the case of exoplanets, each exoplanet represents a distinct atmospheric system, with its own temperature (mostly externally driven for transiting exoplanets), gravity, chemistry, and aerosol structure. Consequently, it is not immediately obvious whether stacking spectra of different planets will yield a physically meaningful result. While stacking can reveal trends in noisy datasets, the interpretation of stacked exoplanet spectra requires a careful understanding of how variations in atmospheric parameters propagate into the resulting average spectrum.

This challenge is particularly pressing for long-period planets, for which repeated transit observations to build high signal-to-noise spectra are observationally expensive. For example, it would take JWST four years (2025--2029) to observe 10 transits of the 32\,d period K2-18b but only three months to observe one transit of 10 similar planets\footnote{The remaining planets from Table 1 of \cite{Madhusudhan2021}, excluding K2-18b.}.  In such cases, stacking spectra across planets could be the only viable strategy for detecting atmospheric features on a reasonable timescale. However, without a clear understanding of the conditions under which stacking is appropriate, there is a risk of introducing biases or obscuring physically relevant information. For example, if the stacked planets have significantly different temperature structures or cross chemical transitions (e.g., the CO/CH$_4$ boundary), the resulting spectrum may not accurately reflect the atmosphere of any physical planet. As we later demonstrate, this occurs because one atmosphere may be CH$_4$-dominated while another is CO-dominated, so stacking combines distinct sets of molecular absorbers and does not yield a physically meaningful representation of any individual atmosphere.


In this work, we aim to rigorously establish the mathematical and practical framework for stacking exoplanet transmission spectra. We demonstrate that, under well-defined conditions, the stacked spectrum is mathematically equivalent to a spectrum derived from the geometric mean of the abundance ratios of each planet. Using grids of forward models spanning surface gravity, equilibrium temperature, and chemical composition, we quantify the extent to which the stacked spectra approximate the geometric mean in different regimes. We explore both simple atmospheres dominated by two species (H$_2$O and CO$_2$) and more complex cases that include additional molecules such as CH$_4$, CO, H$_2$S, and SO$_2$. 

Our analysis focusses on the wavelength range of the JWST's NIRSpec G395H instrument mode (2.8--5.2\,$\mu$m), which is sensitive to several key molecular bands (e.g., H$_2$O, CH$_4$, CO$_2$ and CO), and has become a workhorse in the measurement of exoplanet atmospheres by JWST \citep[e.g.,][]{Alderson2023,Alderson2024,May2023,Moran2023,Wallack2024,Wallack2026,Adams2025,Meech2026}. This choice is further motivated by its central role in the BOWIE-ALIGN programme \citep{Kirk2024,Kirk2025,Ahrer2025_K7,Ahrer2025,Meech2025,Claringbold2026,Fairman2026}, for which stacking analyses of this bandpass are directly relevant. 

We first develop the mathematical formalism and illustrate it with toy models (Section \ref{sec:fundamentals}), then apply the stacking prescription to realistic transmission spectra to identify the parameter space where stacking is valid and where it breaks down (Section \ref{sec:realistic_model_transmission_spectra}). We conclude with a discussion and summary in Sections \ref{sec:discussion} and \ref{sec:summary}.

\section{Fundamentals}
\label{sec:fundamentals}

Unlike the traditional stacking of astronomical observations, it is not obvious how to stack exoplanet transmission spectra, nor is it obvious in doing so what exactly this will achieve. The primary motivation for stacking different targets in astronomical observations is to improve the signal-to-noise ratio, revealing signals that were undetectable or uncertain in a single object. Ultimately, stacking should provide an understanding of the typical properties of the class of objects studied. 

\subsection{What do transmission spectra actually probe?}

In order to understand exactly how one might stack transmission spectra, here we review what transmission spectra measure. These insights are not new and are based on various important contributions in the literature \citep[e.g.][]{Seager2000,Brown2001,Hubbard2001,Fortney2005,Miller-Ricci2009,deWit2013,Griffith2014,Heng2017}.

Transmission spectra are dominated by extinction along a slant optical path through the atmosphere \citep[e.g.][]{Brown2001,Fortney2005}, where the slant optical depth, at an impact parameter $b$, is given by:
\begin{equation}
    \tau_\nu(b) = \int_{-\infty}^{\infty} \sigma_\nu(x)n(x)\,{\rm d}x
\end{equation}
where $x$ is the optical path between the star and the observer such that the distance from the centre of the planet is given by $r=\sqrt{b^2+x^2}$, $\sigma_\nu$ is the extinction cross-section at a frequency $\nu$, and $n$ is the number density of gas. Assuming a hydrostatic density profile $n(r)$ with a density, $n_0$, at some radius\footnote{Many authors choose $r_0$ to be the reference radius; however, its choice is more general than this and can simply be thought of as a scale radius of the atmospheric density structure where the density scale is set.}, $r_0$, in a constant gravity atmosphere (i.e., $H/R_p\ll 1$, with $H$ the planet's atmospheric scale height and $R_p$ its radius). In Appendix~\ref{sec:appendix_1}, we show that the slant optical depth may be written generally as:
\begin{equation}
    \tau_\nu = A n_0\sigma_\nu(r_0)\sqrt{2\pi H(r_0)r_0}
\end{equation}
where $A$ depends on the shape of the abundance profile in the atmosphere and any temperature gradient present. In the case of constant opacity per unit mass in an isothermal atmosphere, $A=1$ is the well-known result \citep[e.g.][]{Fortney2005}. However, important for us with regard to stacking different atmospheric structures, $A$ is constant between planets if the temperature gradient is the same and if the extinction cross-section profile is self-similar, a likely outcome for quenched, major species. This means that if the shape of an absorber's abundance profile is the same\footnote{Provided the extinction cross-section is not strongly pressure and temperature dependent.}, but is just shifted around a different reference density ($n_0$), $A$ remains fixed. 
\subsection{Isothermal atmospheres}
For an isothermal atmosphere, under the reasonable assumption that (i) at a given wavelength a single species is the dominant absorber \citep[e.g.][]{deWit2013} and (ii) that the extinction cross-section of that species is slowly varying with pressure, such that its extinction cross-section can be approximated as a power-law $(n/n_0)^l$ with $l$ being a constant in the transmission region \citep[e.g.][]{deWit2013}\footnote{Since transmission spectroscopy only probes a narrow range of pressures, this is a fairly robust approximation.}, then the measured transit radius at a given frequency may be written as (Appendix~\ref{sec:appendix_1}):
\begin{equation}
    R_{t,\nu}=R_0+\frac{H}{l_\nu+1}\left(\gamma+\log \tau_{\nu,0}\right)
\end{equation}
relative to some optically thick (arbitrary) reference radius $R_0$, where $l_\nu$ is the power-law index for the species that dominates the absorption at a frequency $\nu$, $\tau_{\nu,0}$ is the slant optical depth at the reference radius, and $\gamma$ is the Euler-Mascheroni constant. Now, comparing the transit radii between two different frequencies ($\nu_1$ \& $\nu_2$), we find:
\begin{equation}
    \frac{R_{t,\nu_1}-R_{t,\nu_2}}{H}= \log\left(\frac{\tau_{\nu_1,0}^{1/(l_{\nu_1}+1)}}{\tau_{\nu_2,0}^{1/(l_{\nu_2}+1)}}\right)+\gamma\left(\frac{1}{l_{\nu_1}+1}-\frac{1}{l_{\nu_2}+1}\right)\label{eqn:delta_R1}
\end{equation}
Equation~\ref{eqn:delta_R1} provides insight into how we might stack the transmission spectra of different planets. If a collection of planets were to have the same ratio of extinction co-efficient between the two frequencies $\nu_1$ and $\nu_2$ with altitude then the RHS of Equation~\ref{eqn:delta_R1} would provide an identical result for every planet. The first term encapsulates the well known result that the difference in transit radii between two frequencies is just the natural logarithm of the ratio of the cross-sections (since the optical paths are essentially identical for $H/R_p\ll 1$). The second term encodes the fact that if the dominant species in the two different species have a different vertical distribution, the effective scale height of these separate species is different, and this introduces an offset in the transit radii. However, for the major species, where $l$ is a positive order unity, this correction factor is $\ll 0.3 \Delta R_t/H$, smaller than the errors in recent JWST observations \citep[e.g.][]{Meech2025}. 

Since the reference radius is arbitrary, we should interpret the ratio of the limb optical depths as the ratio of the limb optical depths in the region where transmission spectroscopy is probing the atmosphere. Therefore, for planets with identical profiles of the ratio of extinction co-efficient in the region of the atmosphere probed by transmission spectroscopy (approximately those regions where $\tau \sim \exp(-\gamma)$, \citealt{Heng2017}, e.g. $\sim$mbar pressures) the quantity $\Delta R_t / H$ will be identical between these planets. Thus, this is the quantity we propose to stack. It is important to note that different planets with identical (but varying) extinction profiles with pressure will not give the same spectrum when measured in $\Delta R_t /H$. This is because (as shown in Equation~\ref{eqn:limb_tau}) the specific pressures probed by transmission spectroscopy are sensitive to both the radius and the scale height of the planet. Thus, unless the planets have the same $H\times R_p\propto T/(\mu\rho)$ (with $T$ the temperature, $\mu$ the mean molecular weight, and $\rho$ the gas density), the probed pressure ranges will be different between the planets and, as such, will result in a different $\Delta R_t/ H$ spectrum. This is similar to other astronomical fields, where identical objects with different observed geometries can sometimes give different spectra \citep{Lorenz2023}. As such, when exoplanetary transmission spectra are stacked, one must bear in mind the exact goal and remember that stacking planets with identical, but pressure dependent, extinction profiles would give rise to a stacked spectrum that is not representative of the planet's profile. Rather, stacking planetary spectra with identical extinction profiles in the transmission region but occurring at different pressures would yield a representative spectrum. 

\subsubsection{Toy model}
\begin{figure*}
    \centering
    \includegraphics[width=\textwidth]{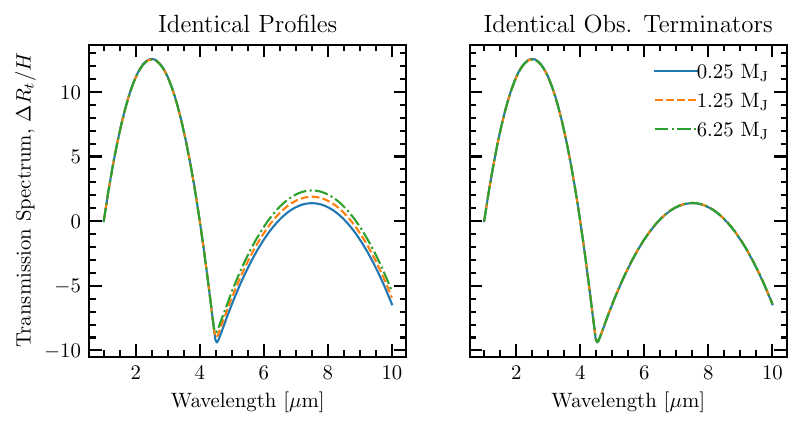}
    \caption{The transmission spectrum, shown as $[R_t-R_t(\lambda=1~\mu{\rm m})]/H$ of three planets with masses of 0.25, 1.25 \& 6.25 M$_{\rm J}$. The atmosphere contains two (arbitrary) species with Gaussian cross-sections. The left-panel shows a scenario where the abundance profiles as a function of pressure are fixed in the atmosphere. In this case because the transmission spectra of the different mass planets probe different pressure levels the abundance ratios of the two species are different at the observable pressures. Thus, despite the fact the abundance profiles are identical the different planets produce different transmission spectra. However, in the right-panel the abundances of species 2 are adjusted so that they are identical at the different pressures probed by transmission spectra for the different mass planets. This results in identical transmission spectra. This toy model highlights that different planets will only have identical transmission spectra (in terms of $\Delta R_t/H$) if the extinction profiles are identical in the region probed by transmission spectra, not if the extinction profiles are identical. }
    \label{fig:toy1}
\end{figure*}

To demonstrate our previous insights explicitly, we consider a toy model, consisting of two species ($i=\{1,2\}$) which have extinction cross-sections ($\sigma_{\rm e}$) of the following Gaussian form:
\begin{equation}
\sigma^i_{\rm e} = {\mathcal{A}}(P) \exp\left[-\frac{\left(\lambda-\lambda_i\right)^2}{2w_i^2}\right] \label{eqn:toy_extinc}
\end{equation}
where the amplitude, $\mathcal{A}$, varies with pressure. In Figure~\ref{fig:toy1}, we show the transmission spectra (in $\Delta R_t / H$) for three solar composition 1000~K Jupiter radii planets with masses of 0.25, 1.25 and 6.25 M$_{\rm J}$. Species 1 has $\lambda_1=2.5~\mu$m and $w_1=0.3~\mu$m, while species 2 has $\lambda_2=7.5~\mu$m and $w_2=0.4~\mu$m. We set the mixing ratio of species 1 to a constant with height ($l_1=0$), while for species 2 we let it vary with a power-law such that $l_2=1.5$. In the left-hand panel the mixing ratio between the two species is 1 at 10 bar, namely, we are adopting a case where the abundance profiles with pressure are identical between the three planets. However, because the scale height of the most massive planet is a factor of 25 smaller, the transmission spectrum of the heaviest planets is probing a pressure level $\sqrt{25}$ times higher than the lowest mass planet where the abundance ratio is $25^{3/4}$ larger. Therefore, in the right-hand panel we reduce the abundance of species 2 at 10 bar by a factor of $5^{3/4}$ for the 1.25 M$_{\rm J}$ and $25^{3/4}$ for the 6.25 M$_{\rm J}$ planets, respectively. In this case, we have adjusted the abundances of species 2 so that the abundance ratios in the observable terminators are identical, resulting in identical transmission spectra.

This toy example explicitly demonstrates that when stacking exoplanet transmission spectra we should only expect identical results if the extinction profiles are identical in the transmission region, rather than if the extinction profiles are identical with pressure (or density), and we will return to the point in Section~\ref{sec:measure} when we discuss what stacked spectra actually encode. 

\subsection{Stacking different profiles of the same species}

\begin{figure}
    \centering
    \includegraphics[width=\columnwidth]{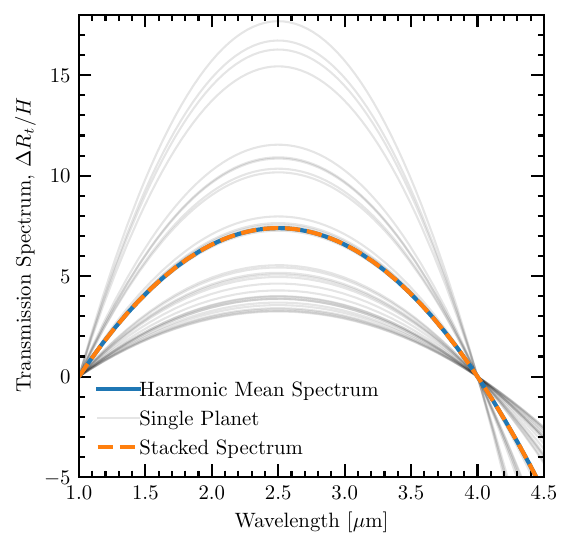}
    \caption{The transmission spectrum, shown as $[R_t-R_t(\lambda=1~\mu{\rm m})]/H$, for 30 planets where the extinction profile with pressure $l$ has been randomly drawn between -0.3 and 3. The stacked spectrum is shown as the solid blue line. A spectrum calculated assuming the $l$ is given by the harmonic mean of all the ${l_j+1}$ is shown as the orange dashed line, perfectly agreeing with the stacked spectrum. This demonstrates that for a given species the stacked spectrum is representative of the harmonic mean of each planet's individual abundance profile.  }
    \label{fig:toy_c}
\end{figure}

Before we expand our discussion to stacking general transmission spectra, we will first explore what happens when you stack transmission spectra in a wavelength range where the spectrum is dominated by a single species (e.g. a H$_2$O or CO$_2$ band). At the typical resolution of space-based observations (e.g. Hubble, JWST and ARIEL) one is not sensitive to the pressure dependence of {\it individual} molecular lines, and as such the overall pressure sensitivity of the molecular bands is not strongly wavelength dependent \citep[e.g.][]{deWit2013}. Similarly, the same is true for the cores of atomic lines, such as sodium and potassium; this does not hold for the line wings that are pressure sensitive, but these typically require higher resolution ground-based data to detect \citep[e.g.][]{Nikolov2018}. Therefore, within a single molecular band we can assume that our power-law dependence for the extinction $l_\nu$ is not wavelength dependent and is fixed to a single value $l_j$ for each planet $j$. Under this simplification, Equation~\ref{eqn:delta_R1} becomes:
\begin{equation}
    \frac{R_{t,\nu_1}-R_{t,\nu_2}}{H} = \frac{1}{l+1}\log\left(\frac{\tau_{\nu_1}}{\tau_{\nu_2}}\right)
\end{equation}
where we have now dropped the (``$0$'') index on the optical depths for simplicity. Thus, stacking $N_{\rm p}$ transmission spectra, we find:
\begin{equation}
\begin{split}
    \frac{1}{N_{\rm p}}\sum_{j=1}^{N_{\rm p}} \frac{\left(R_{t,\nu_1}-R_{t,\nu_2}\right)_j} {H} &= \frac{1}{N_{\rm p}}\sum_{j=1}^{N_{\rm p}} \frac{1}{l_j+1}\log\left(\frac{\tau_{\nu_1}}{\tau_{\nu_2}}\right)_j \\&= \frac{1}{N_{\rm p}}\sum_{j=1}^{N_{\rm p}} \frac{1}{l_j+1}\log\left(\frac{\sigma_{\nu_1}}{\sigma_{\nu_2}}\right)_j 
\end{split}
\end{equation}
where the last result follows from the fact that for each planet the geometric factors in each optical path-length are essentially identical (under the assumption $H\ll R_P$). Now, since the cross-section ratio is a fundamental property of the species, not of the planet, the cross-section ratio is constant for each planet (again highlighting that this is appropriate at low resolution where individual molecular lines are unresolved). This means that the stacked spectra become:

\begin{equation}
\begin{split}
    \frac{1}{N_{\rm p}}\sum_{j=1}^{N_{\rm p}} \frac{\left(R_{t,\nu_1}-R_{t,\nu_2}\right)_j} {H} &= \log\left(\frac{\sigma_{\nu_1}}{\sigma_{\nu_2}}\right)\left[\frac{1}{N_{\rm p}} \sum_{j=1}^{N_{\rm p}} \frac{1}{l_j+1}\right] \\&=  \log\left(\frac{\sigma_{\nu_1}}{\sigma_{\nu_2}}\right) \frac{1}{\mathcal{H}(\{l_j+1\})}
    \end{split}
\end{equation}
where the term $\mathcal{H(.)}$, is the harmonic mean. Therefore, the stacked spectra behave as planetary spectra where the effective amplitude of the species is controlled by the harmonic mean of the abundance profiles. We can demonstrate this explicitly with our toy model. Focusing on species 1 for now, we randomly draw $l_{j}$ uniformly between $-0.3$ and $3$ for 30, Jupiter mass, Jupiter radii planets with an isothermal atmosphere at a temperature of 1000~K. In Figure~\ref{fig:toy_c} we show the 25 individual spectra, along with the stacked transmission spectrum and a single transmission spectrum computed where $l$ is set to the harmonic mean of $\{l_j\}$ for the 30 individual planets. 

Note that while in this toy model we have kept the abundances at the 10\,bar level fixed, although varying these does not make any difference, as plotting in $\Delta R_t$ removes any abundance variations between planets within an individual species. This highlights the well-known degeneracy that transmission spectra do not directly probe absolute abundances, rather relative abundances \citep[e.g.][]{Benneke2012,Kreidberg2018}. Such a result is what one would expect from any stacking procedure: objects with similar radiative properties should be combined in a manner that preserves this similarity. 

Furthermore, since we are scaling with respect to the theoretically computed scale height (computed, one would imagine, from some estimate of the equilibrium temperature) that may not represent the true planetary scale height (e.g. due to the fact that the planet's temperature may not be at the equilibrium temperature, or that our assumed mean molecular weight is incorrect), then the effective scale height of our species, as measured directly from the stacked transmission spectrum, is the harmonic mean of all these variations. Thus, the stacked spectrum provides an estimate (through the harmonic mean) of the typical scale height of these species\footnote{Note the scale height of the species, not definitively, the atmospheric scale height.} in the planet's atmosphere. 

\subsection{Stacking transmission spectra}
\label{sec:stacking_transmission_spectra}

\begin{figure}
    \centering
    \includegraphics[width=\columnwidth]{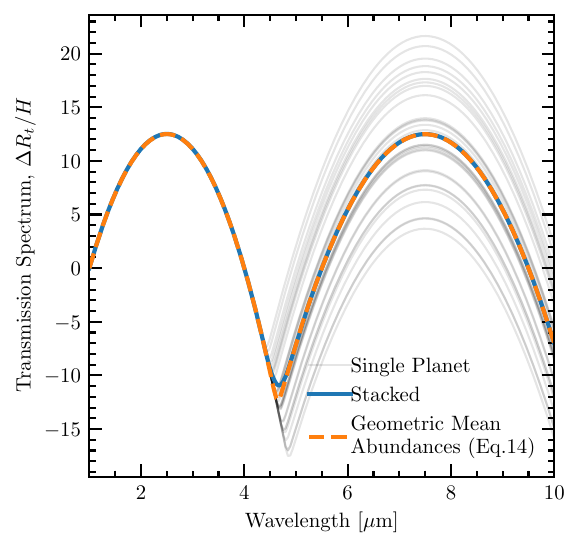}
    \caption{Demonstration that for $l=0$, the stacked spectra is given by the natural logarithm of the geometric mean of the opacity ratio.}
    \label{fig:toy_czero}
\end{figure}

Taking forward the result that within an individual species the stacked spectrum is described by the harmonic mean of their vertical profile distributions, we can now consider what stacking spectra implies for multi-species atmospheres. Again, stacking transmission spectra, now with two different species that dominate at $\nu_1$ and $\nu_2$ respectively, we find:
\begin{equation}
\begin{split}
    \frac{1}{N_{\rm p}}\sum_{j=1}^{N_{\rm p}} \frac{\left(R_{t,\nu_1}-R_{t,\nu_2}\right)_j} {H} &= \frac{1}{N_{\rm p}}\sum_{j=1}^{N_{\rm p}}\log\left(\frac{\sqrt{l_{\nu_2}+1}\sigma_{\nu_1}^{1/(l_{\nu_1}+1)}}{\sqrt{l_{\nu_1}+1}\sigma_{\nu_2}^{1/(l_{\nu_2}+1)}}\right)\\&+\gamma\left(\frac{1}{\mathcal{H}(\{l_{\nu_1,j}+1\})}-\frac{1}{\mathcal{H}(\{l_{\nu_2,j}+1\})}\right)\label{eqn:general1}
    \end{split}
\end{equation}
Now we know from the previous section that an individual molecular band in the spectrum behaves as if it has an effective scale height given by the harmonic mean. Thus, we wish to express the first term on the RHS of Equation~\ref{eqn:general1} in the form:
\begin{equation}
\begin{split}
    \frac{1}{N_{\rm p}}\sum_{j=1}^{N_{\rm p}}\log\left(\frac{\sqrt{l_{\nu_2}+1}\sigma_{\nu_1}^{1/(l_{\nu_1}+1)}}{\sqrt{l_{\nu_1}+1}\sigma_{\nu_2}^{1/(l_{\nu_2}+1)}}\right) = &\\\log\left(\frac{\sqrt{\mathcal{H}(\{l_{\nu_2,j}+1\})}\tilde{\sigma}_{\nu_1}^{1/\mathcal{H}(\{l_{\nu_1,j}+1\})}}{\sqrt{\mathcal{H}(\{l_{\nu_1,j}+1\})}\tilde{\sigma}_{\nu_2}^{1/\mathcal{H}(\{l_{\nu_2,j}+1\})}}\right)
\end{split}
\end{equation}
where $\tilde{\sigma}$ represents some average cross-section, given by:
\begin{equation}
    \tilde{\sigma}_{\nu_i} = \prod_{j=1}^{N_p}\left[\frac{\sqrt{\mathcal{H}(\{l_{\nu_i,j}+1\})}}{\sqrt{l_{\nu_i,j}+1}}\sigma_{\nu_i}^{\mathcal{H}(\{l_{\nu_i,j}+1\})/(l_{\nu_i,j}+1)}\right]^{1/N_p}
\end{equation}
This average is a weighted geometric mean of the cross-sections, which in its present form is not particularly useful. However, noting that the weights are of order unity coefficients for typical molecules, which are already scaled to their harmonic means; therefore, the dispersion in the weights is very small compared to the variation in the cross sections (which can vary over many orders of magnitude). As the logarithm of the ratio between a weighted geometric mean and the unweighted geometric mean is only proportional to the standard deviations in the weights \citep[e.g.][]{Siegel1942}, this means that a suitable approximation would be to replace this weighted geometric mean by the regular geometric mean such that:
\begin{equation}
    \tilde{\sigma_{\nu_i}}\approx \prod_{j=1}^{N_p}\left[\sigma_{\nu_i}\right]^{1/N_p}
\end{equation}
where this approximation is exact in the case where the ${l_j}$ are the same. Therefore, our stacked-spectra is approximated by:
\begin{eqnarray}
    &&\frac{1}{N_{\rm p}}\sum_{j=1}^{N_{\rm p}} \left(\frac{R_{t,\nu_1}-R_{t,\nu_2}} {H}\right)_j \approx \nonumber \\ &&  \log\left(\frac{\sqrt{\mathcal{H}(\{l_{\nu_2,j}+1\})}\mathcal{G}(\{{\sigma}_{\nu_1,j}\})^{1/\mathcal{H}(\{l_{\nu_1,j}+1\})}}{\sqrt{\mathcal{H}(\{l_{\nu_1,j}+1\})}\mathcal{G}(\{{\sigma}_{\nu_2,j}\})^{1/\mathcal{H}(\{l_{\nu_2,j}+1\})}}\right)\nonumber \\ &+&\gamma\left(\frac{1}{\mathcal{H}(\{l_{\nu_1,j}+1\})}-\frac{1}{\mathcal{H}(\{l_{\nu_2,j}+1\})}\right) \label{eqn:approx_formula}
\end{eqnarray}
where $\mathcal{G}(\{.\})$ is the geometric mean. In many cases the species of interest are quenched or have constant abundance profiles in the region of interest such that $l=0$ and the stacked spectra reduce to a more intuitive form:
\begin{equation}
\frac{1}{N_{\rm p}}\sum_{j=1}^{N_{\rm p}} \left(\frac{R_{t,\nu_1}-R_{t,\nu_2}} {H}\right)_j = \log\left[\prod_{j=1}^{N_p}\left(\frac{\sigma_{\nu_1}}{\sigma_{\nu_2}}\right)\right]^{1/N_p}    \label{eqn:geometric_mean_abund}
\end{equation}
That is, the stacked spectra are sensitive to the geometric mean of the opacity ratios in the observable region. In Figure~\ref{fig:toy_czero}, we demonstrate that Equation~\ref{eqn:geometric_mean_abund} is correct, using our toy model. We uniformly draw Jupiter-sized planets with masses between 0.3 and 2 Jupiter masses, equilibrium temperatures between 500 and 2000~K and abundance ratios between species 1 and 2 in a log-uniform fashion between $10^{-5}$ and $10^5$ for a sample of 30 planets. We then stack the individual spectra and compare the result to a spectrum computed with a geometric mean of the opacity ratio.   

This result demonstrates that the stacked transmission spectrum is described exactly by the geometric mean of the opacity ratio, under the approximation that a single species dominates the opacity. In the overlap region this is not true, as in the stacked spectra the exact wavelength where you transition from being dominated by one species to the other is blurred; however, our approximation assumes that it occurs at a specific frequency. This fact will become important later when we consider realistic transmission spectra. 

\begin{figure}
    \centering
    \includegraphics[width=\columnwidth]{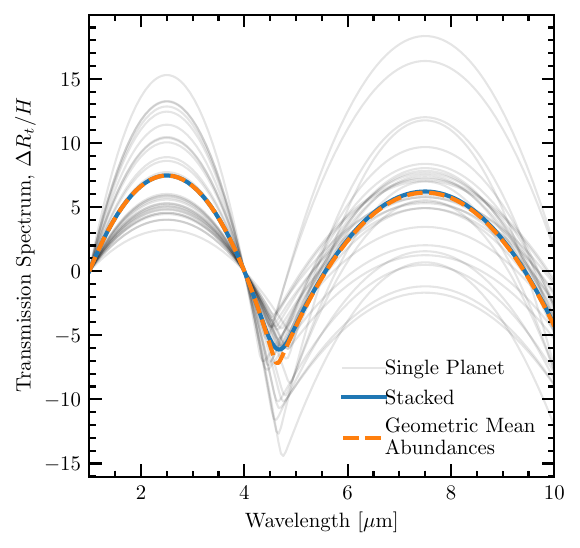}
    \caption{Demonstration that varying all the stacked spectra is given by the harmonic mean of the scale height and geometric mean of the opacities.}
    \label{fig:approx}
\end{figure}

Finally, in Figure~\ref{fig:approx} we demonstrate that the approximate approach in Equation~\ref{eqn:approx_formula} is accurate. 
In this toy experiment we vary $l_1$ uniformly between -0.3 and 3 and $l_2$ uniformly between 0 and 2\footnote{We choose different ranges to ensure the expectation value for $l_1$ and $l_2$ is different, demonstrating the robustness of our result when the average values of the coefficients are different.}. This toy experiment does indeed demonstrate that the stacked spectrum is well described by a scenario where the effective scale height is the harmonic mean and the opacities are the geometric mean of all the individual planets (i.e. Equation~\ref{eqn:approx_formula}). In order to assess the accuracy of this approach we repeat this experiment 50 times, finding a typical maximum error in the estimated stacked spectrum (outside the region where two species overlap) of $\sim 0.1 \Delta R_t/H$, an error that becomes smaller as the planet sample size increases.

\subsection{Non-isothermal atmospheres}
Highly irradiated exoplanets generally have upper atmosphere's that are close to isothermal because of the strong external radiative forcing \citep[e.g.][]{Hansen2008}. In an atmosphere where:
\begin{equation}
    B=\frac{1}{H}\frac{{\rm d}r}{{\rm d}\log H}
\end{equation}
is a measure of the strength of the temperature gradient, this gives the number of scale heights over which the temperature changes by order $e$. Since transmission spectroscopy only probes a few scale heights and most irradiated exoplanets have $|B|\gg10$ \citep{Heng2017}, any temperature gradient only slightly modifies the measured transit radius, such that \citep{Heng2017}:
\begin{equation}
    R_{t,\nu}=R_0+\frac{H\tau_{\nu,0}^{1/B}}{l_\nu+1}\left(\gamma+\log \tau_{\nu,0}\right)
\end{equation}
Given $|B|$ is large for highly irradiated planets, this can be thought of as a small correction to the $l_\nu+1$ terms, and as such a non-isothermal atmosphere does not change the picture that the difference in transit radii, normalised to the scale height is the quantity to consider for stacking. The non-isothermality of the atmospheres of the different planets becomes encoded in the harmonic mean of the $l+1$ correction to the atmospheric scale height. However, we note that for more weakly irradiated planets, in particular temperate sub-Neptunes, which are being studied with JWST \citep[e.g.][]{Benneke2024,Holmberg2024,Rigby2025}, do not necessarily have strongly isothermal atmospheres due to their weaker irradiation levels.

\subsection{What are the representative planetary parameters?}
\label{sec:representative_planet}

So far we have focused on how to combine the data and how it could be interpreted in an abstract manner. In any realistic survey, the planetary parameters will not be identical. Our analysis so far has indicated that stacking the relative transit radii, normalised to the scale height ($\Delta R_t/H$) is an appropriate choice and that this will be representative of the geometric mean of the opacity ratios. However, for it to be useful, we need to determine what the representative planetary parameters should be to compare models with observations. Thus, given a set of planets with individual masses, radii, and temperatures, we need to determine the representative mass, radius, and temperature corresponds to the ``representative planet'' described by the stacked spectra.

In the above, we have been agnostic of the exact opacity source and assumed that it only had a simple form. In reality, the opacity is sensitive to both pressure and temperature. Therefore, our stacked spectra will be representative of the pressure and temperature level probed in the representative planet, and given that the pressure probed by transmission does vary with planetary parameters, we need to take this into account. Thus, making use of the result that the stacked spectra have spectral amplitudes controlled by the harmonic mean of the abundance profiles (which follow the scale height) and the geometric mean of the cross-sections we can require that the representative planetary parameters result in the same means. Specifically, the representative mass ($M_R$), radius ($R_R$) and temperature ($T_R$) should result in the harmonic mean of the scale heights of all the planets in the sample and the geometric mean of the transmission pressure and temperature (under our self-similar opacity structure requirement) of all the planets in the sample. Assuming an isothermal atmosphere, under these requirements, the representative temperature becomes:
\begin{equation}
\label{eqn:T_R}
    T_R = \mathcal{G}(\{T_i\})
\end{equation}
and the representative radius becomes:
\begin{equation}
\label{eqn:R_R}
    R_R =\frac{T_R^2}{\mathcal{H}\left(\left\{T_iR_i^2/M_i\right\}\right)\left[\mathcal{G}\left(\left\{\sqrt{T_iM_i/R_i^3}\right\}\right)\right]^2}
\end{equation}
and the representative mass becomes:
\begin{equation}
\label{eqn:M_R}
    M_R = \frac{T_R^5}{\left[\mathcal{H}\left(\left\{T_iR_i^2/M_i\right\}\right)\right]^3\left[\mathcal{G}\left(\left\{\sqrt{T_iM_i/R_i^3}\right\}\right)\right]^4}
\end{equation}
Using these scalings, the representative planet is now scaled so that its pressure in the transmission region is representative of the appropriate averaged opacity in the stacked spectra. A standard choice for the temperature would be to assume that $T_i$ is given by the equilibrium temperature $T_{\rm eq}$ (or at least scales with it).

\subsection{What do stacked spectra physically measure?}\label{sec:measure}

Finally, now that we have laid out the methodology and assumptions underlying stacking transmission spectra from different planets, we can now discuss what a stacked spectrum might be physically interpreted as representing.  Noting that the cross-sections are typically linearly related to the abundances $X_i$ of an individual species, we finally arrive at an intuitive form for what stacked spectra constrain:
\begin{equation}
    \frac{1}{N_{\rm p}}\sum_{j=1}^{N_{\rm p}}\left(\frac{R_{t,\nu_1}-R_{t,\nu_2}}{H}\right)_j\approx \log\left[\left(\prod_{j=1}^{N_{\rm p}}\frac{X_{\nu_1}}{X_{\nu_2}}\right)^{1/N_{\rm p}}\right] \label{eqn:stack_gma}
\end{equation}
Namely, stacked transmission spectra are representative of the geometric mean of the abundance ratios in the transmission region. This result implies that stacking transmission spectra can provide useful insights and representative information of a population of planets with similar dominant absorbing species. Therefore, this suggests that stacking of transmission spectra will become a useful tool in the analysis of populations of exoplanets. This is particularly valuable in exoplanet spectroscopy, as highlighted in the introduction, and unlike many other areas of astronomy where stacking is common, stacking will always be a faster way to reach a given signal-to-noise ratio compared to repeated observations of an individual target. However, as we shall demonstrate in the next sections, one has to be careful about how wide a population of planets one combines. This is because in reaching Equation~\ref{eqn:stack_gma}, we have made an number of assumptions and simplifications which, while physically motivated, become less accurate as the spread in the planetary population increases.

Obviously, combining different planets one has no expectation to be similar is likely to lead to meaningless results. This is even more important, as the exoplanet field will still be operating in the low $N_p$ limit for the foreseeable future with JWST, and this is where we focus. The large $N_p$ limit is likely to arrive with the ARIEL mission \citep[e.g.][]{Tinetti2018} and further work would be valuable studying stacking in this limit to see if it can overcome the potential shortcomings we identify. 

\section{Realistic model transmission spectra}
\label{sec:realistic_model_transmission_spectra}

In this section, we apply the stacking formalism to realistic model transmission spectra, generated with the exoplanet atmosphere modelling software \texttt{POSEIDON} \citep{MacDonald2017,MacDonald2023}. We begin by stacking few simple model atmospheres comprising of only H$_2$, He, H$_2$O and CO$_2$ before moving onto more numerous and complicated atmospheres, following chemical equilibrium abundances. 

The forward models were calculated over a wavelength range of 2.8--5.2\,$\mu$m to be consistent with the JWST NIRSpec/G395H instrument mode, which has been used for many exoplanet transmission spectroscopy observations \cite[e.g.,][]{Alderson2023,May2023,Gressier2024,Ahrer2025,Kirk2025}. We assumed an isothermal temperature--pressure profile between $10^2-10^{-7}$ bar and a reference pressure of 10\,bar. The models were generated at a spectral resolution of $R=$ 10,000 before being binned down to $R=600$ to match the resolution presented in the Early Release Science study of WASP-39b \citep{Alderson2023}.

\subsection{Stacking identical planets with different compositions}

Firstly, we test how well stacking works in the case of identical planets with different compositions ($X_{\rm{H_2O}}$ and $X_{\rm{CO_2}}$) but the same abundance ratios ($X_{\rm{H_2O}}/X_{\rm{CO_2}}$). As we argued in Section \ref{sec:measure}, maintaining a constant abundance ratio between the planets should result in small differences between the stacked spectrum and the spectrum resulting from the geometric mean of the abundance ratios. For these initial tests, we choose to focus on hot Jupiters to directly inform the results of the BOWIE-ALIGN programme \citep{Kirk2024}. Therefore, we assume vertically constant volume mixing ratios of 85\% H$_2$ and 15\% He as our bulk species, and H$_2$O and CO$_2$ as our trace species. The abundances of H$_2$O and CO$_2$ are varied between $0.01\times, 1\times$ and $10\times$ the solar volume mixing ratios \citep{Asplund2009}, to encompass sub-solar to super-solar metallicities expected from planet formation \citep[e.g.,][]{Madhusudhan2014,Penzlin2024} and consistent with values inferred from existing observations \citep[e.g.,][]{Welbanks2019,Fu2025}. We used the line lists of \cite{H2O} and \cite{CO2} for H$_2$O and CO$_2$, respectively. 

For our planetary parameters, we assume a Jupiter radius, Jupiter mass planet with an equilibrium temperature of 1500\,K that defines the temperature of our isothermal atmosphere. We define the star to have a solar radius.

\begin{figure}
    \centering
    \includegraphics[width=0.9\linewidth]{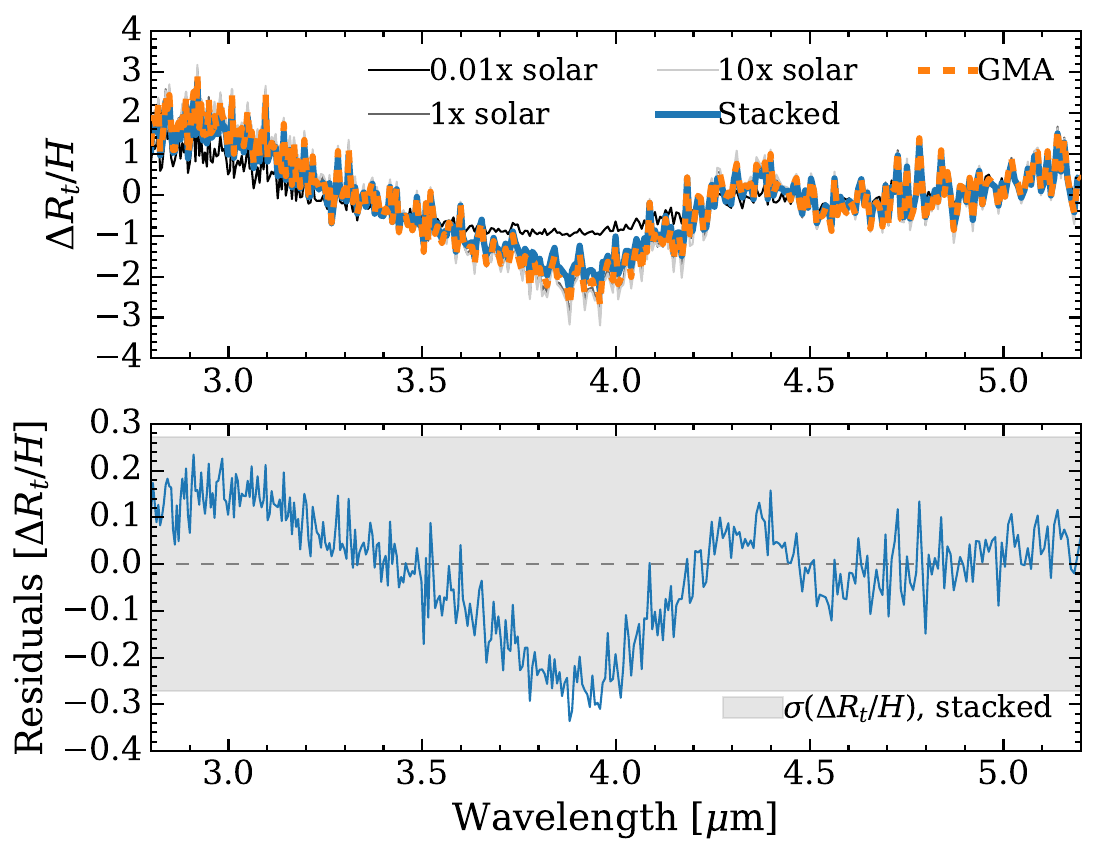}
    \caption{Stacking model transmission spectra of identical planets with different H$_2$O and CO$_2$ abundances ($0.01,1,10\times$ solar) but constant H$_2$O/CO$_2$ abundance ratios. Top panel: the individual planets' model spectra are shown as black/grey lines. The stacked spectrum is in blue and the spectrum derived from the geometric mean of the three planets' abundances (labelled `GMA') is shown in orange. Bottom panel: the difference between the stacked and GMA spectra. This difference is within the uncertainty in the stacked spectrum (shown as the grey shaded region) which assumes WASP-39b-like uncertainties (from \protect\citealt{Alderson2023}) for each of the three planets.}
    \label{fig:stacking_test_1}
\end{figure}

Figure \ref{fig:stacking_test_1} shows the results of this test. The bottom panel demonstrates that the spectrum derived from the representative planetary parameters ($T_R, R_R, M_R$ as defined in equations \ref{eqn:T_R}, \ref{eqn:R_R} and \ref{eqn:M_R}) and the geometric mean of the individual planets' abundance profiles (labelled `GMA' on the figure) provides a good match to the stacked spectrum. Although there is structure within the residuals of Figure \ref{fig:stacking_test_1}, the RMS of these (0.12 \,$\Delta R_t/H$) is comfortably within the error of the stacked spectrum (0.27$\,\Delta R_t/H$), and the largest residuals tend to be in regions where both H$_2$O and CO$_2$ contribute to the opacity, as expected. This error was calculated by adding each planet's errorbars in quadrature and assuming that each planet has WASP-39b-like uncertainties in its transmission spectrum (0.47$\,\Delta R_t/H$ at $R = 600$, calculated from \citealt{Alderson2024}). We adopt WASP-39b as a representative high signal-to-noise benchmark system due to its large scale height, which makes it among the most favourable targets for transmission spectroscopy. Therefore, by choosing WASP-39b, we probe the regime in which deviations between the geometric mean abundance spectrum and the stacked spectrum become most apparent. In practice, the larger uncertainties typical of observed exoplanet spectra would increase the parameter space over which the geometric mean spectrum remains consistent with the stacked spectrum. From this simple three planet test of varying abundances but constant abundance ratios, the geometric mean of the abundances is an appropriate representation of the stacked spectrum. Random noise is not included in this test, since the focus is on demonstrating the robustness of the mathematical formalism with realistic opacities.

\subsection{Stacking different planets with identical compositions}
\label{sec:stacking_different_planets_identical_compositions}

Having demonstrated the appropriateness of stacking the same planet with different compositions in the simplified two-species case (H$_2$O and CO$_2$), here we stack different planets with identical compositions in the same two-species framework. Keeping the volume mixing ratios of water and carbon dioxide fixed to $\log X_{\mathrm{H_2O}} = -3.49$ and $\log X_{\mathrm{CO_2}} = -7.24$ respectively\footnote{These values are the solar volume mixing ratios of water and carbon dioxide at a temperature of 1500\,K.}, we now sequentially vary by $\pm50\%$ the planet's radius (0.5, 1, 1.5\,R$_{\rm{J}}$), mass (0.5, 1, 1.5\,M$_{\rm{J}}$) and temperature (750, 1500, 2250\,K). Unlike the first test, the representative planet's parameters ($T_R, R_R, M_R$) are not the same as the individual planets' parameters, since these are no longer constant. These representative parameters are what we use when constructing the GMA spectrum. Here we also set the reference pressure to the geometric means of the pressures of $\tau=1$ of the individual planets. The individual planets, on the other hand, have reference pressures set to 10\,bar. Ultimately, this choice made a negligible difference in the resulting spectrum as $H\ll R_p$. 

\begin{figure}
    \centering
    \includegraphics[width=1\linewidth]{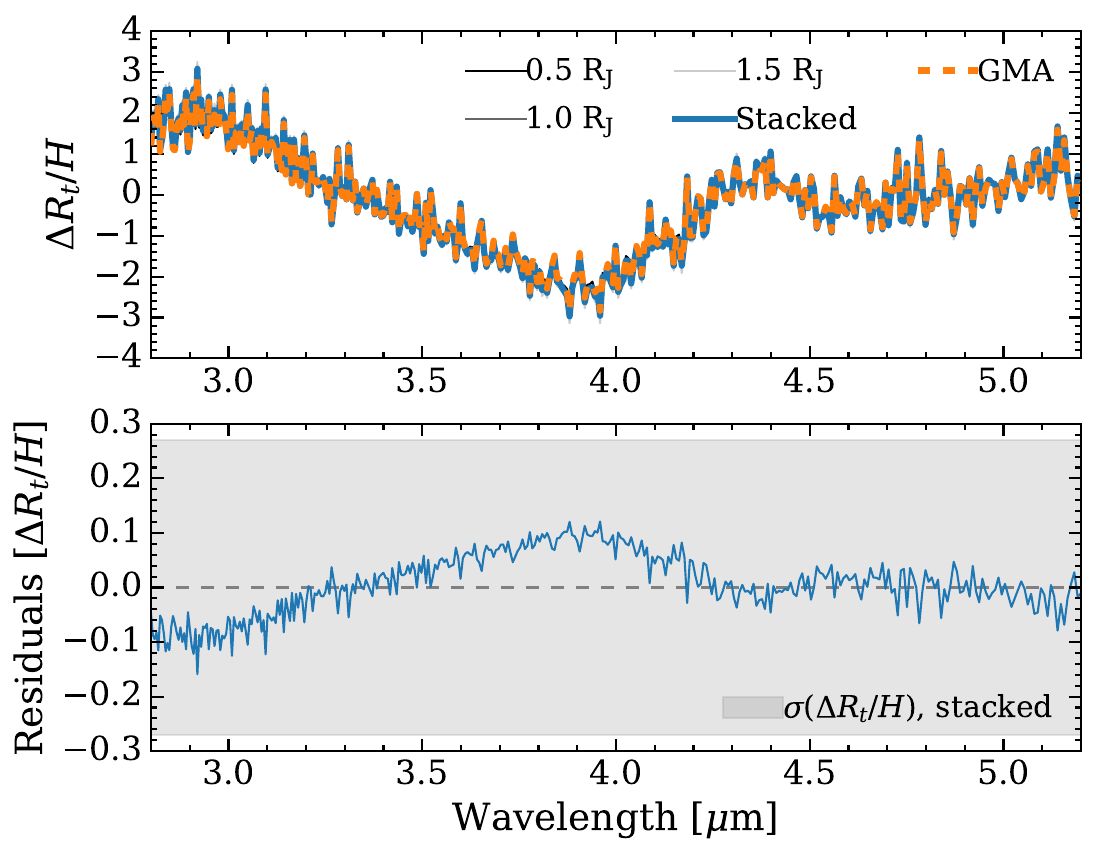}
    \includegraphics[width=1\linewidth]{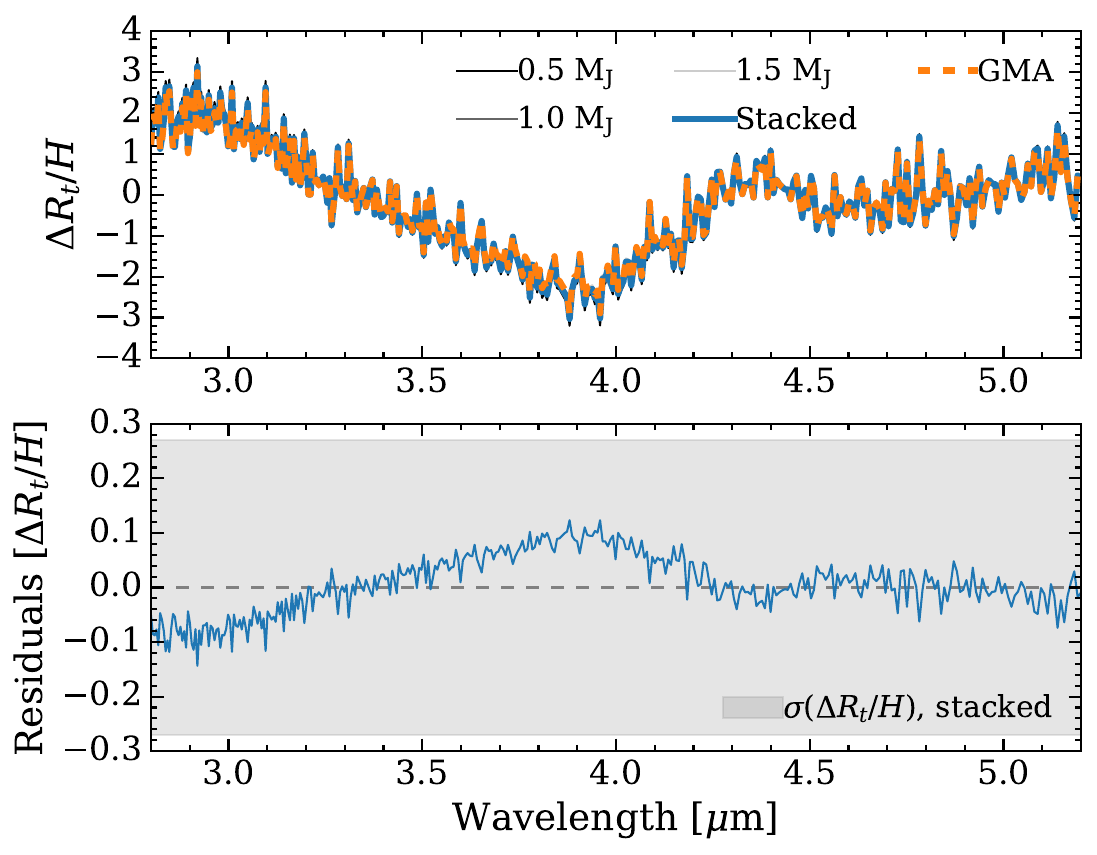}
    \includegraphics[width=1\linewidth]{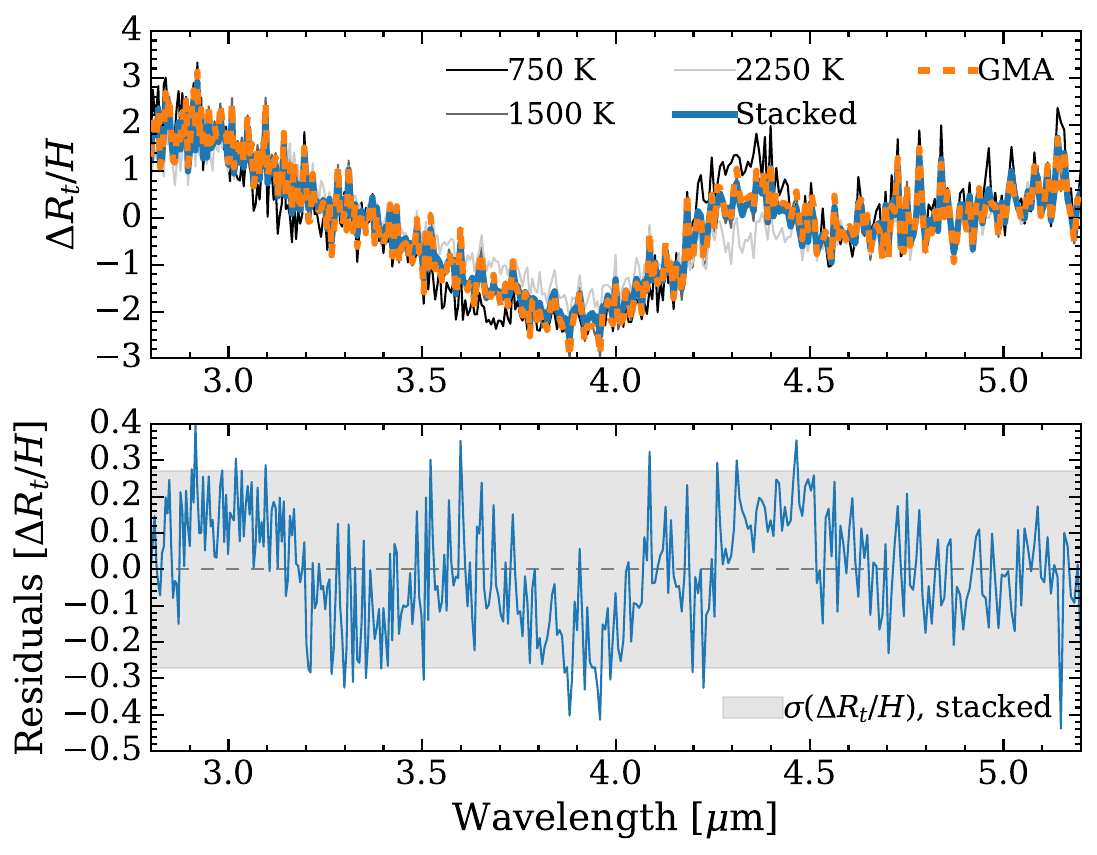}
    \caption{Testing the impact of varying a planet's radius (top plot), mass (middle plot) and temperature (bottom plot) by $\pm 50\%$ while keeping the planets' atmospheric compositions the same (solar abundances of H$_2$O and CO$_2$ as the only species). This demonstrates that variations in the temperature, not radius or mass, leads to the largest difference between the geometric mean abundance spectrum (orange) and the stacked spectrum (blue).}
    \label{fig:stacking_test_2}
\end{figure}

Figure \ref{fig:stacking_test_2} shows the impact of varying each of radius, mass and temperature in turn. Variations in radius and mass lead to similar small differences between the stacked spectrum and the geometric mean abundance spectrum, with a residual RMS of 0.05\,$\Delta R_t/H$. However, changing temperature leads to a larger difference between the geometric mean abundance spectrum and the stacked spectrum of 0.15\,$\Delta R_t/H$, but it is still smaller than the typical uncertainty in the stacked spectrum. In our assumptions, we took the opacities to be temperature insensitive; however, realistic opacities do show a temperature dependence, resulting in a difference. This highlights a fact we will demonstrate throughout the following tests, that stacking too wide a range of temperatures can lead to biases that are too large. 

\begin{figure}
    \centering
    \includegraphics[width=1\linewidth]{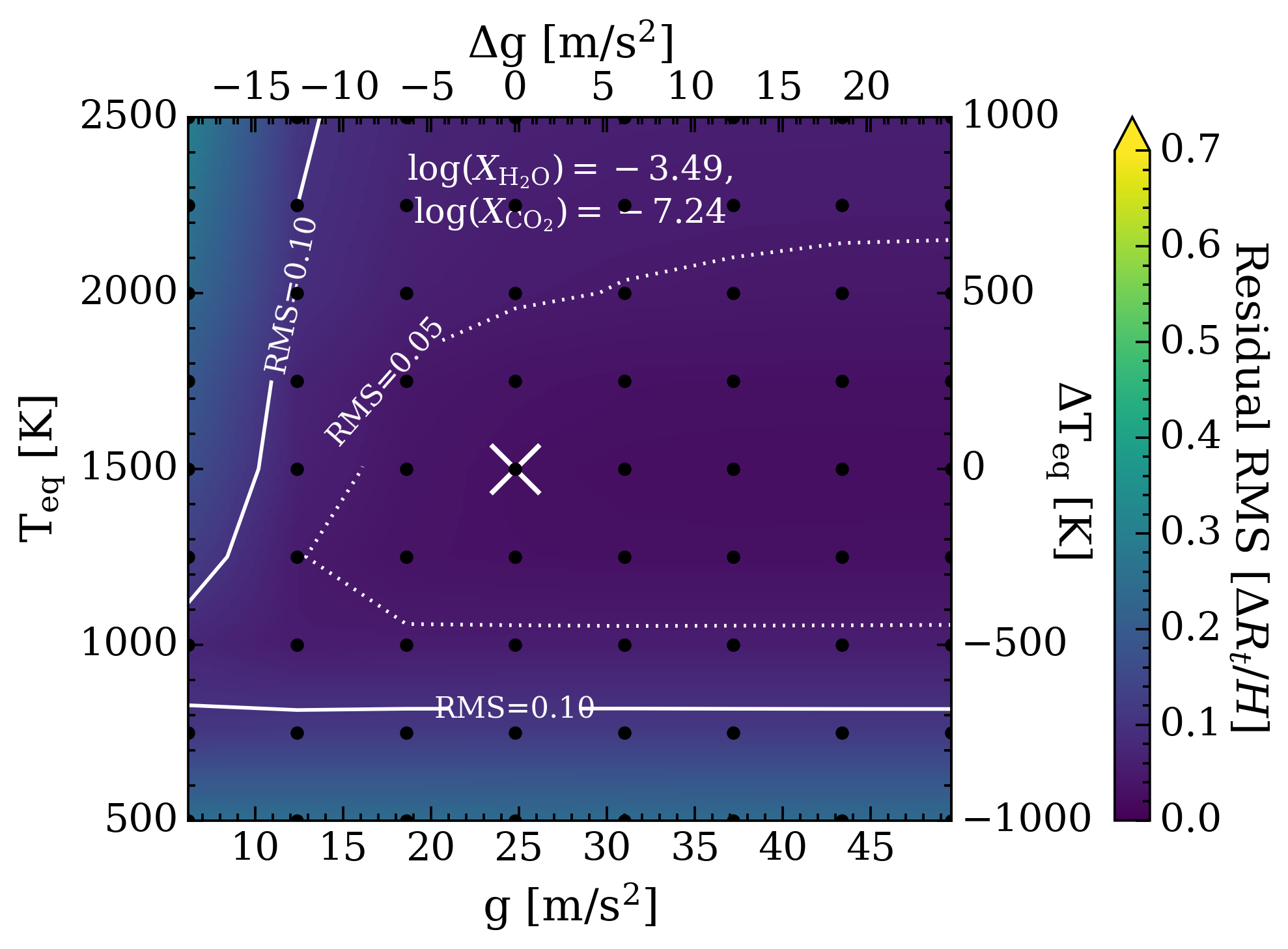}
    \caption{Appropriateness of stacking two planets with identical chemical abundances (H$_2$O and CO$_2$ as the only active species) across surface gravity ($g$) and equilibrium temperature ($\mathrm{T_{eq}}$). The reference planet is marked by the white cross and is paired with a second planet shown as black dots. The background colour shows the residual RMS between the stacked spectra and the geometric mean abundance spectra at $R=600$. White contours indicate RMS $=0.05$ (dotted) and $=0.1$\,$ \Delta R_t /H$ (solid). The RMS is never larger than the average uncertainty in the stacked spectrum ($ =0.3$ $\Delta R_t/H$). This illustrates that the geometric mean abundance spectrum closely reproduces the stacked spectrum for planets with identical abundances and two dominant opacity sources within this range of surface gravity and temperature.}
    \label{fig:two_planet_grid_identical}
\end{figure}

To investigate the range of planetary parameter space over which it is appropriate to stack planets, we explore a grid of masses (0.25 to 2.0\,M$_\mathrm{J}$, in steps of 0.25\,M$_\mathrm{J}$) and temperatures (500 to 2500\,K in steps of 250\,K), all with a fixed radius of 1\,R$_\mathrm{J}$. For each grid point, we stack two planets, one at the grid point's values of mass and temperature, and one which is a fixed reference planet at 1\,M$_\mathrm{J}$ and 1500\,K. For this test, the planets' atmospheres comprise only of H$_2$ and He, as bulk species, with the volume mixing ratios of H$_2$O and CO$_2$ held fixed to the same values used previously. Figure \ref{fig:two_planet_grid_identical} shows the result of this analysis. Within the range of planet parameter space explored, the RMS of the difference between the stacked spectrum and the geometric mean abundance spectrum is $<0.2$\,$\Delta R_t/H$, which is less than the uncertainty in the stacked spectrum (0.3\,$\Delta R_t/H$). 

In this test, since both stacked planets have identical compositions, the geometric mean abundance spectrum has identical compositions to both planets. Therefore, the difference between the stacked spectrum and the geometric mean abundance spectrum is purely driven by differences in the planets' masses and temperatures, and thus their scale heights. Therefore, this test demonstrates that the representative planetary parameters derived in Section \ref{sec:representative_planet} are the appropriate values to use when constructing the geometric mean abundance spectrum. In Appendix \ref{sec:cool_reference_planet} we re-run this test with a cooler reference planet (T$_{\mathrm{eq}} = 500$\,K), and a correspondingly wider range of temperature between the two stacked planets, and find that differences in temperature $> 1250$\,K lead to $>1\sigma$ differences between the geometric mean abundance spectrum and stacked spectrum.

\subsection{Stacking different planets with different compositions}
\label{sec:stacking_different_planets_different_compositions}

We now explore what happens when you stack two planets with different masses, temperatures, and chemical abundances. In addition to H$_2$O and CO$_2$, we now include CH$_4$ \citep{CH4}, CO \citep{CO}, H$_2$S \citep{H2S} and SO$_2$ \citep{SO2} as spectrally active species. We define the same reference planet as used above, which again always forms one planet in a two planet stack. Unlike the previous test where each planet in the grid had an identical H$_2$O and CO$_2$ abundance, on this occasion, we vary each planet's chemical abundances according to chemical equilibrium, which is computed using \texttt{FastChem} \citep{Stock2018,Stock2022,Kitzmann2024} that takes, as default, the solar values from \cite{Asplund2009}. Now, each grid point's temperature, and pressure, defines its chemical abundances. All planets within the grid have solar metallicity and solar C/O \citep[=0.59,][]{Asplund2021}\footnote{In \texttt{POSEIDON}, the abundances of all elements except carbon are determined by the metallicity, which is scaled from the solar composition of \cite{Asplund2009}. The carbon abundance [C/H] is then derived from the corresponding oxygen abundance [O/H] and the specified C/O ratio.}. 

\begin{figure}
    \centering
    \includegraphics[width=1\linewidth]{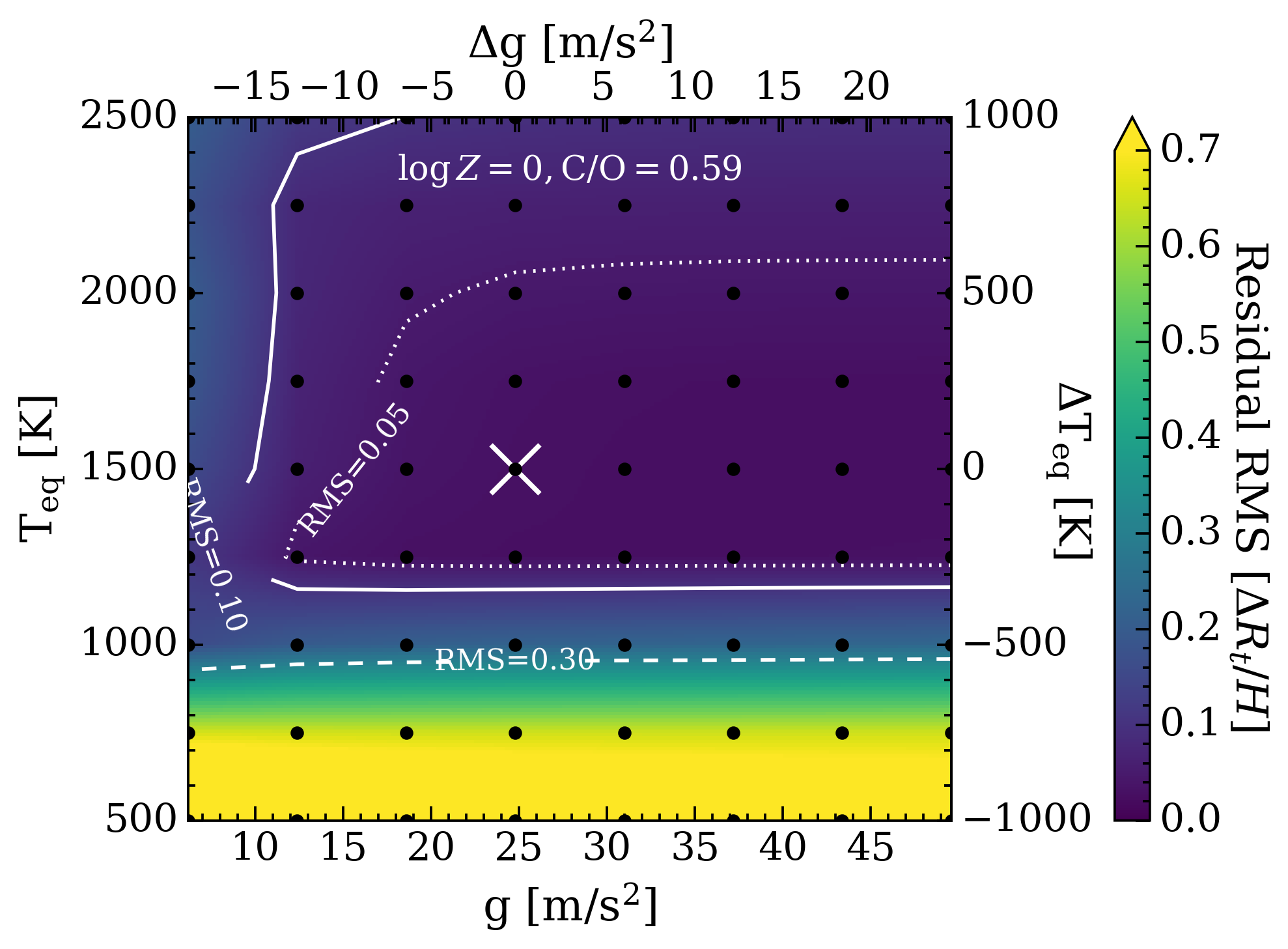}
    \includegraphics[width=1\linewidth]{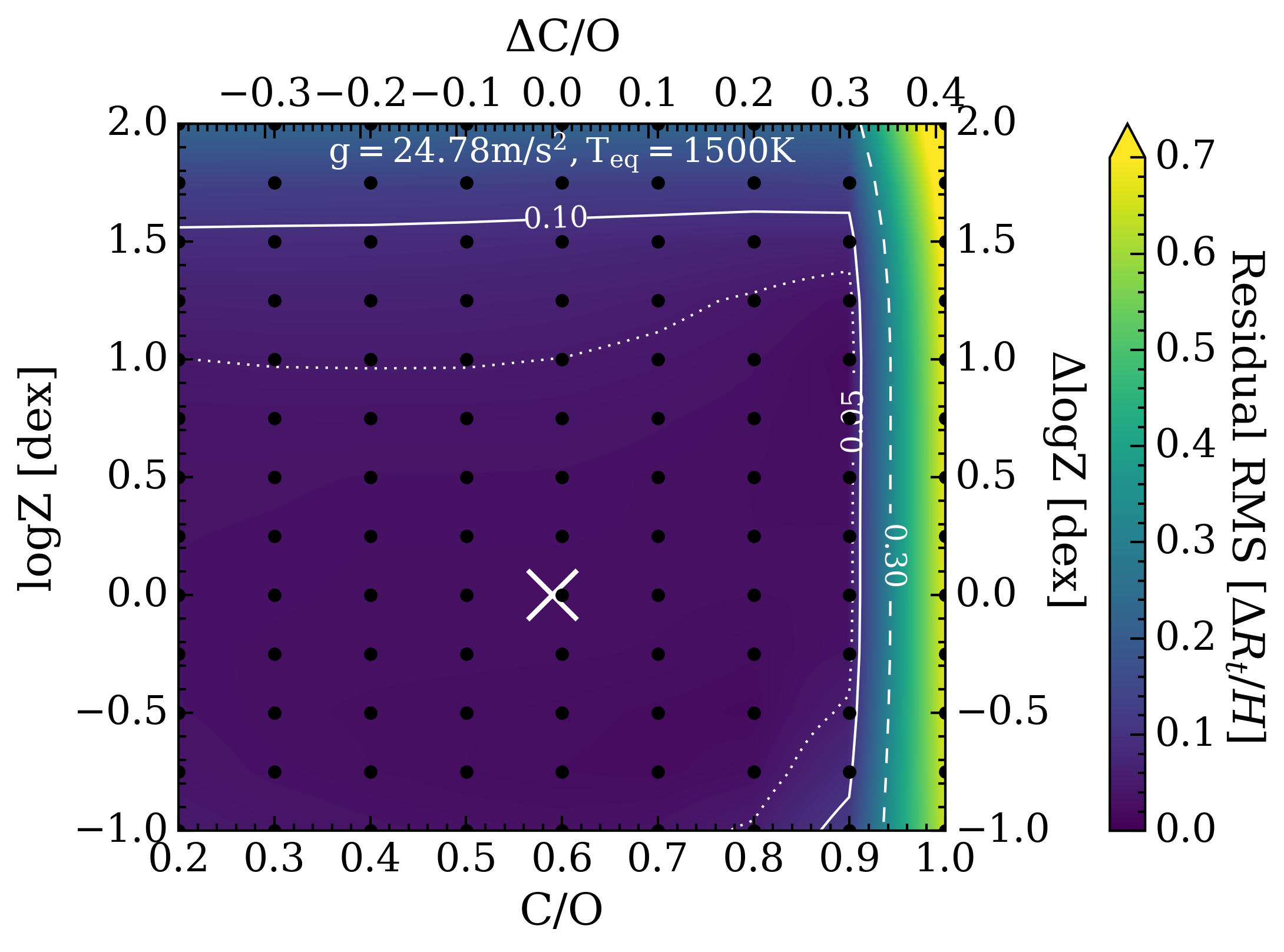}
    \caption{Appropriateness of stacking two planets with different chemical compositions. Top panel: Planets with solar $\log Z$ and C/O, including H$_2$O, CO$_2$, CH$_4$, CO, H$_2$S, and SO$_2$ as active species, varying in $g$ and $\mathrm{T_{eq}}$. Bottom panel: Planets with identical $g$ and $\mathrm{T_{eq}}$ but varying $\log Z$ and C/O. In both panels, the reference planet (white cross) is paired with a second planet (black dots), and colours indicate residual RMS between stacked spectra and geometric mean abundance spectra at $R=600$. White contours show RMS $=0.05$ (dotted), $=0.1$ (solid), and $=0.3$\,$\Delta R_t/H$ (dashed), representing the typical uncertainty in the stacked spectrum. Differences in dominant opacity species--particularly the rising prominence of methane at low temperatures or high C/O--lead to larger residuals when stacking with the reference planet ($\mathrm{T_{eq}}=1500$\,K, C/O$=0.59$).}
    \label{fig:two_planet_grid_variable}
\end{figure}

The results of this test are shown in Figure \ref{fig:two_planet_grid_variable} (top panel). As expected, the differences between the geometric mean abundance spectrum and stacked spectrum are larger than in the identical abundance test, particularly along the temperature axis. The most notable deviations (shown by the largest RMS values in the figure) occur at the lowest temperatures ($\lesssim 1000$\,K) where CH$_4$ plays an increasingly prominent role within the second planet's spectrum. Indeed, Figure \ref{fig:two_planet_grid_variable} shows that even if two planets have identical metallicity and C/O, it is inappropriate to stack their spectra if their temperatures span the CH$_4$/CO boundary. 

Similarly to what was seen above, the appropriateness of stacking two planets' spectra is relatively insensitive to differences in their surface gravities. In this case, none of the surface gravities that were tested -- which correspond to planet masses between 0.25--2\,M$_\mathrm{J}$ at fixed radius (1\,R$_\mathrm{J}$) -- resulted in differences between the geometric mean abundance spectrum and stacked spectrum that were larger than the uncertainties in the stacked spectrum (0.3\,$\Delta R_t/H$). 

Our third grid-based test was to fix surface gravity (24.78\,m\,s$^{-2}$) and equilibrium temperature (1500\,K) and vary $\log Z$ from -1 to 2 in steps of 0.25 and C/O from 0.2 to 1 in steps of 0.1. The results of this test are shown in the bottom panel of Figure \ref{fig:two_planet_grid_variable}. In this case, the residual RMS between the stacked spectrum and the geometric mean abundance spectrum is $<0.1 \Delta R_t/H$ across the majority of $\log Z$--C/O space other than where C/O approaches unity, whereby the H$_2$O and CO$_2$ abundances drop sharply in favour of CH$_4$ and CO. Like the temperature test, therefore, stacking breaks down when one planet has prominent CH$_4$ absorption and the other does not.  However, given that planet formation theory predicts $\mathrm{C/O} < 1$ for close-in planets in the majority of cases \citep[e.g.,][]{Penzlin2024}, changes in temperature, and not C/O, should be the primary consideration when stacking different exoplanets' spectra.

\begin{figure}
    \centering
    \includegraphics[width=1\linewidth]{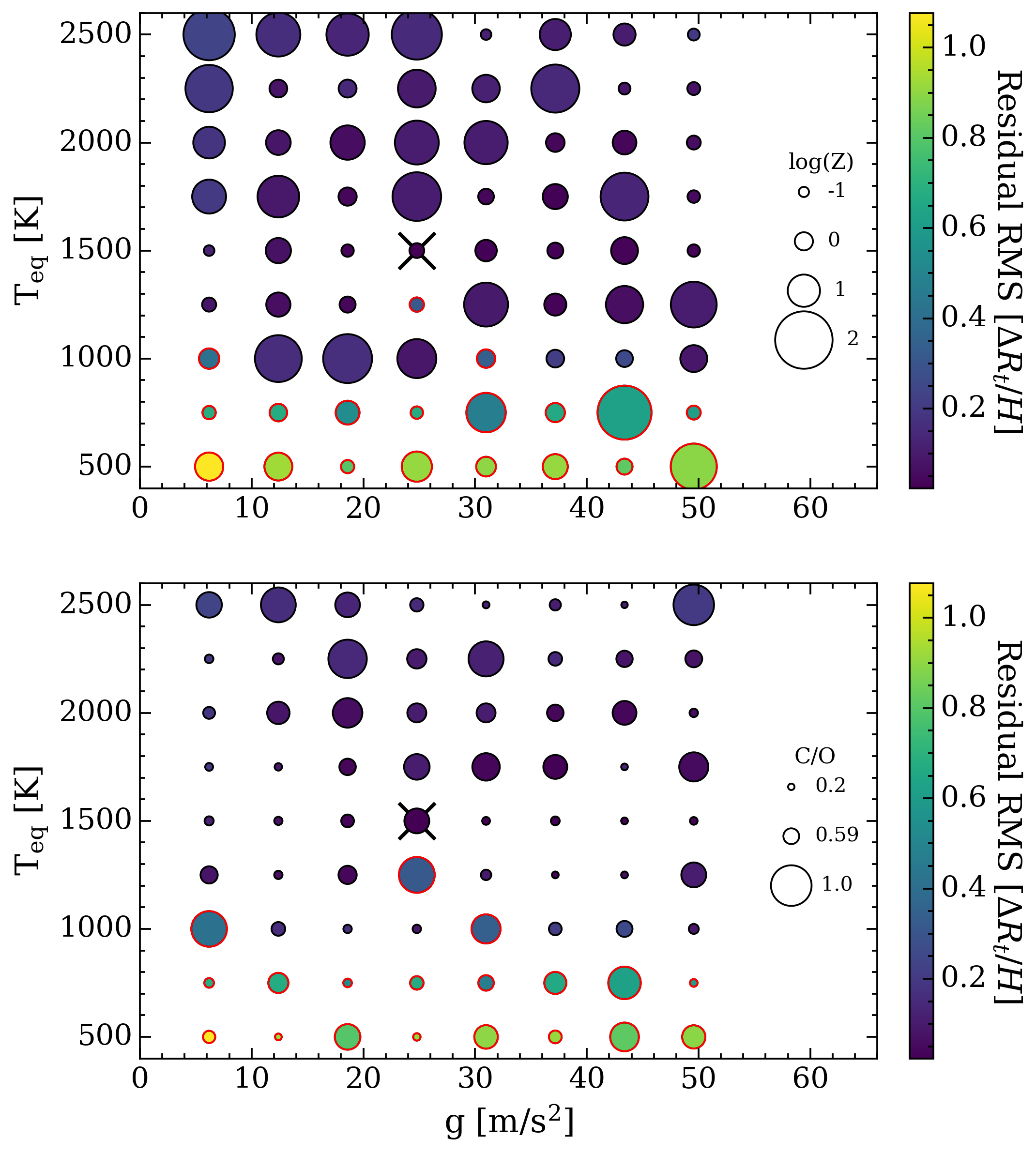}
    \caption{Figures showing the appropriateness of stacking two planets with different temperatures, surface gravities and different compositions. The difference to the bottom panel of Figure \ref{fig:two_planet_grid_variable}, is that one of the planets no longer has a solar Z and C/O but instead these values are randomly selected. The black cross shows the surface gravity and temperature of the planet that always forms the first planet in the stack, with a solar Z and C/O. The sizes of the symbols show the randomly selected $\log Z$ (top panel) and C/O (bottom panel) of the second planet, whose surface gravities and temperatures are sampled across the grid. The colours of the circles indicate the RMS of the differences between the two-planet stacked spectrum and the spectrum generated from the geometric mean of the abundances. Circles with a red outline have an RMS value larger than the uncertainty in the stacked spectrum (0.3\,$\Delta R_t/H$) and thus represent where the geometric mean abundance spectrum breaks down.}
    \label{fig:two_planet_grids_different_Z_and_CtoO}
\end{figure}

In Figure \ref{fig:two_planet_grids_different_Z_and_CtoO} we add further flexibility by allowing the second planet's gravity, temperature, metallicity and C/O to all vary. Gravity and equilibrium temperature were varied across the same grid defined above while $\log Z$ was randomly selected from -1 to 2 and C/O from 0.2 to 1 for each grid point in $g - \mathrm{T_{eq}}$ space. This range of compositions spans a broad set of atmospheric chemistries, including those that may be inherited from variations in host star abundances. This figure reiterates the importance of temperature when stacking planets, while also showing that large differences in C/O increase the RMS between the stacked and geometric mean abundance spectra. Again, this figure demonstrates that differences in surface gravity play a smaller role. Taken together, in the small-$N_p$ limit, these results indicate that stacking is only effective when confined to regions of parameter space where the chemistry produces the same dominant absorbers; otherwise, the stacked spectrum does not provide a physically meaningful representation of the individual planets. We emphasize that these results depend on our assumption of equilibrium chemistry. We discuss the potential effects of disequilibrium processes in Section \ref{sec:when_is_stacking_useful}.

\subsection{Extending to additional planets}
\label{sec:additional_planets}

So far we have considered the simple cases of stacking two and three planets with regularly spaced parameters and compositions (Figures \ref{fig:stacking_test_1}, \ref{fig:stacking_test_2}, \ref{fig:two_planet_grid_identical}, \ref{fig:two_planet_grid_variable}) and two planets with compositions drawn at random (Figure \ref{fig:two_planet_grids_different_Z_and_CtoO}). Now we extend to stacking 5 planets, each with a randomly drawn surface gravity $g$, T$_{\mathrm{eq}}$, $\log Z$ and C/O from the same range of parameter space defined above (hot Jupiters). This represents the more likely scenario whereby, in reality, one would stack multiple planets with different parameters and compositions, the latter of which are not known \textit{a priori}. 

\begin{figure*}
    \centering
    \includegraphics[width=0.49\linewidth]{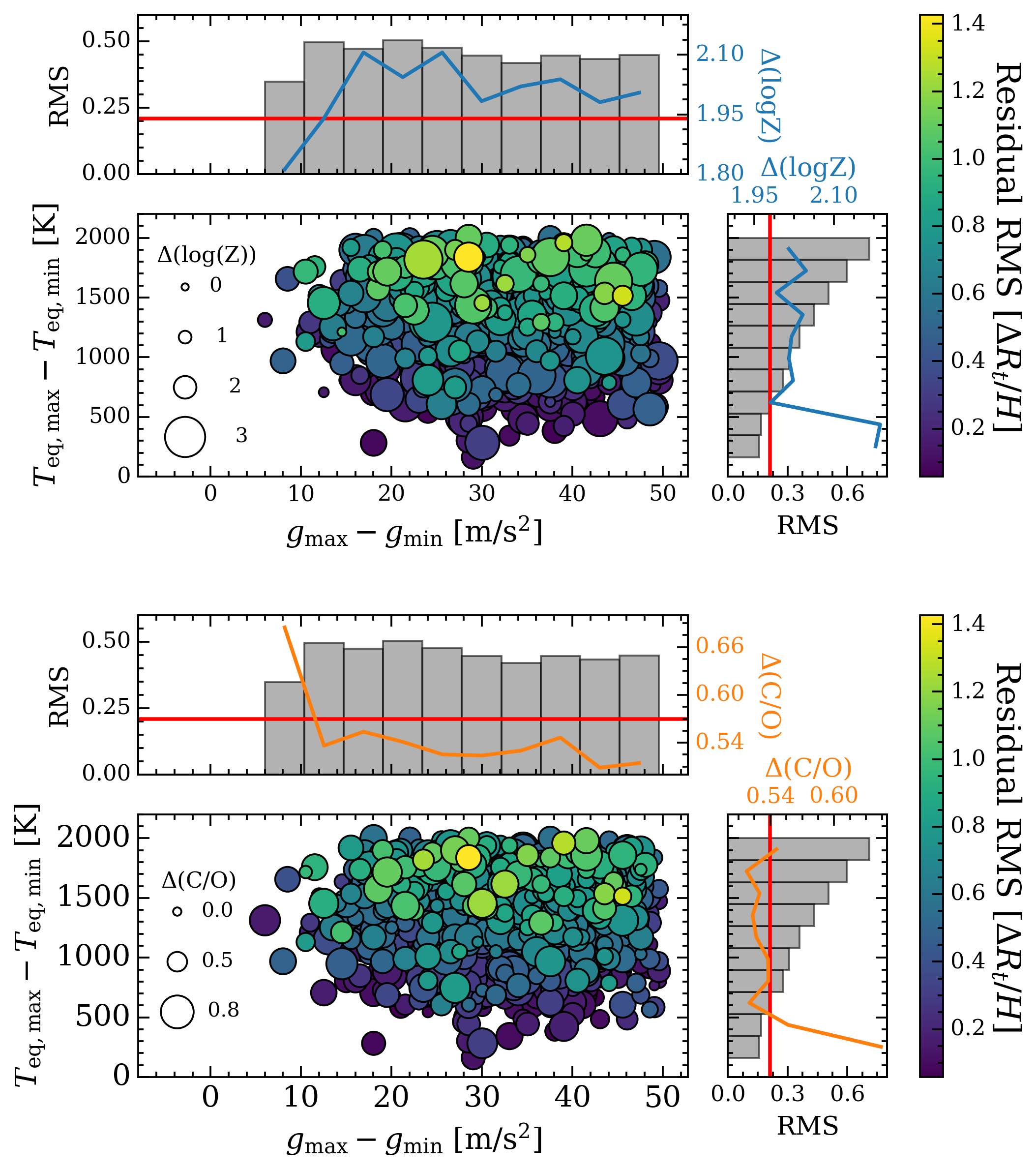}
    \includegraphics[width=0.49\linewidth]{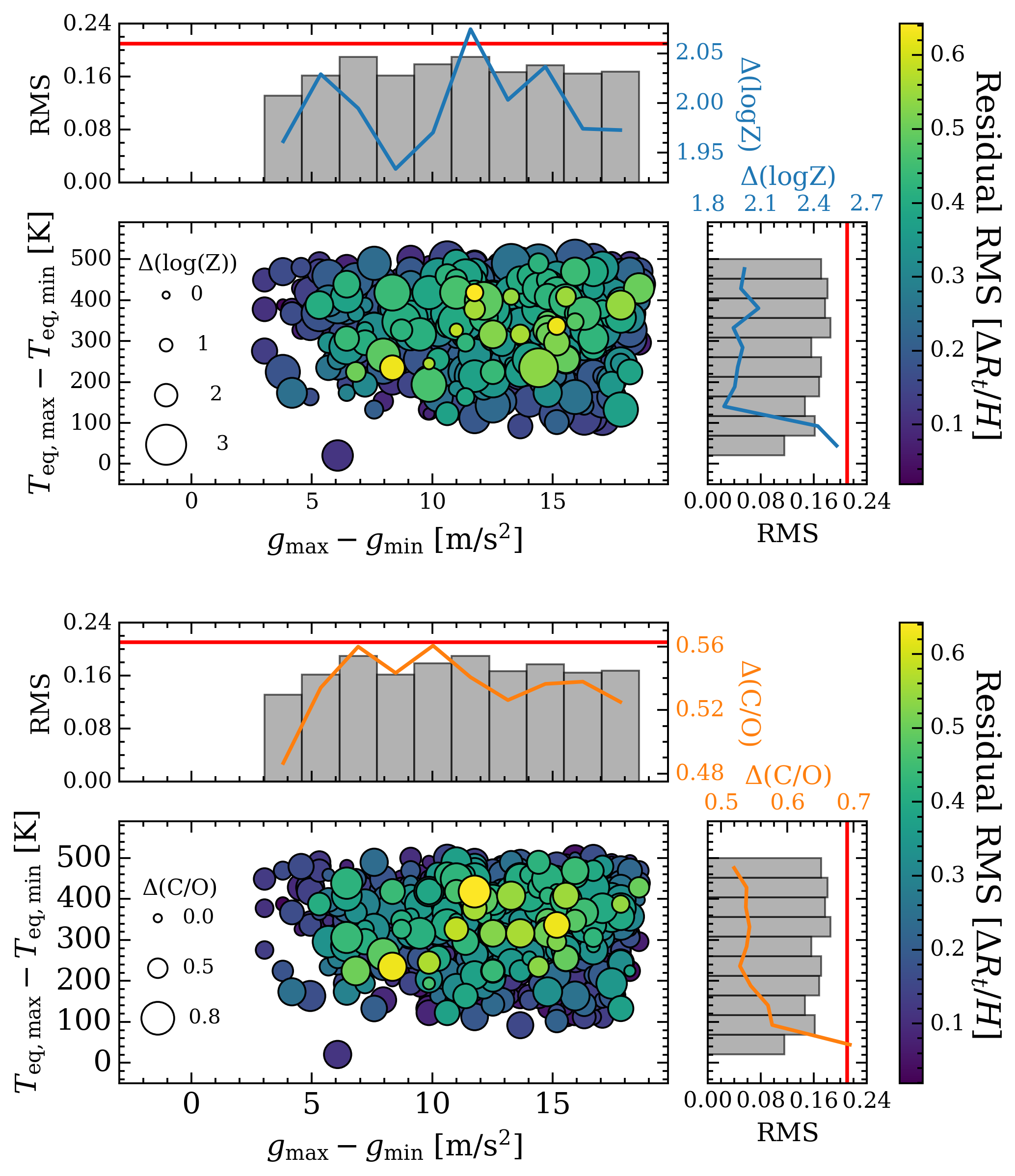}
    \caption{The result of stacking five randomly selected planets 1000 times, drawn from either a wide temperature range (500--2500 K; left column) or a narrow range (1200--1700 K; right column). For each draw, planets have randomly assigned values of surface gravity ($g$), equilibrium temperature ($T_\mathrm{eq}$), metallicity ($\log Z$), and C/O ratio. The x- and y-axes show the range in $g$ and $T_\mathrm{eq}$ within each five-planet sample. The symbol size indicates the range in $\log Z$ (top row) and C/O (bottom row), while the colours correspond to the RMS difference between the five-planet stacked spectrum and the spectrum generated from the geometric mean of their abundances, in units of $\Delta R_t/H$. The grey histograms show the mean RMS collapsed along $g$ (top) and $T_\mathrm{eq}$ (right). Blue and orange lines show the mean $\Delta(\log Z)$ and $\Delta$(C/O), respectively, along the same axes. The red line marks the uncertainty in the five-planet stacked spectrum ($= 0.21 \Delta R_t/H$). These results demonstrate that the geometric-mean abundance spectrum accurately reproduces the stacked spectrum when the temperature range within the sample is $\lesssim600$ K and the variation in C/O is moderate.}
    \label{fig:5_planet_grid}
\end{figure*}

For this test, we draw 5 random planets 1000 times and compute the RMS between the stacked spectrum and geometric mean abundance spectrum for each of the 1000 iterations. We show the results in Figure \ref{fig:5_planet_grid} (left column), where the accompanying histograms are particularly informative. These histograms represent the RMS of the difference between the five planet stacked spectra and geometric mean abundance spectra, collapsed along surface gravity and temperature. The larger RMS values when collapsing along surface gravity arise because each $g$ bin includes planets spanning a wide range of $\mathrm{T_{eq}}$, which, as we showed above, can cause large residuals between the geometric mean abundance spectrum and the stacked spectrum. When we instead collapse along $\Delta(\mathrm{T_{eq}})$, we recover a clear positive correlation between RMS and $\Delta(\mathrm{T_{eq})}$, despite each $\Delta(\mathrm{T_{eq}})$ bin spanning a range of surface gravities, highlighting the sensitivity of stacking to $\mathrm{T_{eq}}$ and its relative insensitivity to $g$.

This test reveals that the RMS is only less than the uncertainty in the stacked spectrum ($=0.21 \Delta R_t/H$, shown by the red lines on the histograms) when the range of the five planets' temperatures is $\lesssim 600$\,K. To verify this, we repeated this test with a narrower range of allowed temperatures (1200--1700\,K) and found that the RMS between the geometric mean abundance spectrum and the stacked spectrum was smaller than the spectral uncertainties in all cases (Figure \ref{fig:5_planet_grid}, right column). This confirms the results of our simplified models in a more realistic context: stacking can be highly effective, but only within sensible ranges of temperature and composition.

\subsection{Stacking sub-Neptunes}

Another motivation behind stacking different exoplanets' spectra will be to determine the overall population-level characteristics of sub-Neptune exoplanets, whose transmission spectra to date are often featureless or muted and consistent with either high-metallicities, aerosols at low pressures, or some combination of the two (e.g., \citealt{kempton2023reflective,schlawin2024possible,Wallack2024,Wallack2026,Ahrer2025_GJ3090b,Ohno2025,Teske2025}; although \citealt{Owen2026} propose that clumpy aerosol distributions can result in flat spectra in low-metallicity atmospheres without requiring large aerosol particles at low pressures -- a scenario which isn't modelled here). With 22 sub-Neptunes (1.8--3.5\,R$_\oplus$) already observed or scheduled to be observed in transmission with JWST, the stacking of sub-Neptune spectra could begin in earnest.

To test the suitability of stacking sub-Neptune spectra, we began by performing a similar test to that done in Section \ref{sec:additional_planets}. Specifically, we stacked five random sub-Neptunes 1000 times across $g$--T$_{\rm eq}$ space to determine the phase space over which stacking five planets is appropriate. For each planet, we randomly drew its mass between 3--10\,M$_\oplus$ and calculated the corresponding radius using the sub-Neptune mass-radius relation from \cite{Rogers2023} (their equation 5). This range corresponds to radii of approximately 1.8--2.5\,R$\oplus$, placing the planets just above the radius valley where thin H$_2$/He envelopes are expected \citep[e.g.,][]{Fulton2017,Owen2017} and atmospheric signals are comparatively weak, providing a natural motivation for stacking. To model this scenario, we defined H$_2$ and He as the bulk species, fixed the metallicity to $1000\times$ solar, C/O to 0.59 and added an opaque, grey cloud deck at a pressure of 10$^{-4}$\,bar to enforce small amplitude features\footnote{We also experimented with a cloud deck at 10$^{-6}$ bar but this gave such flat spectra that the geometric mean abundance spectrum and stacked spectrum were always approximately equal.}. We drew the planet's temperatures at random between 400--1000\,K, meaning this test again spans the CO/CH$_4$ transition. 

\begin{figure}
    \centering
    \includegraphics[width=1\linewidth]{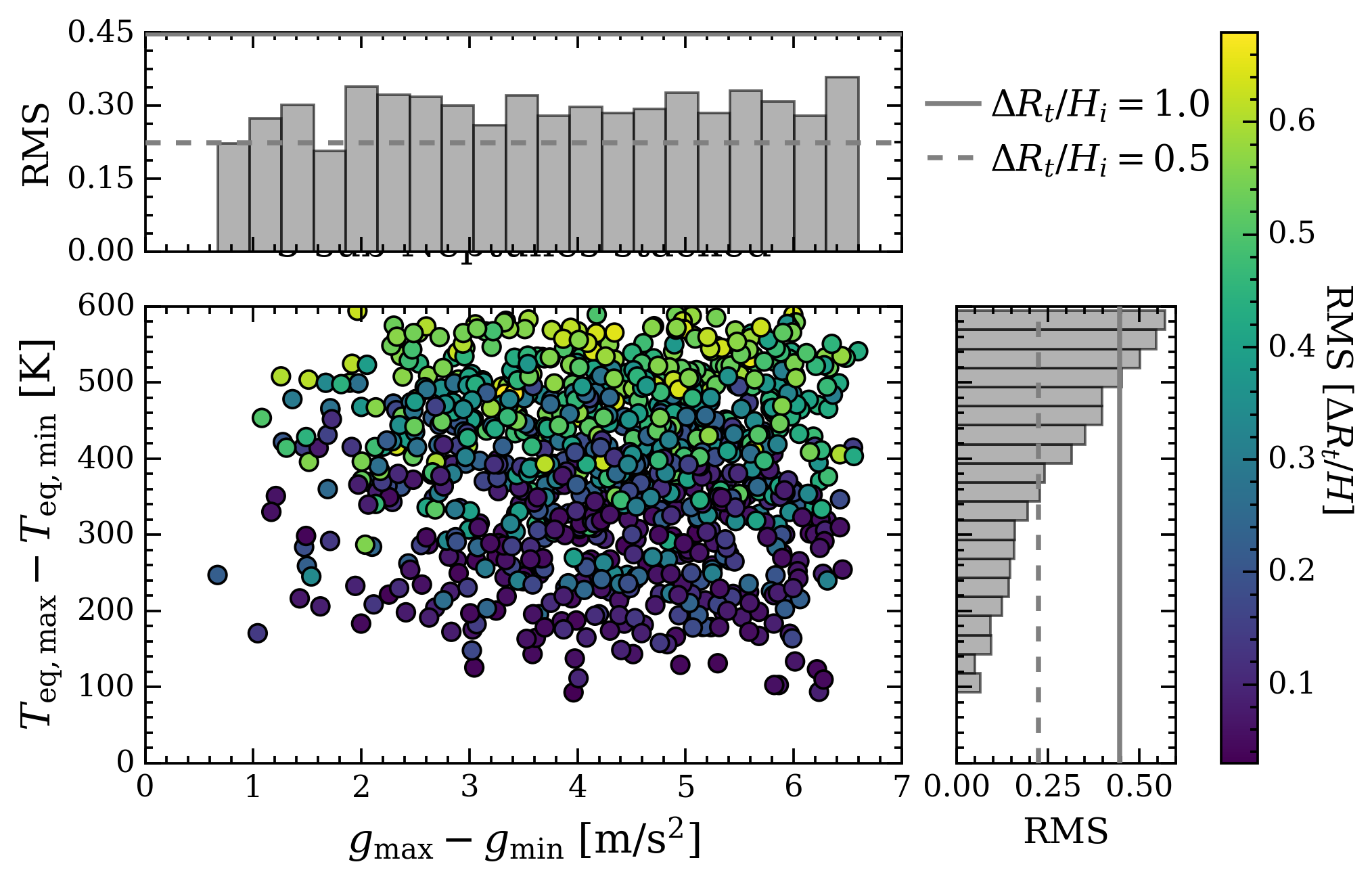}
    \caption{The result of stacking five randomly selected sub-Neptunes 1000 times. For each draw, planets have randomly assigned values of surface gravity ($g$) and equilibrium temperature ($T_\mathrm{eq}$), but fixed high metallicity atmospheres ($1000\times$ solar) with low pressure cloud decks ($10^{-4}$ bar) to ensure small amplitude features in the individual spectra. The x- and y-axes show the range in $g$ and $T_\mathrm{eq}$ within each five-planet sample. Colours indicate the RMS difference between the five-planet stacked spectrum and the spectrum generated from the geometric mean of their abundances, in units of $\Delta R_t/H$. The grey histograms show the mean RMS collapsed along $g$ (top) and $T_\mathrm{eq}$ (right). The grey dashed/solid lines marks the uncertainty in the five-planet \textit{stacked} spectrum when the uncertainties in the \textit{individual} spectra are 0.5/1\,$\Delta R_t/H$, where 1\,$\Delta R_t/H \approx 20$\,ppm.}
    \label{fig:5_planet_grid_sub-Neptunes}
\end{figure}

We show the results of this experiment in Figure \ref{fig:5_planet_grid_sub-Neptunes}. This figure demonstrates that the geometric mean abundance spectrum accurately reflects the stacked spectrum when the range of temperatures within the five planet sample is $\lesssim$ 600 \,K if the individual planets' spectra are measured to a precision of 1\,$\Delta R_t/H$ at a resolution of $R=200$, which was chosen to match typical observations \citep[e.g.,][]{Teske2025}. For context, across all 5000 generated sub-Neptunes, the average scale height corresponded to a transit depth of 20\,ppm for these H$_2$/He-dominated,  $1000\times$ solar metallicity atmospheres with a mean molecular weight of 5.0\,g\,mol$^{-1}$. For a comparison with observed data, the spectrum of the sub-Neptune TOI-776c was measured with JWST NIRSpec/G395H to an average precision of 25\,ppm at a resolution of $R=200$ \citep{Teske2025}. Thus, an uncertainty of 1\,$\Delta R_t/H$ is commensurate with typical measured precisions of sub-Neptune spectra. We note that in this test we did not include the effect of random noise (since, again we wish to assess the accuracy of using the geometric mean spectrum), although we explore some of the effects of this in the following.  

Since a primary objective of stacking muted sub-Neptune spectra will be to make confident detections of atmospheric features, we next tested how many planets would need to be stacked to rule out a flat line to a given sigma confidence. For this test, we performed 100 iterations of sequentially stacking 1--100 planets with masses again drawn at random between 3--10\,$M_\oplus$, radii drawn from the mass-radius relation described above, stellar radii fixed to 0.5\,$R_\odot$ and equilibrium temperatures drawn at random from 400--1000\,K. The use of wide temperature bounds in this test means that the geometric mean abundance spectrum will not always be an accurate reflection of the stacked spectrum, as we demonstrated above. However, since this accuracy depends on the precision of the individual planetary spectra and the, as yet untested, size of the sample we proceed with the wide temperature bound in this test since here we are interested in whether any absorption features are visible at $>5\sigma$. For this experiment, we again fixed metallicity to $1000\times$ solar and C/O to 0.59 but on this occasion, we ran three tests with differing cloud top pressures of $10^{-4}$, $10^{-5}$ and $10^{-6}$ bar. For this test, we also included random noise in the generation of our planetary spectra, varying noise between 0.5, 1.0 and 2.0 $\Delta R_t/H$ ($\sim 10, 20, 40$\,ppm at $R=200$). 

\begin{figure}
    \centering
    \includegraphics[width=1\linewidth]{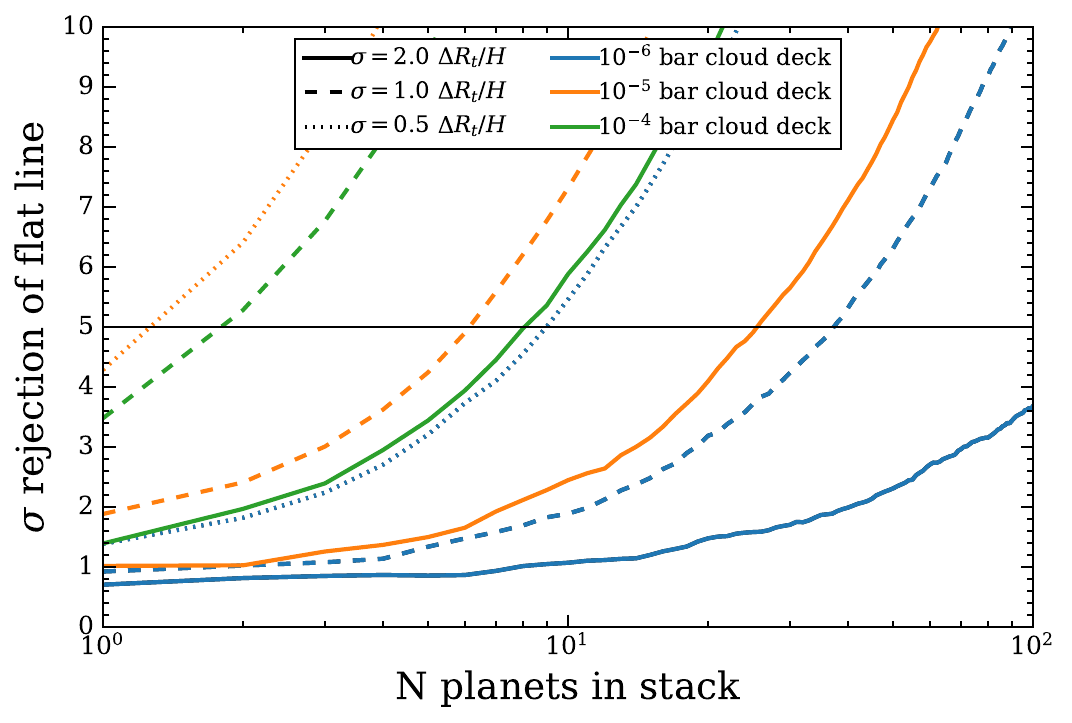}
    \caption{The number of \textit{different }sub-Neptunes that would need to be stacked in order to rule out a flat line at a given $\sigma$ confidence with the JWST NIRSpec/G395H instrument mode. This is shown for different pressures of a grey, opaque cloud deck ($10^{-6}$, blue; $10^{-5}$, orange; $10^{-4}$ bar, green) and different uncertainties in the individual planets' transmission spectra ($\Delta R_t/H = 0.5$, dotted; 1, dashed; 2, solid; $\approx 10$, 20, 40\,ppm).}
    \label{fig:sigma_rejection}
\end{figure}

Figure \ref{fig:sigma_rejection} shows the sigma rejection of a flat line as a function of the number of planets that are stacked. For a $10^{-4}$ bar cloud top pressure, a flat line can be rejected to $>5\sigma$ with nine planets if each planet has an uncertainty in its transmission spectrum of 2\,$\Delta R_t/H$, or with two planets if each has an uncertainty of 1\,$\Delta R_t/H$. For smaller uncertainties of 0.5 $\Delta R_t/H$, $\approx 10$\,ppm, a single planet would reject the flat line at $> 5 \sigma$ negating the need to stack. For a 10$^{-5}$ bar cloud top pressure, seven sub-Neptunes would need to be stacked to rule out a flat line to $>5\sigma$ if the uncertainty on each is 1\,$\Delta R_t/H$, while for a 10$^{-6}$ bar cloud top pressure, 38 planets would need to be stacked. Figure \ref{fig:sigma_rejection} also shows how many planets would need to be stacked for per-planet spectral uncertainties of 0.5 and 2\,$\Delta R_t/H$. We note that if sub-Neptunes have lower metallicities than the $1000\times$ solar considered in this test, fewer planets would need to be stacked even if they have low pressure cloud decks. 

\begin{figure}
    \centering
    \includegraphics[width=1\linewidth]{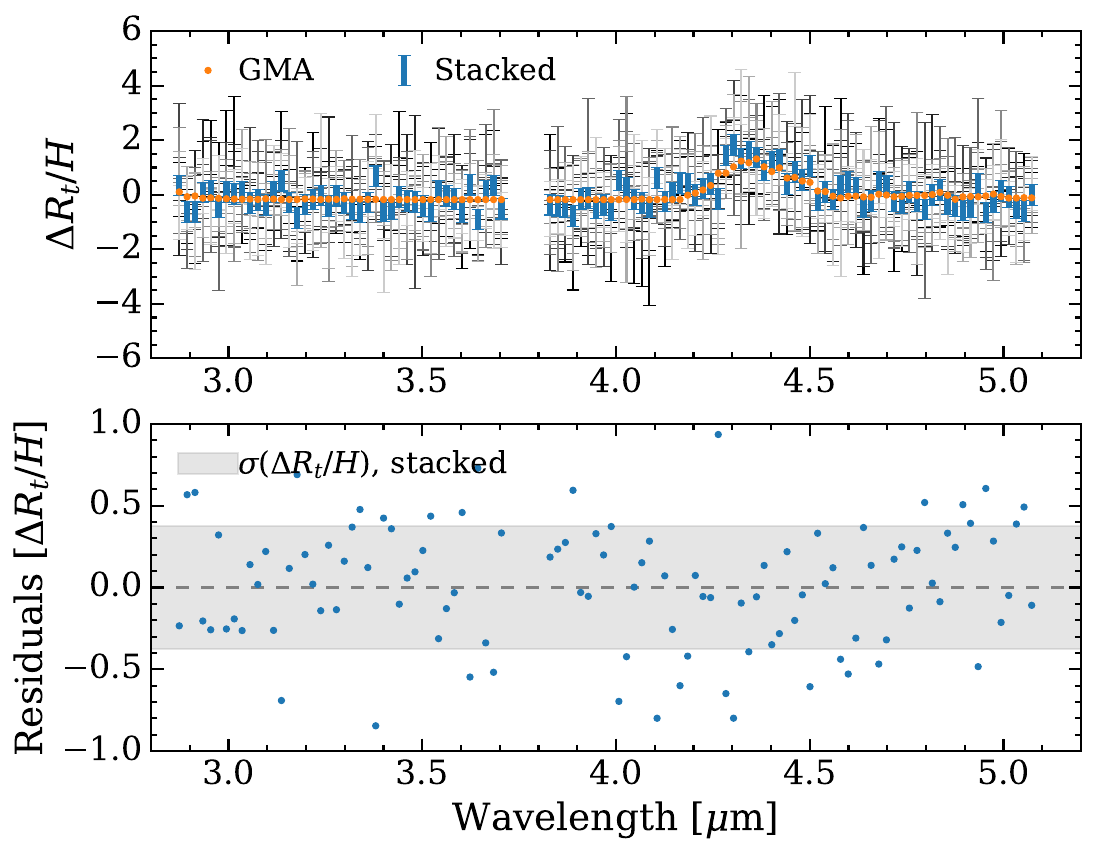}
    \caption{The result of stacking \textit{different} sub-Neptunes' spectra. Top panel: the seven randomly generated sub-Neptunes' spectra are shown in grey, each with a $10^{-5}$ bar grey cloud deck, a $1000\times$ solar metallicity atmosphere, and a per-spectral-bin uncertainty of 1\,$\Delta R_t/H$ at a resolution of $R=200$. The stacked spectrum is shown in blue, with the spectrum generated from the geometric mean of the seven planets' abundances (`GMA') in orange. The stacked spectrum reveals significant CO$_2$ absorption that is able to rule out a flat line at $6.5\sigma$. Bottom panel: the difference between the stacked and geometric mean abundance spectra (blue points) has a smaller RMS than the size of the stacked spectrum's uncertainties (grey shaded region).}
    \label{fig:stacked_sub_Neptune_spectrum}
\end{figure}

To demonstrate the utility of stacking \textit{different} sub-Neptunes, we apply the findings above to show a stacked sub-Neptune spectrum in Figure \ref{fig:stacked_sub_Neptune_spectrum}. Here, we stack seven randomly generated sub-Neptune spectra each with a cloud deck at $10^{-5}$ bar and a per-spectrum uncertainty of  1\,$\Delta R_t/H$, since we showed in Figure \ref{fig:sigma_rejection} that this would allow a flat line to be rejected to $>5\sigma$. We also include the effects of random noise. We draw the planets' temperatures at random between 450--950\,K, since we showed a 500\,K range is appropriate for $\Delta R_t/H = 1$ in Figure \ref{fig:5_planet_grid_sub-Neptunes}. In Figure \ref{fig:stacked_sub_Neptune_spectrum}, the seven planet stacked spectrum reveals a significant CO$_2$ absorption feature that is able to rule out the flat line to $6.5\sigma$. Furthermore, the RMS of the difference between the geometric mean abundance spectrum and the stacked spectrum is $< 1\sigma$. Thus, Figure \ref{fig:stacked_sub_Neptune_spectrum} is an example of where stacking \textit{different} sub-Neptunes is both appropriate and beneficial to making an atmospheric detection in the presence of low pressure clouds ($10^{-5}$ bar) and a high metallicity atmosphere ($1000 \times$ solar).

These results suggest, then, that stacking sub-Neptune spectra will provide additional constraints on the high-altitude aerosol/high metallicity degeneracy. We emphasize, however, that our tests only consider the wavelength range of  NIRSpec/G395H (3--5\,$\mu$m) and grey clouds, as are commonly assumed in the JWST literature due to the precision of existing spectra \citep[e.g.,][]{Lustig-Yaeger2023,Meech2026,Wallack2024,Wallack2026}. More realistic cloud microphysics or photochemical hazes may introduce wavelength-dependent spectral structure that is not captured by a grey cloud model, that could increase spectral diversity across a population of planets and impact the phase space over which stacking is appropriate. The inclusion of short-wavelength data, for example from NIRISS/SOSS (0.6--2.8\,$\mu$m), would provide sensitivity to such non-grey behaviour while also allowing for a more robust means of disentangling cloud properties from atmospheric composition in stacked spectra. A detailed exploration of non-grey clouds and photochemical hazes is beyond the scope of this work, but represents an important direction for future studies, particularly in the context of sub-Neptune atmospheres.

Furthermore, while we have focussed on CO$_2$ as a representative trace species in our sub-Neptune tests, the same framework can be applied to other molecular absorbers such as CH$_4$, NH$_3$, OCS and CS$_2$. In these cases, stacking may enhance the detectability of weak spectral features that provide limited statistical significance in individual spectra, such as OCS in GJ\,1214b \citep{Ohno2025} and CS$_2$ in TOI-270d \citep{Benneke2024,Holmberg2024}, provided that the planets being combined occupy a sufficiently similar temperature and chemical regime for the relevant opacity sources to remain consistent across the sample.

\section{Discussion}
\label{sec:discussion}

The ability to use stacking of different planet's transmission spectra to increase the signal-to-noise ratio significantly more rapidly than could be done by stacking multiple transits of an individual planet makes it a compelling tool. However, one must be careful that stacking different planets provides a physically meaningful result. This is particularly important because, at least with JWST, stacking is likely to take place amongst a handful of planets. In this work, we have shown that there is a pathway to stacking that provides useful, population-level inferences. That is, a stacked spectrum in relative transit depths, scaled by the atmospheric scale height ($\Delta R_t/H$), is described by the geometric mean of the opacity ratios in the region of the atmosphere probed in transmission. Given the direct connection between chemical abundances and opacities, in low-resolution spectra, the stacked spectrum is representative of a spectrum given by a ``representative planet'' containing a geometric mean of the abundance ratios. The representative planetary parameters are defined so that the pressures and temperatures are the same in the region of the atmosphere probed by transmission spectra. Thus, stacking is potentially powerful because it allows us to combine planetary spectra and make more precise inferences. 

\subsection{When is stacking useful?}
\label{sec:when_is_stacking_useful}

Since stacking in the small number of planets limit will always be sensitive to outliers, stacking spectra where the dominant absorbers at a specific wavelength change is ultimately going to lead to meaningless results. Therefore, one must be careful to only stack in regions where {\it a priori} one expects the dominant absorbers at each frequency to be the same. An example we have explicitly demonstrated is that the stacked spectrum is no longer well described by the geometric mean abundance spectrum once you cross the CO/CH$_4$ transition. 

We have also shown that the geometric mean abundance approximation works well for wide ranges in planet mass and radius; however, for wide ranges in temperatures it works less well. We have identified that this likely arises from the temperature dependence of the molecular bands, where the approximation that opacities scale linearly with abundances at a given wavelength begins to break down (e.g., Figure~\ref{fig:stacking_test_2} and Appendix \ref{sec:cool_reference_planet}). Therefore, while the stacked spectrum is still representative of the geometric mean of the opacity ratios, directly interpreting that in terms of a geometric mean of the abundance ratios becomes more biased. One possible avenue for improvement would be to incorporate temperature-dependent opacities within a Bayesian hierarchical framework, in which atmospheric temperatures and compositions are jointly inferred across a population of planets. In such an approach, the mapping between abundances and opacities could be treated self-consistently across heterogeneous planetary samples. 

The drawbacks we have discussed above are particularly problematic in the small $N_p$ limit of stacking, where an individual outlier can significantly bias the results. However, as we approach {\it ARIEL}, whose much larger sample size of 100s to a possible 1000 exoplanets \citep[e.g.][]{Tinetti2018} would possibly enable stacking to overcome such biases. However, the lower spectral resolution and signal-to-noise of ARIEL, relative to JWST, will influence the effectiveness of stacking. Lower signal-to-noise reduces sensitivity to differences between individual spectra, potentially allowing stacking to remain valid over a broader region of parameter space, as the geometric mean abundance spectrum is more likely to be consistent with the stacked spectrum within the noise. Conversely, the lower spectral resolution reduces the ability to resolve individual opacity sources, blending neighbouring molecular bands and potentially introducing larger residuals between the stacked and geometric mean spectra (as illustrated in Figure \ref{fig:approx}). This may act to narrow the region of parameter space over which stacking is appropriate. The impact of these competing effects, together with ARIEL's target selection, will ultimately determine the utility of stacking in this context and should be explored in future work.

Atmospheric variability and stellar contamination are additional considerations that may impact the stacking of transmission spectra. Our analysis assumes time-invariant spectra and does not explicitly account for variability arising from atmospheric dynamics or stellar heterogeneity. Such effects can introduce additional spectral structure or slopes, which may bias stacked spectra. In the context of stacking, their impact will depend on whether these signals are coherent across the sample: correlated stellar contamination or variability could imprint systematic features in the stacked spectrum, while uncorrelated effects are more likely to average out and contribute as additional noise. The extent to which these processes affect stacking will therefore depend on the properties of the planetary sample (e.g., host star type and activity level).

Another important consideration is the role of disequilibrium chemistry. Notably, JWST has detected SO$_2$ in a handful of planets to date \citep[e.g.,][]{Tsai2023,Beatty2024,Dyrek2024,Gressier2024,Sing2024,Welbanks2024}, demonstrating the presence of photochemical by-products. Furthermore, photodissociation could deplete a planet's atmosphere of CH$_4$ \citep[e.g.,][]{Hu2021} which may affect the importance of the CH$_4$/CO transition that we identify with chemical equilibrium models. Thus, disequilibrium chemistry is expected to increase the diversity of spectral signatures across a population of planets, reducing the degree to which individual spectra are self-similar. This, in turn, can limit the regimes over which the geometric mean abundance spectrum provides a robust approximation to the stacked spectrum. In this context, disequilibrium chemistry does not invalidate the geometric mean formulation itself, but it does affect its practical applicability when photochemical products introduce species that are not consistently present across the sample. As with temperature-dependent effects, this highlights the importance of forward modelling in assessing when stacking assumptions are expected to hold.

\subsection{Retrievals}

Retrievals on exoplanet spectra to derive statistical constraints on abundances is a powerful tool \citep[e.g.,][]{Madhusudhan2009}; however, challenges remain \citep[e.g.,][]{Barstow2020} even with JWST \citep[e.g.,][]{Lueber2024}. Therefore, performing a retrieval on a stacked spectrum may provide more accurate constraints. In this work, we have deliberately focused on (i) what quantity to stack; (ii) what the stacked spectrum physically represents; and (iii) how well it performs on realistic spectra; rather than investigating retrievals. Our insights have revealed that there might be an approach to performing retrievals on stacked transmission spectra that should be investigated in further work.

Any retrieval requires an underlying planet model (e.g., atmospheric structure) to model. We have shown that this model for stacked spectra is one in which the transmission region is the geometric mean of the temperatures and opacity ratios. Thus, this representative planet should provide the initial setup for such investigations. However, one important insight from our analysis as borne out from our experiments, is that the stacked spectrum is only well described by the geometric mean of the opacity ratios when one opacity dominates over the other. Modelled wavelengths where different opacity species contribute equally to the total extinction are not representative of the stacked spectrum. This result arises both in our simple models of temperature- and pressure-independent opacities as well as more realistic models (e.g., where H$_2$O and CO$_2$ are both important). Therefore, we suspect that one key question that will face any retrieval on a stacked spectrum will be which wavelengths to include, and which to not include, and whether this should be dependent on the individual sampled model.

\section{Summary}
\label{sec:summary}

In this work, we have derived what stacking different exoplanets' transmission spectra physically represent: when the abundance ratios are self-similar, the stacked spectrum in $\Delta R_t / H$ is approximately the geometric mean of the individual planets' abundance ratios.

We explored the regions of parameter space over which the stacked spectrum is accurately reproduced by the geometric mean abundance spectrum, constructed using representative planetary parameters (Section \ref{sec:representative_planet}). These representative parameters could be useful for retrievals on stacked spectra, which is beyond the scope of this work. Our tests focus on the wavelength range of JWST's NIRSpec/G395H mode (2.8--5.2\,$\mu$m), motivated by its sensitivity to several key molecular absorbers (e.g., H$_2$O, CO$_2$, CO, and CH$_4$) and its central role in current observational programmes such as BOWIE-ALIGN \citep{Kirk2024}.

We find that the geometric mean spectrum accurately reproduces the stacked spectrum when the temperature variations across the stacked planets are modest and is relatively insensitive to variations in surface gravity, metallicity, and C/O (for C/O $<1$). For two-planet stacks, the discrepancy between stacked and geometric mean spectra is smaller than the spectral uncertainties over a wide range of temperatures, provided that the planets do not straddle the CO/CH$_4$ chemical transition. Thus, stacking becomes inappropriate when different planets have different dominant opacity species, consistent with our toy model expectations.

Extending the analysis to samples of five randomly drawn planets, we find that the stacked spectrum deviates from the geometric mean by more than the spectral uncertainties when the range of planets' equilibrium temperatures exceeds $\sim 600$\,K for Jupiters with $T_{\rm eq}$ between 500--2500\,K.

Finally, we investigated stacking muted sub-Neptune spectra arising from high-metallicity atmospheres with low-pressure, grey clouds, as a function of cloud top pressure and per planet spectral precision. If each planet's spectrum is measured to a precision of 1\,$\Delta R_t/H$ at a resolution $R=200$, then the geometric mean spectrum is an accurate reflection of the stacked spectrum when the range of temperatures within a 5 planet sample is $\lesssim 500$\,K. We also investigated how many planets would need to be stacked to rule out a flat line to $>5 \sigma$. Namely, 38, 7 and 2 sub-Neptunes with $1000\times$ solar metallicity atmospheres and cloud decks at pressures of $10^{-6}$, $10^{-5}$ and $10^{-4}$ bar when each planet is measured to a precision of $\Delta R_t/H = 1$. These results demonstrate that stacking spectra from different sub-Neptunes--when performed within appropriate parameter ranges--can yield statistically robust detections of atmospheric absorption even for muted, cloudy, and metal-rich atmospheres.

Beyond improving signal-to-noise, stacking can also mitigate astrophysical scatter between planets, helping to isolate genuine population-level trends. Individual spectra are shaped by a combination of factors--such as metallicity, cloud structure, and formation history--that introduce intrinsic variability even among otherwise similar planets. By averaging over this astrophysical scatter, stacked spectra can reveal systematic differences between sub-populations when grouped by physically motivated parameters (e.g., equilibrium temperature or orbital obliquity). The sensitivity of our results to temperature range and cloud properties provides a framework for assessing whether observed differences between stacked spectra of distinct planet populations are significant or simply reflect underlying astrophysical diversity. In this way, stacking offers a means not only to enhance precision, but also to uncover genuine atmospheric contrasts that trace differences in atmospheric chemistry, or planet formation and evolution.

Our framework demonstrates that stacked transmission spectra can be given a clear physical interpretation, allowing population-level atmospheric signatures to emerge even when individual spectra are muted or noisy. By defining the conditions under which stacking is physically meaningful using simplified atmospheric models, we provide a foundation for carefully using stacked spectra as a tool to probe atmospheric chemistry and diversity across exoplanets. While the stacking formalism developed here is general and not inherently restricted to a particular wavelength range or spectral resolution, its practical applicability depends on the dominant opacity sources within the observed bandpass and the ability to resolve these. At other wavelengths where different molecules dominate, and at contrasting spectral resolutions, the regions of parameter space over which stacking is valid may differ to those we present here for NIRSpec/G395H using chemical equilibrium forward models.

As the JWST archive continues to expand, stacking different exoplanets' transmission spectra offers a promising avenue for extracting population-level insights from increasingly large datasets, provided that careful attention is paid to the parameter space over which stacking remains valid as a function of wavelength coverage and atmospheric complexity.

\section*{Acknowledgements}

JK acknowledges financial support from Imperial College London through an Imperial College Research Fellowship grant and from the Royal Society through a University Research Fellowship. JEO is supported by a Royal Society University Research Fellowship. This project has received funding from the European Research Council (ERC) under the European Union's Horizon 2020 research and innovation programme (Grant agreement No. 853022).

\section*{Data Availability}

 The data underlying this article will be shared with reasonable request to the corresponding author.



\bibliographystyle{mnras}
\bibliography{example} 

@ARTICLE{Holmberg2024,
       author = {{Holmberg}, M{\r{a}}ns and {Madhusudhan}, Nikku},
        title = "{Possible Hycean conditions in the sub-Neptune TOI-270 d}",
      journal = {\aap},
         year = 2024,
        month = mar,
       volume = {683},
          eid = {L2},
        pages = {L2},
          doi = {10.1051/0004-6361/202348238},
archivePrefix = {arXiv},
       eprint = {2403.03244},
 primaryClass = {astro-ph.EP},
       adsurl = {https://ui.adsabs.harvard.edu/abs/2024A&A...683L...2H}
}

@ARTICLE{Nikolov2018,
       author = {{Nikolov}, N. and {Sing}, D.~K. and {Fortney}, J.~J. and {Goyal}, J.~M. and {Drummond}, B. and {Evans}, T.~M. and {Gibson}, N.~P. and {De Mooij}, E.~J.~W. and {Rustamkulov}, Z. and {Wakeford}, H.~R. and {Smalley}, B. and {Burgasser}, A.~J. and {Hellier}, C. and {Helling}, Ch. and {Mayne}, N.~J. and {Madhusudhan}, N. and {Kataria}, T. and {Baines}, J. and {Carter}, A.~L. and {Ballester}, G.~E. and {Barstow}, J.~K. and {McCleery}, J. and {Spake}, J.~J.},
        title = "{An absolute sodium abundance for a cloud-free `hot Saturn' exoplanet}",
      journal = {\nat},
         year = 2018,
        month = may,
       volume = {557},
       number = {7706},
        pages = {526-529},
          doi = {10.1038/s41586-018-0101-7},
archivePrefix = {arXiv},
       eprint = {1806.06089},
 primaryClass = {astro-ph.EP},
       adsurl = {https://ui.adsabs.harvard.edu/abs/2018Natur.557..526N}
}

@ARTICLE{Rigby2025,
       author = {{Rigby}, Frances E. and {Madhusudhan}, Nikku and {Sarkar}, Subhajit and {Pica-Ciamarra}, Lorenzo and {Holmberg}, M{\r{a}}ns and {Moses}, Julianne I.},
        title = "{A JWST Transmission Spectrum of the Temperate Sub-Neptune TOI-732 c}",
      journal = {\apjl},
         year = 2025,
        month = dec,
       volume = {995},
       number = {2},
          eid = {L70},
        pages = {L70},
          doi = {10.3847/2041-8213/ae247d},
archivePrefix = {arXiv},
       eprint = {2512.15844},
 primaryClass = {astro-ph.EP},
       adsurl = {https://ui.adsabs.harvard.edu/abs/2025ApJ...995L..70R}
}

@ARTICLE{Benneke2024,
       author = {{Benneke}, Bj{\"o}rn and {Roy}, Pierre-Alexis and {Coulombe}, Louis-Philippe and {Radica}, Michael and {Piaulet}, Caroline and {Ahrer}, Eva-Maria and {Pierrehumbert}, Raymond and {Krissansen-Totton}, Joshua and {Schlichting}, Hilke E. and {Hu}, Renyu and {Yang}, Jeehyun and {Christie}, Duncan and {Thorngren}, Daniel and {Young}, Edward D. and {Pelletier}, Stefan and {Knutson}, Heather A. and {Miguel}, Yamila and {Evans-Soma}, Thomas M. and {Dorn}, Caroline and {Gagnebin}, Anna and {Fortney}, Jonathan J. and {Komacek}, Thaddeus and {MacDonald}, Ryan and {Raul}, Eshan and {Cloutier}, Ryan and {Acuna}, Lorena and {Lafreni{\`e}re}, David and {Cadieux}, Charles and {Doyon}, Ren{\'e} and {Welbanks}, Luis and {Allart}, Romain},
        title = "{JWST Reveals CH$_4$, CO$_2$, and H$_2$O in a Metal-rich Miscible Atmosphere on a Two-Earth-Radius Exoplanet}",
      journal = {arXiv e-prints},
         year = 2024,
        month = mar,
          eid = {arXiv:2403.03325},
        pages = {arXiv:2403.03325},
          doi = {10.48550/arXiv.2403.03325},
archivePrefix = {arXiv},
       eprint = {2403.03325},
 primaryClass = {astro-ph.EP},
       adsurl = {https://ui.adsabs.harvard.edu/abs/2024arXiv240303325B}
}

@article{Siegel1942,
author = {{Siegel}, Irving H.},
title = {Index-Number Differences: Geometric Means},
journal = {Journal of the American Statistical Association},
volume = {37},
number = {218},
pages = {271--274},
year = {1942},
publisher = {ASA Website},
doi = {10.1080/01621459.1942.10500636},


URL = { 
    
    
        https://www.tandfonline.com/doi/abs/10.1080/01621459.1942.10500636
    

},
eprint = { 
    
    
        https://www.tandfonline.com/doi/pdf/10.1080/01621459.1942.10500636
    

}

}

@ARTICLE{Seager2000,
       author = {{Seager}, S. and {Sasselov}, D.~D.},
        title = "{Theoretical Transmission Spectra during Extrasolar Giant Planet Transits}",
      journal = {\apj},
         year = 2000,
        month = jul,
       volume = {537},
       number = {2},
        pages = {916-921},
          doi = {10.1086/309088},
archivePrefix = {arXiv},
       eprint = {astro-ph/9912241},
 primaryClass = {astro-ph},
       adsurl = {https://ui.adsabs.harvard.edu/abs/2000ApJ...537..916S}
}

@ARTICLE{Miller-Ricci2009,
       author = {{Miller-Ricci}, Eliza and {Seager}, Sara and {Sasselov}, Dimitar},
        title = "{The Atmospheric Signatures of Super-Earths: How to Distinguish Between Hydrogen-Rich and Hydrogen-Poor Atmospheres}",
      journal = {\apj},
         year = 2009,
        month = jan,
       volume = {690},
       number = {2},
        pages = {1056-1067},
          doi = {10.1088/0004-637X/690/2/1056},
archivePrefix = {arXiv},
       eprint = {0808.1902},
 primaryClass = {astro-ph},
       adsurl = {https://ui.adsabs.harvard.edu/abs/2009ApJ...690.1056M}
}

@ARTICLE{Griffith2014,
       author = {{Griffith}, C.~A.},
        title = "{Disentangling degenerate solutions from primary transit and secondary eclipse spectroscopy of exoplanets}",
      journal = {Philosophical Transactions of the Royal Society of London Series A},
         year = 2014,
        month = mar,
       volume = {372},
       number = {2014},
        pages = {20130086-20130086},
          doi = {10.1098/rsta.2013.0086},
archivePrefix = {arXiv},
       eprint = {1312.3988},
 primaryClass = {astro-ph.EP},
       adsurl = {https://ui.adsabs.harvard.edu/abs/2014RSPTA.37230086G}
}

@ARTICLE{Hubbard2001,
       author = {{Hubbard}, W.~B. and {Fortney}, J.~J. and {Lunine}, J.~I. and {Burrows}, A. and {Sudarsky}, D. and {Pinto}, P.},
        title = "{Theory of Extrasolar Giant Planet Transits}",
      journal = {\apj},
         year = 2001,
        month = oct,
       volume = {560},
       number = {1},
        pages = {413-419},
          doi = {10.1086/322490},
archivePrefix = {arXiv},
       eprint = {astro-ph/0101024},
 primaryClass = {astro-ph},
       adsurl = {https://ui.adsabs.harvard.edu/abs/2001ApJ...560..413H}
}

@ARTICLE{Brown2001,
       author = {{Brown}, Timothy M.},
        title = "{Transmission Spectra as Diagnostics of Extrasolar Giant Planet Atmospheres}",
      journal = {\apj},
         year = 2001,
        month = jun,
       volume = {553},
       number = {2},
        pages = {1006-1026},
          doi = {10.1086/320950},
archivePrefix = {arXiv},
       eprint = {astro-ph/0101307},
 primaryClass = {astro-ph},
       adsurl = {https://ui.adsabs.harvard.edu/abs/2001ApJ...553.1006B}
}

@ARTICLE{Heng2017,
       author = {{Heng}, Kevin and {Kitzmann}, Daniel},
        title = "{The theory of transmission spectra revisited: a semi-analytical method for interpreting WFC3 data and an unresolved challenge}",
      journal = {\mnras},
         year = 2017,
        month = sep,
       volume = {470},
       number = {3},
        pages = {2972-2981},
          doi = {10.1093/mnras/stx1453},
archivePrefix = {arXiv},
       eprint = {1702.02051},
 primaryClass = {astro-ph.EP},
       adsurl = {https://ui.adsabs.harvard.edu/abs/2017MNRAS.470.2972H}
}

@ARTICLE{Fortney2005,
       author = {{Fortney}, Jonathan J.},
        title = "{The effect of condensates on the characterization of transiting planet atmospheres with transmission spectroscopy}",
      journal = {\mnras},
         year = 2005,
        month = dec,
       volume = {364},
       number = {2},
        pages = {649-653},
          doi = {10.1111/j.1365-2966.2005.09587.x},
archivePrefix = {arXiv},
       eprint = {astro-ph/0509292},
 primaryClass = {astro-ph},
       adsurl = {https://ui.adsabs.harvard.edu/abs/2005MNRAS.364..649F}
}

@ARTICLE{deWit2013,
       author = {{de Wit}, Julien and {Seager}, Sara},
        title = "{Constraining Exoplanet Mass from Transmission Spectroscopy}",
      journal = {Science},
         year = 2013,
        month = dec,
       volume = {342},
       number = {6165},
        pages = {1473-1477},
          doi = {10.1126/science.1245450},
archivePrefix = {arXiv},
       eprint = {1401.6181},
 primaryClass = {astro-ph.EP},
       adsurl = {https://ui.adsabs.harvard.edu/abs/2013Sci...342.1473D}
}

@ARTICLE{Benneke2012,
       author = {{Benneke}, Bjoern and {Seager}, Sara},
        title = "{Atmospheric Retrieval for Super-Earths: Uniquely Constraining the Atmospheric Composition with Transmission Spectroscopy}",
      journal = {\apj},
         year = 2012,
        month = jul,
       volume = {753},
       number = {2},
          eid = {100},
        pages = {100},
          doi = {10.1088/0004-637X/753/2/100},
archivePrefix = {arXiv},
       eprint = {1203.4018},
 primaryClass = {astro-ph.EP},
       adsurl = {https://ui.adsabs.harvard.edu/abs/2012ApJ...753..100B}
}

@ARTICLE{Hansen2008,
       author = {{Hansen}, Brad M.~S.},
        title = "{On the Absorption and Redistribution of Energy in Irradiated Planets}",
      journal = {\apjs},
         year = 2008,
        month = dec,
       volume = {179},
       number = {2},
        pages = {484-508},
          doi = {10.1086/591964},
archivePrefix = {arXiv},
       eprint = {0801.2972},
 primaryClass = {astro-ph},
       adsurl = {https://ui.adsabs.harvard.edu/abs/2008ApJS..179..484H}
}

@ARTICLE{Lueber2024,
       author = {{Lueber}, Anna and {Novais}, Aline and {Fisher}, Chloe and {Heng}, Kevin},
        title = "{Information content of JWST spectra of WASP-39b}",
      journal = {\aap},
         year = 2024,
        month = jul,
       volume = {687},
          eid = {A110},
        pages = {A110},
          doi = {10.1051/0004-6361/202348802},
archivePrefix = {arXiv},
       eprint = {2405.02656},
 primaryClass = {astro-ph.EP},
       adsurl = {https://ui.adsabs.harvard.edu/abs/2024A&A...687A.110L}
}

@ARTICLE{Barstow2020,
       author = {{Barstow}, Joanna K. and {Heng}, Kevin},
        title = "{Outstanding Challenges of Exoplanet Atmospheric Retrievals}",
      journal = {\ssr},
         year = 2020,
        month = jun,
       volume = {216},
       number = {5},
          eid = {82},
        pages = {82},
          doi = {10.1007/s11214-020-00666-x},
archivePrefix = {arXiv},
       eprint = {2003.14311},
 primaryClass = {astro-ph.EP},
       adsurl = {https://ui.adsabs.harvard.edu/abs/2020SSRv..216...82B}
}

@ARTICLE{Madhusudhan2009,
       author = {{Madhusudhan}, N. and {Seager}, S.},
        title = "{A Temperature and Abundance Retrieval Method for Exoplanet Atmospheres}",
      journal = {\apj},
         year = 2009,
        month = dec,
       volume = {707},
       number = {1},
        pages = {24-39},
          doi = {10.1088/0004-637X/707/1/24},
archivePrefix = {arXiv},
       eprint = {0910.1347},
 primaryClass = {astro-ph.EP},
       adsurl = {https://ui.adsabs.harvard.edu/abs/2009ApJ...707...24M}
}

@ARTICLE{Tinetti2018,
       author = {{Tinetti}, Giovanna and {Drossart}, Pierre and {Eccleston}, Paul and {Hartogh}, Paul and {Heske}, Astrid and {Leconte}, J{\'e}r{\'e}my and {Micela}, Giusi and {Ollivier}, Marc and {Pilbratt}, G{\"o}ran and {Puig}, Ludovic and {Turrini}, Diego and {Vandenbussche}, Bart and {Wolkenberg}, Paulina and {Beaulieu}, Jean-Philippe and {Buchave}, Lars A. and {Ferus}, Martin and {Griffin}, Matt and {Guedel}, Manuel and {Justtanont}, Kay and {Lagage}, Pierre-Olivier and {Machado}, Pedro and {Malaguti}, Giuseppe and {Min}, Michiel and {N{\o}rgaard-Nielsen}, Hans Ulrik and {Rataj}, Mirek and {Ray}, Tom and {Ribas}, Ignasi and {Swain}, Mark and {Szabo}, Robert and {Werner}, Stephanie and {Barstow}, Joanna and {Burleigh}, Matt and {Cho}, James and {Coud{\'e} du Foresto}, Vincent and {Coustenis}, Athena and {Decin}, Leen and {Encrenaz}, Therese and {Galand}, Marina and {Gillon}, Michael and {Helled}, Ravit and {Morales}, Juan Carlos and {Garc{\'\i}a Mu{\~n}oz}, Antonio and {Moneti}, Andrea and {Pagano}, Isabella and {Pascale}, Enzo and {Piccioni}, Giuseppe and {Pinfield}, David and {Sarkar}, Subhajit and {Selsis}, Franck and {Tennyson}, Jonathan and {Triaud}, Amaury and {Venot}, Olivia and {Waldmann}, Ingo and {Waltham}, David and {Wright}, Gillian and {Amiaux}, Jerome and {Augu{\`e}res}, Jean-Louis and {Berth{\'e}}, Michel and {Bezawada}, Naidu and {Bishop}, Georgia and {Bowles}, Neil and {Coffey}, Deirdre and {Colom{\'e}}, Josep and {Crook}, Martin and {Crouzet}, Pierre-Elie and {Da Peppo}, Vania and {Sanz}, Isabel Escudero and {Focardi}, Mauro and {Frericks}, Martin and {Hunt}, Tom and {Kohley}, Ralf and {Middleton}, Kevin and {Morgante}, Gianluca and {Ottensamer}, Roland and {Pace}, Emanuele and {Pearson}, Chris and {Stamper}, Richard and {Symonds}, Kate and {Rengel}, Miriam and {Renotte}, Etienne and {Ade}, Peter and {Affer}, Laura and {Alard}, Christophe and {Allard}, Nicole and {Altieri}, Francesca and {Andr{\'e}}, Yves and {Arena}, Claudio and {Argyriou}, Ioannis and {Aylward}, Alan and {Baccani}, Cristian and {Bakos}, Gaspar and {Banaszkiewicz}, Marek and {Barlow}, Mike and {Batista}, Virginie and {Bellucci}, Giancarlo and {Benatti}, Serena and {Bernardi}, Pernelle and {B{\'e}zard}, Bruno and {Blecka}, Maria and {Bolmont}, Emeline and {Bonfond}, Bertrand and {Bonito}, Rosaria and {Bonomo}, Aldo S. and {Brucato}, John Robert and {Brun}, Allan Sacha and {Bryson}, Ian and {Bujwan}, Waldemar and {Casewell}, Sarah and {Charnay}, Bejamin and {Pestellini}, Cesare Cecchi and {Chen}, Guo and {Ciaravella}, Angela and {Claudi}, Riccardo and {Cl{\'e}dassou}, Rodolphe and {Damasso}, Mario and {Damiano}, Mario and {Danielski}, Camilla and {Deroo}, Pieter and {Di Giorgio}, Anna Maria and {Dominik}, Carsten and {Doublier}, Vanessa and {Doyle}, Simon and {Doyon}, Ren{\'e} and {Drummond}, Benjamin and {Duong}, Bastien and {Eales}, Stephen and {Edwards}, Billy and {Farina}, Maria and {Flaccomio}, Ettore and {Fletcher}, Leigh and {Forget}, Fran{\c{c}}ois and {Fossey}, Steve and {Fr{\"a}nz}, Markus and {Fujii}, Yuka and {Garc{\'\i}a-Piquer}, {\'A}lvaro and {Gear}, Walter and {Geoffray}, Herv{\'e} and {G{\'e}rard}, Jean Claude and {Gesa}, Lluis and {Gomez}, H. and {Graczyk}, Rafa{\l} and {Griffith}, Caitlin and {Grodent}, Denis and {Guarcello}, Mario Giuseppe and {Gustin}, Jacques and {Hamano}, Keiko and {Hargrave}, Peter and {Hello}, Yann and {Heng}, Kevin and {Herrero}, Enrique and {Hornstrup}, Allan and {Hubert}, Benoit and {Ida}, Shigeru and {Ikoma}, Masahiro and {Iro}, Nicolas and {Irwin}, Patrick and {Jarchow}, Christopher and {Jaubert}, Jean and {Jones}, Hugh and {Julien}, Queyrel and {Kameda}, Shingo and {Kerschbaum}, Franz and {Kervella}, Pierre and {Koskinen}, Tommi and {Krijger}, Matthijs and {Krupp}, Norbert and {Lafarga}, Marina and {Landini}, Federico and {Lellouch}, Emanuel and {Leto}, Giuseppe and {Luntzer}, A. and {Rank-L{\"u}ftinger}, Theresa and {Maggio}, Antonio and {Maldonado}, Jesus and {Maillard}, Jean-Pierre and {Mall}, Urs and {Marquette}, Jean-Baptiste and {Mathis}, Stephane and {Maxted}, Pierre and {Matsuo}, Taro and {Medvedev}, Alexander and {Miguel}, Yamila and {Minier}, Vincent and {Morello}, Giuseppe and {Mura}, Alessandro and {Narita}, Norio and {Nascimbeni}, Valerio and {Nguyen Tong}, N. and {Noce}, Vladimiro and {Oliva}, Fabrizio and {Palle}, Enric and {Palmer}, Paul and {Pancrazzi}, Maurizio and {Papageorgiou}, Andreas and {Parmentier}, Vivien and {Perger}, Manuel and {Petralia}, Antonino and {Pezzuto}, Stefano and {Pierrehumbert}, Ray and {Pillitteri}, Ignazio},
        title = "{A chemical survey of exoplanets with ARIEL}",
      journal = {Experimental Astronomy},
         year = 2018,
        month = nov,
       volume = {46},
       number = {1},
        pages = {135-209},
          doi = {10.1007/s10686-018-9598-x},
       adsurl = {https://ui.adsabs.harvard.edu/abs/2018ExA....46..135T}
}

@INCOLLECTION{Kreidberg2018,
       author = {{Kreidberg}, Laura},
        title = "{Exoplanet Atmosphere Measurements from Transmission Spectroscopy and Other Planet Star Combined Light Observations}",
    booktitle = {Handbook of Exoplanets},
         year = 2018,
       editor = {{Deeg}, Hans J. and {Belmonte}, Juan Antonio},
          eid = {100},
        pages = {100},
          doi = {10.1007/978-3-319-55333-7_100},
       adsurl = {https://ui.adsabs.harvard.edu/abs/2018haex.bookE.100K}
}

@ARTICLE{Lorenz2023,
       author = {{Lorenz}, Brian and {Kriek}, Mariska and {Shapley}, Alice E. and {Reddy}, Naveen A. and {Sanders}, Ryan L. and {Barro}, Guillermo and {Coil}, Alison L. and {Mobasher}, Bahram and {Price}, Sedona H. and {Runco}, Jordan N. and {Shivaei}, Irene and {Siana}, Brian and {Weisz}, Daniel R.},
        title = "{An Updated Dust-to-Star Geometry: Dust Attenuation Does Not Depend on Inclination in 1.3 {\ensuremath{\leq}}z {\ensuremath{\leq}}2.6 Star-forming Galaxies from MOSDEF}",
      journal = {\apj},
         year = 2023,
        month = jul,
       volume = {951},
       number = {1},
          eid = {29},
        pages = {29},
          doi = {10.3847/1538-4357/accdd1},
archivePrefix = {arXiv},
       eprint = {2304.08521},
 primaryClass = {astro-ph.GA},
       adsurl = {https://ui.adsabs.harvard.edu/abs/2023ApJ...951...29L}
}

@ARTICLE{Wallack2024,
       author = {{Wallack}, Nicole L. and {Batalha}, Natasha E. and {Alderson}, Lili and {Scarsdale}, Nicholas and {Adams Redai}, Jea I. and {Aguichine}, Artyom and {Alam}, Munazza K. and {Gao}, Peter and {Wolfgang}, Angie and {Batalha}, Natalie M. and {Kirk}, James and {L{\'o}pez-Morales}, Mercedes and {Moran}, Sarah E. and {Teske}, Johanna and {Wakeford}, Hannah R. and {Wogan}, Nicholas F.},
        title = "{JWST COMPASS: A NIRSpec/G395H Transmission Spectrum of the Sub-Neptune TOI-836c}",
      journal = {arXiv e-prints},
         year = 2024,
        month = apr,
          eid = {arXiv:2404.01264},
        pages = {arXiv:2404.01264},
          doi = {10.48550/arXiv.2404.01264},
archivePrefix = {arXiv},
       eprint = {2404.01264},
 primaryClass = {astro-ph.EP},
       adsurl = {https://ui.adsabs.harvard.edu/abs/2024arXiv240401264W}
}

@ARTICLE{MacDonald2017,
       author = {{MacDonald}, Ryan J. and {Madhusudhan}, Nikku},
        title = "{HD 209458b in new light: evidence of nitrogen chemistry, patchy clouds and sub-solar water}",
      journal = {\mnras},
         year = 2017,
        month = aug,
       volume = {469},
       number = {2},
        pages = {1979-1996},
          doi = {10.1093/mnras/stx804},
archivePrefix = {arXiv},
       eprint = {1701.01113},
 primaryClass = {astro-ph.EP},
       adsurl = {https://ui.adsabs.harvard.edu/abs/2017MNRAS.469.1979M}
}

@ARTICLE{MacDonald2023,
       author = {{MacDonald}, Ryan J.},
        title = "{POSEIDON: A Multidimensional Atmospheric Retrieval Code for Exoplanet Spectra}",
      journal = {The Journal of Open Source Software},
         year = 2023,
        month = jan,
       volume = {8},
          eid = {4873},
        pages = {4873},
          doi = {10.21105/joss.04873},
archivePrefix = {arXiv},
       eprint = {2410.18181},
 primaryClass = {astro-ph.IM},
       adsurl = {https://ui.adsabs.harvard.edu/abs/2023JOSS....8.4873M}
}

@ARTICLE{Stock2018,
       author = {{Stock}, Joachim W. and {Kitzmann}, Daniel and {Patzer}, A. Beate C. and {Sedlmayr}, Erwin},
        title = "{FastChem: A computer program for efficient complex chemical equilibrium calculations in the neutral/ionized gas phase with applications to stellar and planetary atmospheres}",
      journal = {\mnras},
         year = 2018,
        month = sep,
       volume = {479},
       number = {1},
        pages = {865-874},
          doi = {10.1093/mnras/sty1531},
archivePrefix = {arXiv},
       eprint = {1804.05010},
 primaryClass = {astro-ph.EP},
       adsurl = {https://ui.adsabs.harvard.edu/abs/2018MNRAS.479..865S}
}

@ARTICLE{Stock2022,
       author = {{Stock}, Joachim W. and {Kitzmann}, Daniel and {Patzer}, A. Beate C.},
        title = "{FASTCHEM 2 : an improved computer program to determine the gas-phase chemical equilibrium composition for arbitrary element distributions}",
      journal = {\mnras},
         year = 2022,
        month = dec,
       volume = {517},
       number = {3},
        pages = {4070-4080},
          doi = {10.1093/mnras/stac2623},
archivePrefix = {arXiv},
       eprint = {2206.08247},
 primaryClass = {astro-ph.EP},
       adsurl = {https://ui.adsabs.harvard.edu/abs/2022MNRAS.517.4070S}
}

@ARTICLE{Kitzmann2024,
       author = {{Kitzmann}, Daniel and {Stock}, Joachim W. and {Patzer}, A. Beate C.},
        title = "{FASTCHEM COND: equilibrium chemistry with condensation and rainout for cool planetary and stellar environments}",
      journal = {\mnras},
         year = 2024,
        month = jan,
       volume = {527},
       number = {3},
        pages = {7263-7283},
          doi = {10.1093/mnras/stad3515},
archivePrefix = {arXiv},
       eprint = {2309.02337},
 primaryClass = {astro-ph.EP},
       adsurl = {https://ui.adsabs.harvard.edu/abs/2024MNRAS.527.7263K}
}

@ARTICLE{Kirk2024,
       author = {{Kirk}, James and {Ahrer}, Eva-Maria and {Penzlin}, Anna B.~T. and {Owen}, James E. and {Booth}, Richard A. and {Alderson}, Lili and {Christie}, Duncan A. and {Claringbold}, Alastair B. and {Esparza-Borges}, Emma and {Fisher}, Chloe E. and {L{\'o}pez-Morales}, Mercedes and {Mayne}, N.~J. and {McCormack}, Mason and {Meech}, Annabella and {Panwar}, Vatsal and {Powell}, Diana and {Sergeev}, Denis E. and {Taylor}, Jake and {Tsai}, Shang-Min and {Valentine}, Daniel and {Wakeford}, Hannah R. and {Wheatley}, Peter J. and {Zamyatina}, Maria},
        title = "{BOWIE-ALIGN: A JWST comparative survey of aligned versus misaligned hot Jupiters to test the dependence of atmospheric composition on migration history}",
      journal = {RAS Techniques and Instruments},
         year = 2024,
        month = jan,
       volume = {3},
       number = {1},
        pages = {691-704},
          doi = {10.1093/rasti/rzae043},
archivePrefix = {arXiv},
       eprint = {2407.03198},
 primaryClass = {astro-ph.EP},
       adsurl = {https://ui.adsabs.harvard.edu/abs/2024RASTI...3..691K}
}

@ARTICLE{H2O,
       author = {{Polyansky}, Oleg L. and {Kyuberis}, Aleksandra A. and {Zobov}, Nikolai F. and {Tennyson}, Jonathan and {Yurchenko}, Sergei N. and {Lodi}, Lorenzo},
        title = "{ExoMol molecular line lists XXX: a complete high-accuracy line list for water}",
      journal = {\mnras},
         year = 2018,
        month = oct,
       volume = {480},
       number = {2},
        pages = {2597-2608},
          doi = {10.1093/mnras/sty1877},
archivePrefix = {arXiv},
       eprint = {1807.04529},
 primaryClass = {astro-ph.EP},
       adsurl = {https://ui.adsabs.harvard.edu/abs/2018MNRAS.480.2597P}
}

@ARTICLE{CO2,
       author = {{Yurchenko}, S.~N. and {Mellor}, Thomas M. and {Freedman}, Richard S. and {Tennyson}, J.},
        title = "{ExoMol line lists - XXXIX. Ro-vibrational molecular line list for CO$_{2}$}",
      journal = {\mnras},
         year = 2020,
        month = aug,
       volume = {496},
       number = {4},
        pages = {5282-5291},
          doi = {10.1093/mnras/staa1874},
archivePrefix = {arXiv},
       eprint = {2007.02122},
 primaryClass = {astro-ph.EP},
       adsurl = {https://ui.adsabs.harvard.edu/abs/2020MNRAS.496.5282Y}
}

@ARTICLE{CH4,
       author = {{Yurchenko}, Sergei N. and {Owens}, Alec and {Kefala}, Kyriaki and {Tennyson}, Jonathan},
        title = "{ExoMol line lists - LVII. High accuracy ro-vibrational line list for methane (CH$_{4}$)}",
      journal = {\mnras},
         year = 2024,
        month = feb,
       volume = {528},
       number = {2},
        pages = {3719-3729},
          doi = {10.1093/mnras/stae148},
       adsurl = {https://ui.adsabs.harvard.edu/abs/2024MNRAS.528.3719Y}
}

@ARTICLE{CO,
       author = {{Li}, Gang and {Gordon}, Iouli E. and {Rothman}, Laurence S. and {Tan}, Yan and {Hu}, Shui-Ming and {Kassi}, Samir and {Campargue}, Alain and {Medvedev}, Emile S.},
        title = "{Rovibrational Line Lists for Nine Isotopologues of the CO Molecule in the X $^{1}${\ensuremath{\Sigma}}$^{+}$ Ground Electronic State}",
      journal = {\apjs},
         year = 2015,
        month = jan,
       volume = {216},
       number = {1},
          eid = {15},
        pages = {15},
          doi = {10.1088/0067-0049/216/1/15},
       adsurl = {https://ui.adsabs.harvard.edu/abs/2015ApJS..216...15L}
}

@ARTICLE{SO2,
       author = {{Underwood}, Daniel S. and {Tennyson}, Jonathan and {Yurchenko}, Sergei N. and {Huang}, Xinchuan and {Schwenke}, David W. and {Lee}, Timothy J. and {Clausen}, S{\o}nnik and {Fateev}, Alexander},
        title = "{ExoMol molecular line lists - XIV. The rotation-vibration spectrum of hot SO$_{2}$}",
      journal = {\mnras},
         year = 2016,
        month = jul,
       volume = {459},
       number = {4},
        pages = {3890-3899},
          doi = {10.1093/mnras/stw849},
archivePrefix = {arXiv},
       eprint = {1603.04065},
 primaryClass = {astro-ph.EP},
       adsurl = {https://ui.adsabs.harvard.edu/abs/2016MNRAS.459.3890U}
}

@ARTICLE{H2S,
       author = {{Azzam}, Ala'a. A.~A. and {Tennyson}, Jonathan and {Yurchenko}, Sergei N. and {Naumenko}, Olga V.},
        title = "{ExoMol molecular line lists - XVI. The rotation-vibration spectrum of hot H$_{2}$S}",
      journal = {\mnras},
         year = 2016,
        month = aug,
       volume = {460},
       number = {4},
        pages = {4063-4074},
          doi = {10.1093/mnras/stw1133},
archivePrefix = {arXiv},
       eprint = {1607.00499},
 primaryClass = {astro-ph.EP},
       adsurl = {https://ui.adsabs.harvard.edu/abs/2016MNRAS.460.4063A}
}

@ARTICLE{Alderson2024,
       author = {{Alderson}, Lili and {Batalha}, Natasha E. and {Wakeford}, Hannah R. and {Wallack}, Nicole L. and {Aguichine}, Artyom and {Teske}, Johanna and {Adams Redai}, Jea and {Alam}, Munazza K. and {Batalha}, Natalie M. and {Gao}, Peter and {Kirk}, James and {L{\'o}pez-Morales}, Mercedes and {Moran}, Sarah E. and {Scarsdale}, Nicholas and {Wogan}, Nicholas F. and {Wolfgang}, Angie},
        title = "{JWST COMPASS: NIRSpec/G395H Transmission Observations of the Super-Earth TOI-836b}",
      journal = {\aj},
         year = 2024,
        month = may,
       volume = {167},
       number = {5},
          eid = {216},
        pages = {216},
          doi = {10.3847/1538-3881/ad32c9},
archivePrefix = {arXiv},
       eprint = {2404.00093},
 primaryClass = {astro-ph.EP},
       adsurl = {https://ui.adsabs.harvard.edu/abs/2024AJ....167..216A}
}

@ARTICLE{Kirk2025,
       author = {{Kirk}, James and {Ahrer}, Eva-Maria and {Claringbold}, Alastair B. and {Zamyatina}, Maria and {Fisher}, Chloe and {McCormack}, Mason and {Panwar}, Vatsal and {Powell}, Diana and {Taylor}, Jake and {Thorngren}, Daniel P. and {Christie}, Duncan A. and {Esparza-Borges}, Emma and {Tsai}, Shang-Min and {Alderson}, Lili and {Booth}, Richard A. and {Fairman}, Charlotte and {L{\'o}pez-Morales}, Mercedes and {Mayne}, N.~J. and {Meech}, Annabella and {Molli{\`e}re}, Paul and {Owen}, James E. and {Penzlin}, Anna B.~T. and {Sergeev}, Denis E. and {Valentine}, Daniel and {Wakeford}, Hannah R. and {Wheatley}, Peter J.},
        title = "{BOWIE-ALIGN: JWST reveals hints of planetesimal accretion and complex sulphur chemistry in the atmosphere of the misaligned hot Jupiter WASP-15b}",
      journal = {\mnras},
         year = 2025,
        month = mar,
       volume = {537},
       number = {4},
        pages = {3027-3052},
          doi = {10.1093/mnras/staf208},
archivePrefix = {arXiv},
       eprint = {2410.08116},
 primaryClass = {astro-ph.EP},
       adsurl = {https://ui.adsabs.harvard.edu/abs/2025MNRAS.537.3027K}
}

@ARTICLE{Asplund2021,
       author = {{Asplund}, M. and {Amarsi}, A.~M. and {Grevesse}, N.},
        title = "{The chemical make-up of the Sun: A 2020 vision}",
      journal = {\aap},
         year = 2021,
        month = sep,
       volume = {653},
          eid = {A141},
        pages = {A141},
          doi = {10.1051/0004-6361/202140445},
archivePrefix = {arXiv},
       eprint = {2105.01661},
 primaryClass = {astro-ph.SR},
       adsurl = {https://ui.adsabs.harvard.edu/abs/2021A&A...653A.141A}
}

@ARTICLE{Meech2025,
       author = {{Meech}, Annabella and {Claringbold}, Alastair B. and {Ahrer}, Eva-Maria and {Kirk}, James and {L{\'o}pez-Morales}, Mercedes and {Taylor}, Jake and {Booth}, Richard A. and {Penzlin}, Anna B.~T. and {Alderson}, Lili and {Christie}, Duncan A. and {Esparza-Borges}, Emma and {Fairman}, Charlotte and {Mayne}, Nathan J. and {McCormack}, Mason and {Owen}, James E. and {Panwar}, Vatsal and {Powell}, Diana and {Sergeev}, Denis E. and {Valentine}, Daniel and {Wakeford}, Hannah R. and {Wheatley}, Peter J. and {Zamyatina}, Maria},
        title = "{BOWIE-ALIGN: substellar metallicity and carbon depletion in the aligned TrES-4b with JWST NIRSpec transmission spectroscopy}",
      journal = {\mnras},
         year = 2025,
        month = may,
       volume = {539},
       number = {2},
        pages = {1381-1403},
          doi = {10.1093/mnras/staf530},
archivePrefix = {arXiv},
       eprint = {2503.24280},
 primaryClass = {astro-ph.EP},
       adsurl = {https://ui.adsabs.harvard.edu/abs/2025MNRAS.539.1381M}
}

@ARTICLE{Ahrer2025,
       author = {{Ahrer}, Eva-Maria and {Gandhi}, Siddharth and {Alderson}, Lili and {Kirk}, James and {Teske}, Johanna and {Booth}, Richard A. and {McDonald}, Catriona H. and {Christie}, Duncan A. and {Claringbold}, Alastair B. and {Nealon}, Rebecca and {Panwar}, Vatsal and {Veras}, Dimitri and {Wakeford}, Hannah R. and {Wheatley}, Peter J. and {Zamyatina}, Maria},
        title = "{Tracing the formation and migration history: molecular signatures in the atmosphere of misaligned hot Jupiter WASP-94 A b using JWST NIRSpec/G395H}",
      journal = {\mnras},
         year = 2025,
        month = jul,
       volume = {540},
       number = {3},
        pages = {2535-2554},
          doi = {10.1093/mnras/staf819},
archivePrefix = {arXiv},
       eprint = {2505.11224},
 primaryClass = {astro-ph.EP},
       adsurl = {https://ui.adsabs.harvard.edu/abs/2025MNRAS.540.2535A}
}

@ARTICLE{Alderson2023,
       author = {{Alderson}, Lili and {Wakeford}, Hannah R. and {Alam}, Munazza K. and {Batalha}, Natasha E. and {Lothringer}, Joshua D. and {Adams Redai}, Jea and {Barat}, Saugata and {Brande}, Jonathan and {Damiano}, Mario and {Daylan}, Tansu and {Espinoza}, N{\'e}stor and {Flagg}, Laura and {Goyal}, Jayesh M. and {Grant}, David and {Hu}, Renyu and {Inglis}, Julie and {Lee}, Elspeth K.~H. and {Mikal-Evans}, Thomas and {Ramos-Rosado}, Lakeisha and {Roy}, Pierre-Alexis and {Wallack}, Nicole L. and {Batalha}, Natalie M. and {Bean}, Jacob L. and {Benneke}, Bj{\"o}rn and {Berta-Thompson}, Zachory K. and {Carter}, Aarynn L. and {Changeat}, Quentin and {Col{\'o}n}, Knicole D. and {Crossfield}, Ian J.~M. and {D{\'e}sert}, Jean-Michel and {Foreman-Mackey}, Daniel and {Gibson}, Neale P. and {Kreidberg}, Laura and {Line}, Michael R. and {L{\'o}pez-Morales}, Mercedes and {Molaverdikhani}, Karan and {Moran}, Sarah E. and {Morello}, Giuseppe and {Moses}, Julianne I. and {Mukherjee}, Sagnick and {Schlawin}, Everett and {Sing}, David K. and {Stevenson}, Kevin B. and {Taylor}, Jake and {Aggarwal}, Keshav and {Ahrer}, Eva-Maria and {Allen}, Natalie H. and {Barstow}, Joanna K. and {Bell}, Taylor J. and {Blecic}, Jasmina and {Casewell}, Sarah L. and {Chubb}, Katy L. and {Crouzet}, Nicolas and {Cubillos}, Patricio E. and {Decin}, Leen and {Feinstein}, Adina D. and {Fortney}, Joanthan J. and {Harrington}, Joseph and {Heng}, Kevin and {Iro}, Nicolas and {Kempton}, Eliza M. -R. and {Kirk}, James and {Knutson}, Heather A. and {Krick}, Jessica and {Leconte}, J{\'e}r{\'e}my and {Lendl}, Monika and {MacDonald}, Ryan J. and {Mancini}, Luigi and {Mansfield}, Megan and {May}, Erin M. and {Mayne}, Nathan J. and {Miguel}, Yamila and {Nikolov}, Nikolay K. and {Ohno}, Kazumasa and {Palle}, Enric and {Parmentier}, Vivien and {Petit dit de la Roche}, Dominique J.~M. and {Piaulet}, Caroline and {Powell}, Diana and {Rackham}, Benjamin V. and {Redfield}, Seth and {Rogers}, Laura K. and {Rustamkulov}, Zafar and {Tan}, Xianyu and {Tremblin}, P. and {Tsai}, Shang-Min and {Turner}, Jake D. and {de Val-Borro}, Miguel and {Venot}, Olivia and {Welbanks}, Luis and {Wheatley}, Peter J. and {Zhang}, Xi},
        title = "{Early Release Science of the exoplanet WASP-39b with JWST NIRSpec G395H}",
      journal = {\nat},
         year = 2023,
        month = feb,
       volume = {614},
       number = {7949},
        pages = {664-669},
          doi = {10.1038/s41586-022-05591-3},
archivePrefix = {arXiv},
       eprint = {2211.10488},
 primaryClass = {astro-ph.EP},
       adsurl = {https://ui.adsabs.harvard.edu/abs/2023Natur.614..664A}
}

@ARTICLE{Shapley2003,
       author = {{Shapley}, Alice E. and {Steidel}, Charles C. and {Pettini}, Max and {Adelberger}, Kurt L.},
        title = "{Rest-Frame Ultraviolet Spectra of z\raisebox{-0.5ex}\textasciitilde3 Lyman Break Galaxies}",
      journal = {\apj},
         year = 2003,
        month = may,
       volume = {588},
       number = {1},
        pages = {65-89},
          doi = {10.1086/373922},
archivePrefix = {arXiv},
       eprint = {astro-ph/0301230},
 primaryClass = {astro-ph},
       adsurl = {https://ui.adsabs.harvard.edu/abs/2003ApJ...588...65S}
}

@ARTICLE{Steidel2016,
       author = {{Steidel}, Charles C. and {Strom}, Allison L. and {Pettini}, Max and {Rudie}, Gwen C. and {Reddy}, Naveen A. and {Trainor}, Ryan F.},
        title = "{Reconciling the Stellar and Nebular Spectra of High-redshift Galaxies}",
      journal = {\apj},
         year = 2016,
        month = aug,
       volume = {826},
       number = {2},
          eid = {159},
        pages = {159},
          doi = {10.3847/0004-637X/826/2/159},
archivePrefix = {arXiv},
       eprint = {1605.07186},
 primaryClass = {astro-ph.GA},
       adsurl = {https://ui.adsabs.harvard.edu/abs/2016ApJ...826..159S}
}

@ARTICLE{Ahrer2025_K7,
       author = {{Ahrer}, Eva-Maria and {Fairman}, Charlotte and {Kirk}, James and {Wakeford}, Hannah R. and {Barstow}, Joanna K. and {Penzlin}, Anna B.~T. and {Alderson}, Lili and {Booth}, Richard A. and {Christie}, Duncan A. and {Claringbold}, Alastair B. and {Esparza-Borges}, Emma and {Gasc{\'o}n}, Carlos and {L{\'o}pez-Morales}, Mercedes and {Mayne}, N.~J. and {McCormack}, Mason and {Meech}, Annabella and {Molli{\`e}re}, Paul and {Owen}, James E. and {Panwar}, Vatsal and {Sergeev}, Denis E. and {Valentine}, Daniel and {Wheatley}, Peter J. and {Zamyatina}, Maria},
        title = "{BOWIE-ALIGN: Weak spectral features in KELT-7b's JWST NIRSpec/G395H transmission spectrum imply a high cloud deck or a low-metallicity atmosphere}",
      journal = {arXiv e-prints},
         year = 2025,
        month = sep,
          eid = {arXiv:2509.12479},
        pages = {arXiv:2509.12479},
          doi = {10.48550/arXiv.2509.12479},
archivePrefix = {arXiv},
       eprint = {2509.12479},
 primaryClass = {astro-ph.EP},
       adsurl = {https://ui.adsabs.harvard.edu/abs/2025arXiv250912479A}
}

@ARTICLE{Teske2025,
       author = {{Teske}, Johanna and {Batalha}, Natasha E. and {Wallack}, Nicole L. and {Kirk}, James and {Wogan}, Nicholas F. and {Gordon}, Tyler A. and {Alam}, Munazza K. and {Aguichine}, Artyom and {Wolfgang}, Angie and {Wakeford}, Hannah R. and {Scarsdale}, Nicholas and {Adams Redai}, Jea and {Moran}, Sarah E. and {L{\'o}pez-Morales}, Mercedes and {Meech}, Annabella and {Gao}, Peter and {Batalha}, Natalie M. and {Alderson}, Lili and {Gagnebin}, Anna},
        title = "{JWST COMPASS: NIRSpec/G395H Transmission Observations of TOI-776 c, a 2 R$_{{\ensuremath{\oplus}}}$ M Dwarf Planet}",
      journal = {\aj},
         year = 2025,
        month = may,
       volume = {169},
       number = {5},
          eid = {249},
        pages = {249},
          doi = {10.3847/1538-3881/adb975},
archivePrefix = {arXiv},
       eprint = {2502.20501},
 primaryClass = {astro-ph.EP},
       adsurl = {https://ui.adsabs.harvard.edu/abs/2025AJ....169..249T}
}

@ARTICLE{Crossfield2017,
       author = {{Crossfield}, Ian J.~M. and {Kreidberg}, Laura},
        title = "{Trends in Atmospheric Properties of Neptune-size Exoplanets}",
      journal = {\aj},
         year = 2017,
        month = dec,
       volume = {154},
       number = {6},
          eid = {261},
        pages = {261},
          doi = {10.3847/1538-3881/aa9279},
archivePrefix = {arXiv},
       eprint = {1708.00016},
 primaryClass = {astro-ph.EP},
       adsurl = {https://ui.adsabs.harvard.edu/abs/2017AJ....154..261C}
}

@ARTICLE{Brande2024,
       author = {{Brande}, Jonathan and {Crossfield}, Ian J.~M. and {Kreidberg}, Laura and {Morley}, Caroline V. and {Barman}, Travis and {Benneke}, Bj{\"o}rn and {Christiansen}, Jessie L. and {Dragomir}, Diana and {Fortney}, Jonathan J. and {Greene}, Thomas P. and {Hardegree-Ullman}, Kevin K. and {Howard}, Andrew W. and {Knutson}, Heather A. and {Lothringer}, Joshua D. and {Mikal-Evans}, Thomas},
        title = "{Clouds and Clarity: Revisiting Atmospheric Feature Trends in Neptune-size Exoplanets}",
      journal = {\apjl},
         year = 2024,
        month = jan,
       volume = {961},
       number = {1},
          eid = {L23},
        pages = {L23},
          doi = {10.3847/2041-8213/ad1b5c},
archivePrefix = {arXiv},
       eprint = {2310.07714},
 primaryClass = {astro-ph.EP},
       adsurl = {https://ui.adsabs.harvard.edu/abs/2024ApJ...961L..23B}
}

@ARTICLE{Sing2016,
       author = {{Sing}, David K. and {Fortney}, Jonathan J. and {Nikolov}, Nikolay and {Wakeford}, Hannah R. and {Kataria}, Tiffany and {Evans}, Thomas M. and {Aigrain}, Suzanne and {Ballester}, Gilda E. and {Burrows}, Adam S. and {Deming}, Drake and {D{\'e}sert}, Jean-Michel and {Gibson}, Neale P. and {Henry}, Gregory W. and {Huitson}, Catherine M. and {Knutson}, Heather A. and {Lecavelier Des Etangs}, Alain and {Pont}, Frederic and {Showman}, Adam P. and {Vidal-Madjar}, Alfred and {Williamson}, Michael H. and {Wilson}, Paul A.},
        title = "{A continuum from clear to cloudy hot-Jupiter exoplanets without primordial water depletion}",
      journal = {\nat},
         year = 2016,
        month = jan,
       volume = {529},
       number = {7584},
        pages = {59-62},
          doi = {10.1038/nature16068},
archivePrefix = {arXiv},
       eprint = {1512.04341},
 primaryClass = {astro-ph.EP},
       adsurl = {https://ui.adsabs.harvard.edu/abs/2016Natur.529...59S}
}

@ARTICLE{Fu2017,
       author = {{Fu}, Guangwei and {Deming}, Drake and {Knutson}, Heather and {Madhusudhan}, Nikku and {Mandell}, Avi and {Fraine}, Jonathan},
        title = "{Statistical Analysis of Hubble/WFC3 Transit Spectroscopy of Extrasolar Planets}",
      journal = {\apjl},
         year = 2017,
        month = oct,
       volume = {847},
       number = {2},
          eid = {L22},
        pages = {L22},
          doi = {10.3847/2041-8213/aa8e40},
archivePrefix = {arXiv},
       eprint = {1709.07385},
 primaryClass = {astro-ph.EP},
       adsurl = {https://ui.adsabs.harvard.edu/abs/2017ApJ...847L..22F}
}

@ARTICLE{Stevenson2016,
       author = {{Stevenson}, Kevin B.},
        title = "{Quantifying and Predicting the Presence of Clouds in Exoplanet Atmospheres}",
      journal = {\apjl},
         year = 2016,
        month = feb,
       volume = {817},
       number = {2},
          eid = {L16},
        pages = {L16},
          doi = {10.3847/2041-8205/817/2/L16},
archivePrefix = {arXiv},
       eprint = {1601.03492},
 primaryClass = {astro-ph.EP},
       adsurl = {https://ui.adsabs.harvard.edu/abs/2016ApJ...817L..16S}
}

@ARTICLE{Gao2020,
       author = {{Gao}, Peter and {Thorngren}, Daniel P. and {Lee}, Elspeth K.~H. and {Fortney}, Jonathan J. and {Morley}, Caroline V. and {Wakeford}, Hannah R. and {Powell}, Diana K. and {Stevenson}, Kevin B. and {Zhang}, Xi},
        title = "{Aerosol composition of hot giant exoplanets dominated by silicates and hydrocarbon hazes}",
      journal = {Nature Astronomy},
         year = 2020,
        month = may,
       volume = {4},
        pages = {951-956},
          doi = {10.1038/s41550-020-1114-3},
archivePrefix = {arXiv},
       eprint = {2005.11939},
 primaryClass = {astro-ph.EP},
       adsurl = {https://ui.adsabs.harvard.edu/abs/2020NatAs...4..951G}
}

@ARTICLE{Nortmann2018,
       author = {{Nortmann}, Lisa and {Pall{\'e}}, Enric and {Salz}, Michael and {Sanz-Forcada}, Jorge and {Nagel}, Evangelos and {Alonso-Floriano}, F. Javier and {Czesla}, Stefan and {Yan}, Fei and {Chen}, Guo and {Snellen}, Ignas A.~G. and {Zechmeister}, Mathias and {Schmitt}, J{\"u}rgen H.~M.~M. and {L{\'o}pez-Puertas}, Manuel and {Casasayas-Barris}, N{\'u}ria and {Bauer}, Florian F. and {Amado}, Pedro J. and {Caballero}, Jos{\'e} A. and {Dreizler}, Stefan and {Henning}, Thomas and {Lamp{\'o}n}, Manuel and {Montes}, David and {Molaverdikhani}, Karan and {Quirrenbach}, Andreas and {Reiners}, Ansgar and {Ribas}, Ignasi and {S{\'a}nchez-L{\'o}pez}, Alejandro and {Schneider}, P. Christian and {Zapatero Osorio}, Mar{\'\i}a R.},
        title = "{Ground-based detection of an extended helium atmosphere in the Saturn-mass exoplanet WASP-69b}",
      journal = {Science},
         year = 2018,
        month = dec,
       volume = {362},
       number = {6421},
        pages = {1388-1391},
          doi = {10.1126/science.aat5348},
archivePrefix = {arXiv},
       eprint = {1812.03119},
 primaryClass = {astro-ph.EP},
       adsurl = {https://ui.adsabs.harvard.edu/abs/2018Sci...362.1388N}
}

@ARTICLE{Zhang2023,
       author = {{Zhang}, Michael and {Dai}, Fei and {Bean}, Jacob L. and {Knutson}, Heather A. and {Rescigno}, Federica},
        title = "{Outflowing Helium from a Mature Mini-Neptune}",
      journal = {\apjl},
         year = 2023,
        month = aug,
       volume = {953},
       number = {2},
          eid = {L25},
        pages = {L25},
          doi = {10.3847/2041-8213/aced51},
archivePrefix = {arXiv},
       eprint = {2308.02002},
 primaryClass = {astro-ph.EP},
       adsurl = {https://ui.adsabs.harvard.edu/abs/2023ApJ...953L..25Z}
}

@ARTICLE{Orell-Miquel2024,
       author = {{Orell-Miquel}, J. and {Murgas}, F. and {Pall{\'e}}, E. and {Mallorqu{\'\i}n}, M. and {L{\'o}pez-Puertas}, M. and {Lamp{\'o}n}, M. and {Sanz-Forcada}, J. and {Nortmann}, L. and {Czesla}, S. and {Nagel}, E. and {Ribas}, I. and {Stangret}, M. and {Livingston}, J. and {Knudstrup}, E. and {Albrecht}, S.~H. and {Carleo}, I. and {Caballero}, J.~A. and {Dai}, F. and {Esparza-Borges}, E. and {Fukui}, A. and {Heng}, K. and {Henning}, Th. and {Kagetani}, T. and {Lesjak}, F. and {de Leon}, J.~P. and {Montes}, D. and {Morello}, G. and {Narita}, N. and {Quirrenbach}, A. and {Amado}, P.~J. and {Reiners}, A. and {Schweitzer}, A. and {Vico Linares}, J.~I.},
        title = "{The MOPYS project: A survey of 70 planets in search of extended He I and H atmospheres: No evidence of enhanced evaporation in young planets}",
      journal = {\aap},
         year = 2024,
        month = sep,
       volume = {689},
          eid = {A179},
        pages = {A179},
          doi = {10.1051/0004-6361/202449411},
archivePrefix = {arXiv},
       eprint = {2404.16732},
 primaryClass = {astro-ph.EP},
       adsurl = {https://ui.adsabs.harvard.edu/abs/2024A&A...689A.179O}
}

@ARTICLE{Rogers2023,
       author = {{Rogers}, James G. and {Schlichting}, Hilke E. and {Owen}, James E.},
        title = "{Conclusive Evidence for a Population of Water Worlds around M Dwarfs Remains Elusive}",
      journal = {\apjl},
         year = 2023,
        month = apr,
       volume = {947},
       number = {1},
          eid = {L19},
        pages = {L19},
          doi = {10.3847/2041-8213/acc86f},
archivePrefix = {arXiv},
       eprint = {2301.04321},
 primaryClass = {astro-ph.EP},
       adsurl = {https://ui.adsabs.harvard.edu/abs/2023ApJ...947L..19R}
}

@ARTICLE{Fu2025,
       author = {{Fu}, Guangwei and {Stevenson}, Kevin B. and {Sing}, David K. and {Mukherjee}, Sagnick and {Welbanks}, Luis and {Thorngren}, Daniel and {Tsai}, Shang-Min and {Gao}, Peter and {Lothringer}, Joshua and {Beatty}, Thomas G. and {Gapp}, Cyril and {Evans-Soma}, Thomas M. and {Allart}, Romain and {Pelletier}, Stefan and {Thao}, Pa Chia and {Mann}, Andrew W.},
        title = "{Statistical Trends in JWST Transiting Exoplanet Atmospheres}",
      journal = {\apj},
         year = 2025,
        month = jun,
       volume = {986},
       number = {1},
          eid = {1},
        pages = {1},
          doi = {10.3847/1538-4357/ad7bb8},
archivePrefix = {arXiv},
       eprint = {2501.02081},
 primaryClass = {astro-ph.EP},
       adsurl = {https://ui.adsabs.harvard.edu/abs/2025ApJ...986....1F}
}

@ARTICLE{Welbanks2019,
       author = {{Welbanks}, Luis and {Madhusudhan}, Nikku and {Allard}, Nicole F. and {Hubeny}, Ivan and {Spiegelman}, Fernand and {Leininger}, Thierry},
        title = "{Mass-Metallicity Trends in Transiting Exoplanets from Atmospheric Abundances of H$_{2}$O, Na, and K}",
      journal = {\apjl},
         year = 2019,
        month = dec,
       volume = {887},
       number = {1},
          eid = {L20},
        pages = {L20},
          doi = {10.3847/2041-8213/ab5a89},
archivePrefix = {arXiv},
       eprint = {1912.04904},
 primaryClass = {astro-ph.EP},
       adsurl = {https://ui.adsabs.harvard.edu/abs/2019ApJ...887L..20W}
}

@ARTICLE{Spake2021,
       author = {{Spake}, Jessica J. and {Sing}, David K. and {Wakeford}, Hannah R. and {Nikolov}, Nikolay and {Mikal-Evans}, Thomas and {Deming}, Drake and {Barstow}, Joanna K. and {Anderson}, David R. and {Carter}, Aarynn L. and {Gillon}, Michael and {Goyal}, Jayesh M. and {Hebrard}, Guillaume and {Hellier}, Coel and {Kataria}, Tiffany and {Lam}, Kristine W.~F. and {Triaud}, A.~H.~M.~J. and {Wheatley}, Peter J.},
        title = "{Abundance measurements of H$_{2}$O and carbon-bearing species in the atmosphere of WASP-127b confirm its supersolar metallicity}",
      journal = {\mnras},
         year = 2021,
        month = jan,
       volume = {500},
       number = {3},
        pages = {4042-4064},
          doi = {10.1093/mnras/staa3116},
archivePrefix = {arXiv},
       eprint = {1911.08859},
 primaryClass = {astro-ph.EP},
       adsurl = {https://ui.adsabs.harvard.edu/abs/2021MNRAS.500.4042S}
}

@ARTICLE{Madhusudhan2021,
       author = {{Madhusudhan}, Nikku and {Piette}, Anjali A.~A. and {Constantinou}, Savvas},
        title = "{Habitability and Biosignatures of Hycean Worlds}",
      journal = {\apj},
         year = 2021,
        month = sep,
       volume = {918},
       number = {1},
          eid = {1},
        pages = {1},
          doi = {10.3847/1538-4357/abfd9c},
archivePrefix = {arXiv},
       eprint = {2108.10888},
 primaryClass = {astro-ph.EP},
       adsurl = {https://ui.adsabs.harvard.edu/abs/2021ApJ...918....1M}
}

@ARTICLE{Gressier2024,
       author = {{Gressier}, Am{\'e}lie and {Espinoza}, N{\'e}stor and {Allen}, Natalie H. and {Sing}, David K. and {Banerjee}, Agnibha and {Barstow}, Joanna K. and {Valenti}, Jeff A. and {Lewis}, Nikole K. and {Birkmann}, Stephan M. and {Challener}, Ryan C. and {Manjavacas}, Elena and {Alves de Oliveira}, Catarina and {Crouzet}, Nicolas and {Beck}, Tracy. L.},
        title = "{Hints of a Sulfur-rich Atmosphere around the 1.6 R $_{{\ensuremath{\oplus}}}$ Super-Earth L98-59 d from JWST NIRspec G395H Transmission Spectroscopy}",
      journal = {\apjl},
         year = 2024,
        month = nov,
       volume = {975},
       number = {1},
          eid = {L10},
        pages = {L10},
          doi = {10.3847/2041-8213/ad73d1},
archivePrefix = {arXiv},
       eprint = {2408.15855},
 primaryClass = {astro-ph.EP},
       adsurl = {https://ui.adsabs.harvard.edu/abs/2024ApJ...975L..10G}
}

@ARTICLE{May2023,
       author = {{May}, E.~M. and {MacDonald}, Ryan J. and {Bennett}, Katherine A. and {Moran}, Sarah E. and {Wakeford}, Hannah R. and {Peacock}, Sarah and {Lustig-Yaeger}, Jacob and {Highland}, Alicia N. and {Stevenson}, Kevin B. and {Sing}, David K. and {Mayorga}, L.~C. and {Batalha}, Natasha E. and {Kirk}, James and {L{\'o}pez-Morales}, Mercedes and {Valenti}, Jeff A. and {Alam}, Munazza K. and {Alderson}, Lili and {Fu}, Guangwei and {Gonzalez-Quiles}, Junellie and {Lothringer}, Joshua D. and {Rustamkulov}, Zafar and {Sotzen}, Kristin S.},
        title = "{Double Trouble: Two Transits of the Super-Earth GJ 1132 b Observed with JWST NIRSpec G395H}",
      journal = {\apjl},
         year = 2023,
        month = dec,
       volume = {959},
       number = {1},
          eid = {L9},
        pages = {L9},
          doi = {10.3847/2041-8213/ad054f},
archivePrefix = {arXiv},
       eprint = {2310.10711},
 primaryClass = {astro-ph.EP},
       adsurl = {https://ui.adsabs.harvard.edu/abs/2023ApJ...959L...9M}
}

@ARTICLE{Asplund2009,
       author = {{Asplund}, Martin and {Grevesse}, Nicolas and {Sauval}, A. Jacques and {Scott}, Pat},
        title = "{The Chemical Composition of the Sun}",
      journal = {\araa},
         year = 2009,
        month = sep,
       volume = {47},
       number = {1},
        pages = {481-522},
          doi = {10.1146/annurev.astro.46.060407.145222},
archivePrefix = {arXiv},
       eprint = {0909.0948},
 primaryClass = {astro-ph.SR},
       adsurl = {https://ui.adsabs.harvard.edu/abs/2009ARA&A..47..481A}
}

@ARTICLE{Penzlin2024,
       author = {{Penzlin}, Anna B.~T. and {Booth}, Richard A. and {Kirk}, James and {Owen}, James E. and {Ahrer}, E. and {Christie}, Duncan A. and {Claringbold}, Alastair B. and {Esparza-Borges}, Emma and {L{\'o}pez-Morales}, M. and {Mayne}, N.~J. and {McCormack}, Mason and {Meech}, Annabella and {Panwar}, Vatsal and {Powell}, Diana and {Sergeev}, Denis E. and {Taylor}, Jake and {Wheatley}, Peter J. and {Zamyatina}, Maria},
        title = "{BOWIE-ALIGN: how formation and migration histories of giant planets impact atmospheric compositions}",
      journal = {\mnras},
         year = 2024,
        month = nov,
       volume = {535},
       number = {1},
        pages = {171-186},
          doi = {10.1093/mnras/stae2362},
archivePrefix = {arXiv},
       eprint = {2407.03199},
 primaryClass = {astro-ph.EP},
       adsurl = {https://ui.adsabs.harvard.edu/abs/2024MNRAS.535..171P}
}

@article{kempton2023reflective,
  title={A reflective, metal-rich atmosphere for GJ 1214b from its JWST phase curve},
  author={Kempton, Eliza M-R and Zhang, Michael and Bean, Jacob L and Steinrueck, Maria E and Piette, Anjali AA and Parmentier, Vivien and Malsky, Isaac and Roman, Michael T and Rauscher, Emily and Gao, Peter and others},
  journal={Nature},
  volume={620},
  number={7972},
  pages={67--71},
  year={2023},
  publisher={Nature Publishing Group UK London}
}

@article{schlawin2024possible,
  title={Possible Carbon Dioxide above the Thick Aerosols of GJ 1214 b},
  author={Schlawin, Everett and Ohno, Kazumasa and Bell, Taylor J and Murphy, Matthew M and Welbanks, Luis and Beatty, Thomas G and Greene, Thomas P and Fortney, Jonathan J and Parmentier, Vivien and Edelman, Isaac R and others},
  journal={The Astrophysical Journal Letters},
  volume={974},
  number={2},
  pages={L33},
  year={2024},
  publisher={IOP Publishing}
}

@ARTICLE{Ohno2025,
       author = {{Ohno}, Kazumasa and {Schlawin}, Everett and {Bell}, Taylor J. and {Murphy}, Matthew M. and {Beatty}, Thomas G. and {Welbanks}, Luis and {Greene}, Thomas P. and {Fortney}, Jonathan J. and {Parmentier}, Vivien and {Edelman}, Isaac R. and {Mehta}, Nishil and {Rieke}, Marcia J.},
        title = "{A Possible Metal-dominated Atmosphere below the Thick Aerosols of GJ 1214 b Suggested by Its JWST Panchromatic Transmission Spectrum}",
      journal = {\apjl},
         year = 2025,
        month = jan,
       volume = {979},
       number = {1},
          eid = {L7},
        pages = {L7},
          doi = {10.3847/2041-8213/ada02c},
archivePrefix = {arXiv},
       eprint = {2410.10186},
 primaryClass = {astro-ph.EP},
       adsurl = {https://ui.adsabs.harvard.edu/abs/2025ApJ...979L...7O}
}

@ARTICLE{Moran2023,
       author = {{Moran}, Sarah E. and {Stevenson}, Kevin B. and {Sing}, David K. and {MacDonald}, Ryan J. and {Kirk}, James and {Lustig-Yaeger}, Jacob and {Peacock}, Sarah and {Mayorga}, L.~C. and {Bennett}, Katherine A. and {L{\'o}pez-Morales}, Mercedes and {May}, E.~M. and {Rustamkulov}, Zafar and {Valenti}, Jeff A. and {Adams Redai}, J{\'e}a I. and {Alam}, Munazza K. and {Batalha}, Natasha E. and {Fu}, Guangwei and {Gonzalez-Quiles}, Junellie and {Highland}, Alicia N. and {Kruse}, Ethan and {Lothringer}, Joshua D. and {Ortiz Ceballos}, Kevin N. and {Sotzen}, Kristin S. and {Wakeford}, Hannah R.},
        title = "{High Tide or Riptide on the Cosmic Shoreline? A Water-rich Atmosphere or Stellar Contamination for the Warm Super-Earth GJ 486b from JWST Observations}",
      journal = {\apjl},
         year = 2023,
        month = may,
       volume = {948},
       number = {1},
          eid = {L11},
        pages = {L11},
          doi = {10.3847/2041-8213/accb9c},
archivePrefix = {arXiv},
       eprint = {2305.00868},
 primaryClass = {astro-ph.EP},
       adsurl = {https://ui.adsabs.harvard.edu/abs/2023ApJ...948L..11M}
}

@ARTICLE{Claringbold2026,
       author = {{Claringbold}, Alastair B. and {Fisher}, Chloe E. and {Kirk}, James and {Ahrer}, Eva-Maria and {Penzlin}, Anna B.~T. and {Thorngren}, Daniel P. and {L{\'o}pez-Morales}, Mercedes and {Wheatley}, Peter J. and {Alderson}, Lili and {Booth}, Richard A. and {Christie}, Duncan A. and {Fairman}, Charlotte and {Mayne}, Nathan J. and {McCormack}, Mason and {Meech}, Annabella and {Owen}, James E. and {Panwar}, Vatsal and {Sergeev}, Denis E. and {Valentine}, Daniel and {Wakeford}, Hannah R. and {Zamyatina}, Maria},
        title = "{BOWIE-ALIGN: Sub-solar C/O ratio and metallicity atmosphere of the misaligned hot Jupiter HAT-P-30 b}",
      journal = {\mnras},
         year = 2026,
        month = mar,
       volume = {546},
       number = {4},
          eid = {stag143},
        pages = {stag143},
          doi = {10.1093/mnras/stag143},
archivePrefix = {arXiv},
       eprint = {2601.13104},
 primaryClass = {astro-ph.EP},
       adsurl = {https://ui.adsabs.harvard.edu/abs/2026MNRAS.546ag143C}
}

@ARTICLE{Fairman2026,
       author = {{Fairman}, Charlotte and {Wakeford}, Hannah R. and {Claringbold}, Alastair B. and {Kirk}, James and {Ahrer}, Eva-Maria and {Thorngren}, Daniel and {Tsai}, Shang-Min and {Booth}, R.~A. and {Penzlin}, Anna B.~T. and {Alderson}, Lili and {Christie}, Duncan A. and {L{\'o}pez-Morales}, M. and {Mayne}, N.~J. and {Meech}, Annabella and {Owen}, James E. and {Panwar}, Vatsal and {Valentine}, Daniel and {Wheatley}, Peter J. and {Zamyatina}, Maria},
        title = "{BOWIE-ALIGN: Exploring degeneracies in the muted transmission spectrum of the aligned hot Jupiter NGTS-2b with NIRSpec/G395H.}",
      journal = {\mnras},
         year = 2026,
        month = mar,
          doi = {10.1093/mnras/stag528},
archivePrefix = {arXiv},
       eprint = {2603.18332},
 primaryClass = {astro-ph.EP},
       adsurl = {https://ui.adsabs.harvard.edu/abs/2026MNRAS.tmp..498F}
}

@ARTICLE{Meech2026,
       author = {{Meech}, Annabella and {Gao}, Peter and {Wallack}, Nicole L. and {L{\'o}pez-Morales}, Mercedes and {Oddo}, Dominic and {Teske}, Johanna and {Dragomir}, Diana and {Wolfgang}, Angie and {Wogan}, Nicholas and {Wakeford}, Hannah R. and {Moran}, Sarah E. and {Kirk}, James and {Gordon}, Tyler A. and {Gagnebin}, Anna and {Batalha}, Natasha E. and {Batalha}, Natalie M. and {Alderson}, Lili and {Alam}, Munazza K. and {Aguichine}, Artyom},
        title = "{JWST COMPASS: A NIRSpec G395H Transmission Spectrum of Radius Valley Dweller TOI-260 b}",
      journal = {\aj},
         year = 2026,
        month = may,
       volume = {171},
       number = {5},
          eid = {274},
        pages = {274},
          doi = {10.3847/1538-3881/ae472f},
archivePrefix = {arXiv},
       eprint = {2602.22329},
 primaryClass = {astro-ph.EP},
       adsurl = {https://ui.adsabs.harvard.edu/abs/2026AJ....171..274M}
}

@ARTICLE{Wallack2026,
       author = {{Wallack}, Nicole L. and {Gao}, Peter and {Greklek-McKeon}, Michael and {Meech}, Annabella and {Aguichine}, Artyom and {Alam}, Munazza K. and {Alderson}, Lili and {Batalha}, Natasha E. and {Batalha}, Natalie M. and {Gagnebin}, Anna and {Gordon}, Tyler A. and {Kirk}, James and {L{\'o}pez-Morales}, Mercedes and {Moran}, Sarah E. and {Redai}, Jea Iyanla and {Scarsdale}, Nicholas and {Teske}, Johanna and {Wakeford}, Hannah R. and {Wogan}, Nicholas F. and {Wolfgang}, Angie},
        title = "{JWST COMPASS: NIRSpec/G395H Transmission Observations of the Sub-Neptune HD 15337 c}",
      journal = {\aj},
         year = 2026,
        month = mar,
       volume = {171},
       number = {3},
          eid = {180},
        pages = {180},
          doi = {10.3847/1538-3881/ae2d12},
archivePrefix = {arXiv},
       eprint = {2602.22327},
 primaryClass = {astro-ph.EP},
       adsurl = {https://ui.adsabs.harvard.edu/abs/2026AJ....171..180W}
}

@ARTICLE{Adams2025,
       author = {{Adams Redai}, Jea and {Wogan}, Nicholas and {Wallack}, Nicole L. and {Alam}, Munazza K. and {Aguichine}, Artyom and {Wolfgang}, Angie and {Wakeford}, Hannah R. and {Teske}, Johanna and {Scarsdale}, Nicholas and {Moran}, Sarah E. and {L{\'o}pez-Morales}, Mercedes and {Meech}, Annabella and {Gao}, Peter and {Gagnebin}, Anna and {Batalha}, Natasha E. and {Batalha}, Natalie M. and {Alderson}, Lili},
        title = "{JWST COMPASS: A NIRSpec G395H Transmission Spectrum of the Super-Earth GJ 357 b}",
      journal = {\aj},
         year = 2025,
        month = oct,
       volume = {170},
       number = {4},
          eid = {219},
        pages = {219},
          doi = {10.3847/1538-3881/adee92},
archivePrefix = {arXiv},
       eprint = {2507.07165},
 primaryClass = {astro-ph.EP},
       adsurl = {https://ui.adsabs.harvard.edu/abs/2025AJ....170..219A}
}

@ARTICLE{Madhusudhan2014,
       author = {{Madhusudhan}, Nikku and {Amin}, Mustafa A. and {Kennedy}, Grant M.},
        title = "{Toward Chemical Constraints on Hot Jupiter Migration}",
      journal = {\apjl},
         year = 2014,
        month = oct,
       volume = {794},
       number = {1},
          eid = {L12},
        pages = {L12},
          doi = {10.1088/2041-8205/794/1/L12},
archivePrefix = {arXiv},
       eprint = {1408.3668},
 primaryClass = {astro-ph.EP},
       adsurl = {https://ui.adsabs.harvard.edu/abs/2014ApJ...794L..12M}
}

@ARTICLE{Fulton2017,
       author = {{Fulton}, Benjamin J. and {Petigura}, Erik A. and {Howard}, Andrew W. and {Isaacson}, Howard and {Marcy}, Geoffrey W. and {Cargile}, Phillip A. and {Hebb}, Leslie and {Weiss}, Lauren M. and {Johnson}, John Asher and {Morton}, Timothy D. and {Sinukoff}, Evan and {Crossfield}, Ian J.~M. and {Hirsch}, Lea A.},
        title = "{The California-Kepler Survey. III. A Gap in the Radius Distribution of Small Planets}",
      journal = {\aj},
         year = 2017,
        month = sep,
       volume = {154},
       number = {3},
          eid = {109},
        pages = {109},
          doi = {10.3847/1538-3881/aa80eb},
archivePrefix = {arXiv},
       eprint = {1703.10375},
 primaryClass = {astro-ph.EP},
       adsurl = {https://ui.adsabs.harvard.edu/abs/2017AJ....154..109F}
}

@ARTICLE{Owen2017,
       author = {{Owen}, James E. and {Wu}, Yanqin},
        title = "{The Evaporation Valley in the Kepler Planets}",
      journal = {\apj},
         year = 2017,
        month = sep,
       volume = {847},
       number = {1},
          eid = {29},
        pages = {29},
          doi = {10.3847/1538-4357/aa890a},
archivePrefix = {arXiv},
       eprint = {1705.10810},
 primaryClass = {astro-ph.EP},
       adsurl = {https://ui.adsabs.harvard.edu/abs/2017ApJ...847...29O}
}

@ARTICLE{Owen2026,
       author = {{Owen}, James E. and {Kirk}, James},
        title = "{Resolving the flat-spectrum conundrum: clumpy aerosol distributions in sub-Neptune atmospheres}",
      journal = {\mnras},
         year = 2026,
        month = jan,
       volume = {545},
       number = {3},
          eid = {staf2066},
        pages = {staf2066},
          doi = {10.1093/mnras/staf2066},
archivePrefix = {arXiv},
       eprint = {2511.19013},
 primaryClass = {astro-ph.EP},
       adsurl = {https://ui.adsabs.harvard.edu/abs/2026MNRAS.545f2066O}
}

@ARTICLE{Ahrer2025_GJ3090b,
       author = {{Ahrer}, Eva-Maria and {Radica}, Michael and {Piaulet-Ghorayeb}, Caroline and {Raul}, Eshan and {Wiser}, Lindsey and {Welbanks}, Luis and {Acu{\~n}a}, Lorena and {Allart}, Romain and {Coulombe}, Louis-Philippe and {Louca}, Amy and {MacDonald}, Ryan and {Saidel}, Morgan and {Evans-Soma}, Thomas M. and {Benneke}, Bj{\"o}rn and {Christie}, Duncan and {Beatty}, Thomas G. and {Cadieux}, Charles and {Cloutier}, Ryan and {Doyon}, Ren{\'e} and {Fortney}, Jonathan J. and {Gagnebin}, Anna and {Gapp}, Cyril and {Innes}, Hamish and {Knutson}, Heather A. and {Komacek}, Thaddeus and {Krissansen-Totton}, Joshua and {Miguel}, Yamila and {Pierrehumbert}, Raymond and {Roy}, Pierre-Alexis and {Schlichting}, Hilke E.},
        title = "{Escaping Helium and a Highly Muted Spectrum Suggest a Metal-enriched Atmosphere on Sub-Neptune GJ 3090 b from JWST Transit Spectroscopy}",
      journal = {\apjl},
         year = 2025,
        month = may,
       volume = {985},
       number = {1},
          eid = {L10},
        pages = {L10},
          doi = {10.3847/2041-8213/add010},
archivePrefix = {arXiv},
       eprint = {2504.20428},
 primaryClass = {astro-ph.EP},
       adsurl = {https://ui.adsabs.harvard.edu/abs/2025ApJ...985L..10A}
}

@ARTICLE{Lustig-Yaeger2023,
       author = {{Lustig-Yaeger}, Jacob and {Fu}, Guangwei and {May}, E.~M. and {Ceballos}, Kevin N. Ortiz and {Moran}, Sarah E. and {Peacock}, Sarah and {Stevenson}, Kevin B. and {Kirk}, James and {L{\'o}pez-Morales}, Mercedes and {MacDonald}, Ryan J. and {Mayorga}, L.~C. and {Sing}, David K. and {Sotzen}, Kristin S. and {Valenti}, Jeff A. and {Redai}, J{\'e}a I. Adams and {Alam}, Munazza K. and {Batalha}, Natasha E. and {Bennett}, Katherine A. and {Gonzalez-Quiles}, Junellie and {Kruse}, Ethan and {Lothringer}, Joshua D. and {Rustamkulov}, Zafar and {Wakeford}, Hannah R.},
        title = "{A JWST transmission spectrum of the nearby Earth-sized exoplanet LHS 475 b}",
      journal = {Nature Astronomy},
         year = 2023,
        month = nov,
       volume = {7},
        pages = {1317-1328},
          doi = {10.1038/s41550-023-02064-z},
archivePrefix = {arXiv},
       eprint = {2301.04191},
 primaryClass = {astro-ph.EP},
       adsurl = {https://ui.adsabs.harvard.edu/abs/2023NatAs...7.1317L}
}

@ARTICLE{Tsai2023,
       author = {{Tsai}, Shang-Min and {Lee}, Elspeth K.~H. and {Powell}, Diana and {Gao}, Peter and {Zhang}, Xi and {Moses}, Julianne and {H{\'e}brard}, Eric and {Venot}, Olivia and {Parmentier}, Vivien and {Jordan}, Sean and {Hu}, Renyu and {Alam}, Munazza K. and {Alderson}, Lili and {Batalha}, Natalie M. and {Bean}, Jacob L. and {Benneke}, Bj{\"o}rn and {Bierson}, Carver J. and {Brady}, Ryan P. and {Carone}, Ludmila and {Carter}, Aarynn L. and {Chubb}, Katy L. and {Inglis}, Julie and {Leconte}, J{\'e}r{\'e}my and {Line}, Michael and {L{\'o}pez-Morales}, Mercedes and {Miguel}, Yamila and {Molaverdikhani}, Karan and {Rustamkulov}, Zafar and {Sing}, David K. and {Stevenson}, Kevin B. and {Wakeford}, Hannah R. and {Yang}, Jeehyun and {Aggarwal}, Keshav and {Baeyens}, Robin and {Barat}, Saugata and {de Val-Borro}, Miguel and {Daylan}, Tansu and {Fortney}, Jonathan J. and {France}, Kevin and {Goyal}, Jayesh M. and {Grant}, David and {Kirk}, James and {Kreidberg}, Laura and {Louca}, Amy and {Moran}, Sarah E. and {Mukherjee}, Sagnick and {Nasedkin}, Evert and {Ohno}, Kazumasa and {Rackham}, Benjamin V. and {Redfield}, Seth and {Taylor}, Jake and {Tremblin}, Pascal and {Visscher}, Channon and {Wallack}, Nicole L. and {Welbanks}, Luis and {Youngblood}, Allison and {Ahrer}, Eva-Maria and {Batalha}, Natasha E. and {Behr}, Patrick and {Berta-Thompson}, Zachory K. and {Blecic}, Jasmina and {Casewell}, S.~L. and {Crossfield}, Ian J.~M. and {Crouzet}, Nicolas and {Cubillos}, Patricio E. and {Decin}, Leen and {D{\'e}sert}, Jean-Michel and {Feinstein}, Adina D. and {Gibson}, Neale P. and {Harrington}, Joseph and {Heng}, Kevin and {Henning}, Thomas and {Kempton}, Eliza M. -R. and {Krick}, Jessica and {Lagage}, Pierre-Olivier and {Lendl}, Monika and {Lothringer}, Joshua D. and {Mansfield}, Megan and {Mayne}, N.~J. and {Mikal-Evans}, Thomas and {Palle}, Enric and {Schlawin}, Everett and {Shorttle}, Oliver and {Wheatley}, Peter J. and {Yurchenko}, Sergei N.},
        title = "{Photochemically produced SO$_{2}$ in the atmosphere of WASP-39b}",
      journal = {\nat},
         year = 2023,
        month = may,
       volume = {617},
       number = {7961},
        pages = {483-487},
          doi = {10.1038/s41586-023-05902-2},
archivePrefix = {arXiv},
       eprint = {2211.10490},
 primaryClass = {astro-ph.EP},
       adsurl = {https://ui.adsabs.harvard.edu/abs/2023Natur.617..483T}
}

@ARTICLE{Dyrek2024,
       author = {{Dyrek}, Achr{\`e}ne and {Min}, Michiel and {Decin}, Leen and {Bouwman}, Jeroen and {Crouzet}, Nicolas and {Molli{\`e}re}, Paul and {Lagage}, Pierre-Olivier and {Konings}, Thomas and {Tremblin}, Pascal and {G{\"u}del}, Manuel and {Pye}, John and {Waters}, Rens and {Henning}, Thomas and {Vandenbussche}, Bart and {Ardevol Martinez}, Francisco and {Argyriou}, Ioannis and {Ducrot}, Elsa and {Heinke}, Linus and {van Looveren}, Gwenael and {Absil}, Olivier and {Barrado}, David and {Baudoz}, Pierre and {Boccaletti}, Anthony and {Cossou}, Christophe and {Coulais}, Alain and {Edwards}, Billy and {Gastaud}, Ren{\'e} and {Glasse}, Alistair and {Glauser}, Adrian and {Greene}, Thomas P. and {Kendrew}, Sarah and {Krause}, Oliver and {Lahuis}, Fred and {Mueller}, Michael and {Olofsson}, Goran and {Patapis}, Polychronis and {Rouan}, Daniel and {Royer}, Pierre and {Scheithauer}, Silvia and {Waldmann}, Ingo and {Whiteford}, Niall and {Colina}, Luis and {van Dishoeck}, Ewine F. and {{\"O}stlin}, G{\"o}ran and {Ray}, Tom P. and {Wright}, Gillian},
        title = "{SO$_{2}$, silicate clouds, but no CH$_{4}$ detected in a warm Neptune}",
      journal = {\nat},
         year = 2024,
        month = jan,
       volume = {625},
       number = {7993},
        pages = {51-54},
          doi = {10.1038/s41586-023-06849-0},
archivePrefix = {arXiv},
       eprint = {2311.12515},
 primaryClass = {astro-ph.EP},
       adsurl = {https://ui.adsabs.harvard.edu/abs/2024Natur.625...51D}
}

@ARTICLE{Sing2024,
       author = {{Sing}, David K. and {Rustamkulov}, Zafar and {Thorngren}, Daniel P. and {Barstow}, Joanna K. and {Tremblin}, Pascal and {Alves de Oliveira}, Catarina and {Beck}, Tracy L. and {Birkmann}, Stephan M. and {Challener}, Ryan C. and {Crouzet}, Nicolas and {Espinoza}, N{\'e}stor and {Ferruit}, Pierre and {Giardino}, Giovanna and {Gressier}, Am{\'e}lie and {Lee}, Elspeth K.~H. and {Lewis}, Nikole K. and {Maiolino}, Roberto and {Manjavacas}, Elena and {Rauscher}, Bernard J. and {Sirianni}, Marco and {Valenti}, Jeff A.},
        title = "{A warm Neptune's methane reveals core mass and vigorous atmospheric mixing}",
      journal = {\nat},
         year = 2024,
        month = jun,
       volume = {630},
       number = {8018},
        pages = {831-835},
          doi = {10.1038/s41586-024-07395-z},
archivePrefix = {arXiv},
       eprint = {2405.11027},
 primaryClass = {astro-ph.EP},
       adsurl = {https://ui.adsabs.harvard.edu/abs/2024Natur.630..831S}
}

@ARTICLE{Welbanks2024,
       author = {{Welbanks}, Luis and {Bell}, Taylor J. and {Beatty}, Thomas G. and {Line}, Michael R. and {Ohno}, Kazumasa and {Fortney}, Jonathan J. and {Schlawin}, Everett and {Greene}, Thomas P. and {Rauscher}, Emily and {McGill}, Peter and {Murphy}, Matthew and {Parmentier}, Vivien and {Tang}, Yao and {Edelman}, Isaac and {Mukherjee}, Sagnick and {Wiser}, Lindsey S. and {Lagage}, Pierre-Olivier and {Dyrek}, Achr{\`e}ne and {Arnold}, Kenneth E.},
        title = "{A high internal heat flux and large core in a warm Neptune exoplanet}",
      journal = {\nat},
         year = 2024,
        month = jun,
       volume = {630},
       number = {8018},
        pages = {836-840},
          doi = {10.1038/s41586-024-07514-w},
archivePrefix = {arXiv},
       eprint = {2405.11018},
 primaryClass = {astro-ph.EP},
       adsurl = {https://ui.adsabs.harvard.edu/abs/2024Natur.630..836W}
}

@ARTICLE{Beatty2024,
       author = {{Beatty}, Thomas G. and {Welbanks}, Luis and {Schlawin}, Everett and {Bell}, Taylor J. and {Line}, Michael R. and {Murphy}, Matthew and {Edelman}, Isaac and {Greene}, Thomas P. and {Fortney}, Jonathan J. and {Henry}, Gregory W. and {Mukherjee}, Sagnick and {Ohno}, Kazumasa and {Parmentier}, Vivien and {Rauscher}, Emily and {Wiser}, Lindsey S. and {Arnold}, Kenneth E.},
        title = "{Sulfur Dioxide and Other Molecular Species in the Atmosphere of the Sub-Neptune GJ 3470 b}",
      journal = {\apjl},
         year = 2024,
        month = jul,
       volume = {970},
       number = {1},
          eid = {L10},
        pages = {L10},
          doi = {10.3847/2041-8213/ad55e9},
archivePrefix = {arXiv},
       eprint = {2406.04450},
 primaryClass = {astro-ph.EP},
       adsurl = {https://ui.adsabs.harvard.edu/abs/2024ApJ...970L..10B}
}

@ARTICLE{Hu2021,
       author = {{Hu}, Renyu},
        title = "{Photochemistry and Spectral Characterization of Temperate and Gas-rich Exoplanets}",
      journal = {\apj},
         year = 2021,
        month = nov,
       volume = {921},
       number = {1},
          eid = {27},
        pages = {27},
          doi = {10.3847/1538-4357/ac1789},
archivePrefix = {arXiv},
       eprint = {2108.04419},
 primaryClass = {astro-ph.EP},
       adsurl = {https://ui.adsabs.harvard.edu/abs/2021ApJ...921...27H}
}




\appendix
\section{Slant optical depth: isothermal, constant gravity atmosphere}  \label{sec:appendix_1}
 Transmission spectroscopy is sensitive to the slant optical depth through the planet's atmosphere \citep[e.g.][]{Seager2000,Brown2001}. As such:
 \begin{equation}
     \tau_\nu(b) = \int_{-\infty}^\infty\sigma_\nu(x)n(x){\rm d}x
 \end{equation}
 Making an assumption of an isothermal atmosphere\footnote{Strictly speaking, we mean a constant isothermal sound-speed for an ideal gas, such that the ratio of pressure to density is constant.}, constant gravity atmosphere, the number density with radius, is given by:
 \begin{equation}
     n(r) = n_0(b)\exp\left(-\frac{r-b}{H}\right)
 \end{equation}
 Now, in general the extinction must consider all species in the atmosphere, and their respective opacities:
 \begin{equation}
     \sigma_\nu(r)n(r) = \sum_{i=1}^{n_s}\sigma_{\nu,i}(r)n_i(r)
 \end{equation}
 Without loss of generality, we make write the cross-sections as explicit functions of number density such that $\sigma_{\nu,i}(r)=\sigma_{\nu,i}(n/n_0)$, and defining abundance ratios $X_i=n_i/n$ we find:
 \begin{equation}
     \sigma_\nu(r)n(r) = n(r)\sum_{i=1}^{n_s}\sigma_{\nu,i}(n/n_0)X_{i}(n/n_0)=n(r)\sum_{i=1}^{n_s}\alpha_{\nu,i}(n/n_0)
 \end{equation}
 where $\alpha$ is the extinction cross-section of each species for a given gas composition. Now, since transmission spectroscopy is only sensitive to a few scale heights, and the extinction cross-section is unlikely to be rapidly varying with altitude, we may locally approximate:
 \begin{equation}
     \alpha_{\nu,i}=\alpha_{\nu,i}(n_0)\left(\frac{n}{n_0}\right)^{l_i}
 \end{equation}
 where, due to the assumption that the extinction cross-section is not rapidly varying $l_i$ is approximately constant. Under this assumption, the slant optical depth can now be analytically approximated:
 \begin{equation}
 \tau_{\nu}=n_0(b)\sum_{i=1}^{n_s}\alpha_{\nu,i}(n_0(b))\int_{-\infty}^\infty\exp\left[-\frac{(l_i+1)(r-b)}{H}\right]{\rm d}x    
 \end{equation}
 making the standard slant path approximation $r-b\approx x^2/2b$, we find:
 \begin{equation}
 \begin{split}
 \tau_{\nu}&=n_0(b)\sum_{i=1}^{n_s}\alpha_{\nu,i}(n_0(b))\int_{-\infty}^\infty\exp\left[-\frac{(l_i+1)x^2}{2bH}\right]{\rm d}x  \\&= \sqrt{2\pi Hb}\,n_0(b)\sum_{i=1}^{n_s}\frac{\alpha_{\nu,i}(n_0(b))}{\sqrt{l_i+1}} = A\sqrt{2\pi Hb}\,n_0(b)\sigma_\nu \label{eqn:limb_tau}
 \end{split}
 \end{equation}
 where $\sigma_\nu$ is the gas' total extinction cross-section and $l_i=0$ is the the well known constant cross-section with altitude result. Now, at a given frequency is likely that a single line of a single species dominates the optical depth, such that the optical depth maybe approximated as:
 \begin{equation}
     \tau_\nu\approx \sqrt{2\pi Hb}\,n_0(b) \frac{\alpha_\nu(n_0(b))}{\sqrt{l_\nu+1}}
 \end{equation}
 where $l_\nu$ is the appropriate power-law index for the species that dominates at the frequency $\nu$. Now further, under the assumption $H<<b$ we may rewrite the optical depth as a function of radius, about some reference radius $R_0$ as:
 \begin{equation}
     \tau_\nu(r)\approx\sqrt{2\pi H R_0}\,n_0(R_0)\frac{\alpha_\nu(R_0)}{\sqrt{l_\nu+1}}\exp\left(-\frac{(l_\nu+1)(r-R_0)}{H}\right)
 \end{equation}
 which is of the form $\tau_\nu(z)=\tau_{\nu,0} \exp(-z/B_\nu)$ where $z=r-R_0$ is the altitude above the reference radius. An optical depth of this form, has a transit radius of (Appendix A of \citealt{deWit2013}):
 \begin{equation}
     R_{t,\nu}=R_0+\frac{H}{l_\nu+1}\left(\gamma+\log \tau_{\nu,0}\right)
 \end{equation}
 with $\gamma$ the Euler-Mascheroni constant.

\section{Stacking with a cooler reference planet}
\label{sec:cool_reference_planet}

\begin{figure}
    \centering
    \includegraphics[width=1\linewidth]{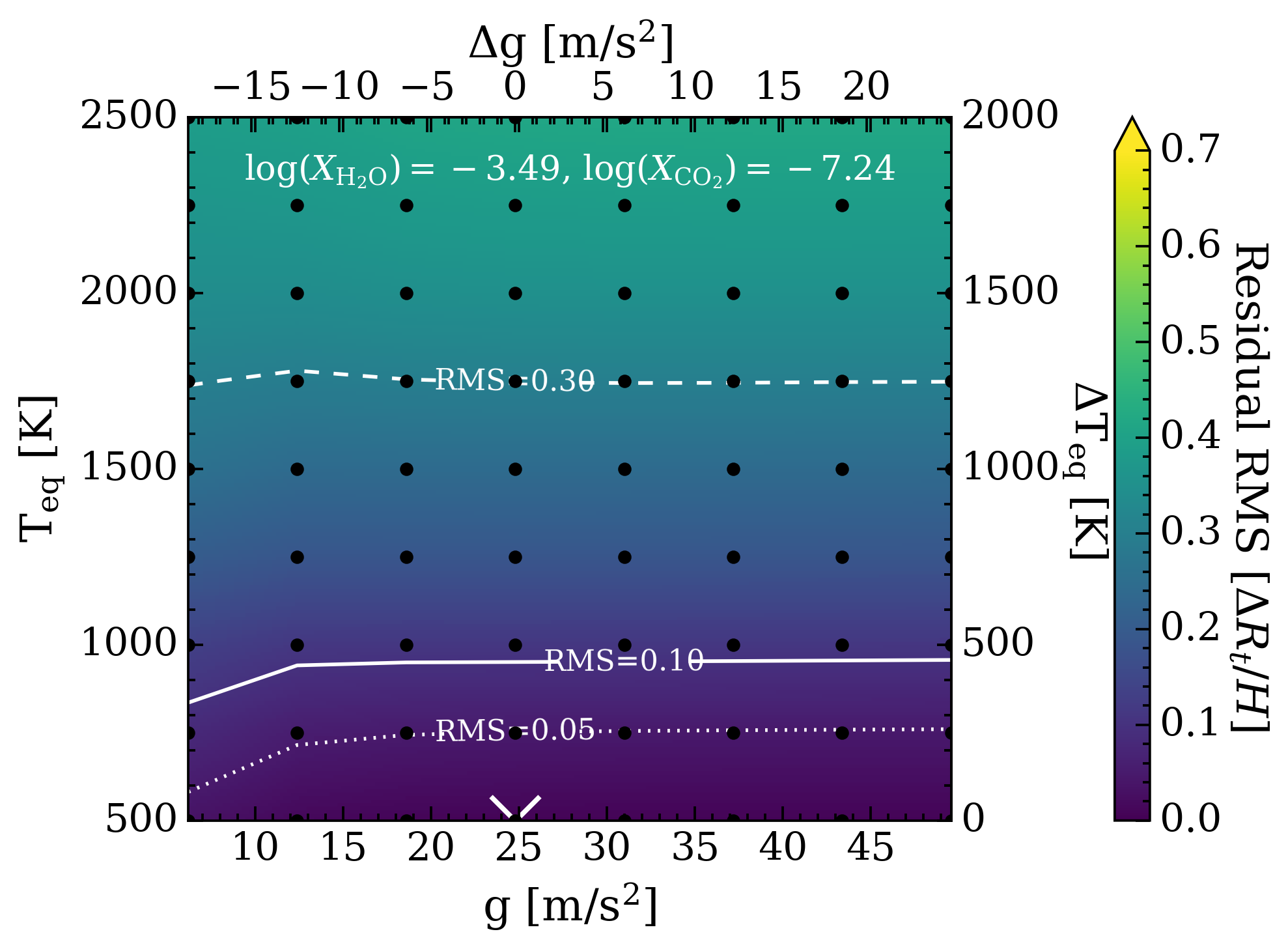}
    \includegraphics[width=1\linewidth]{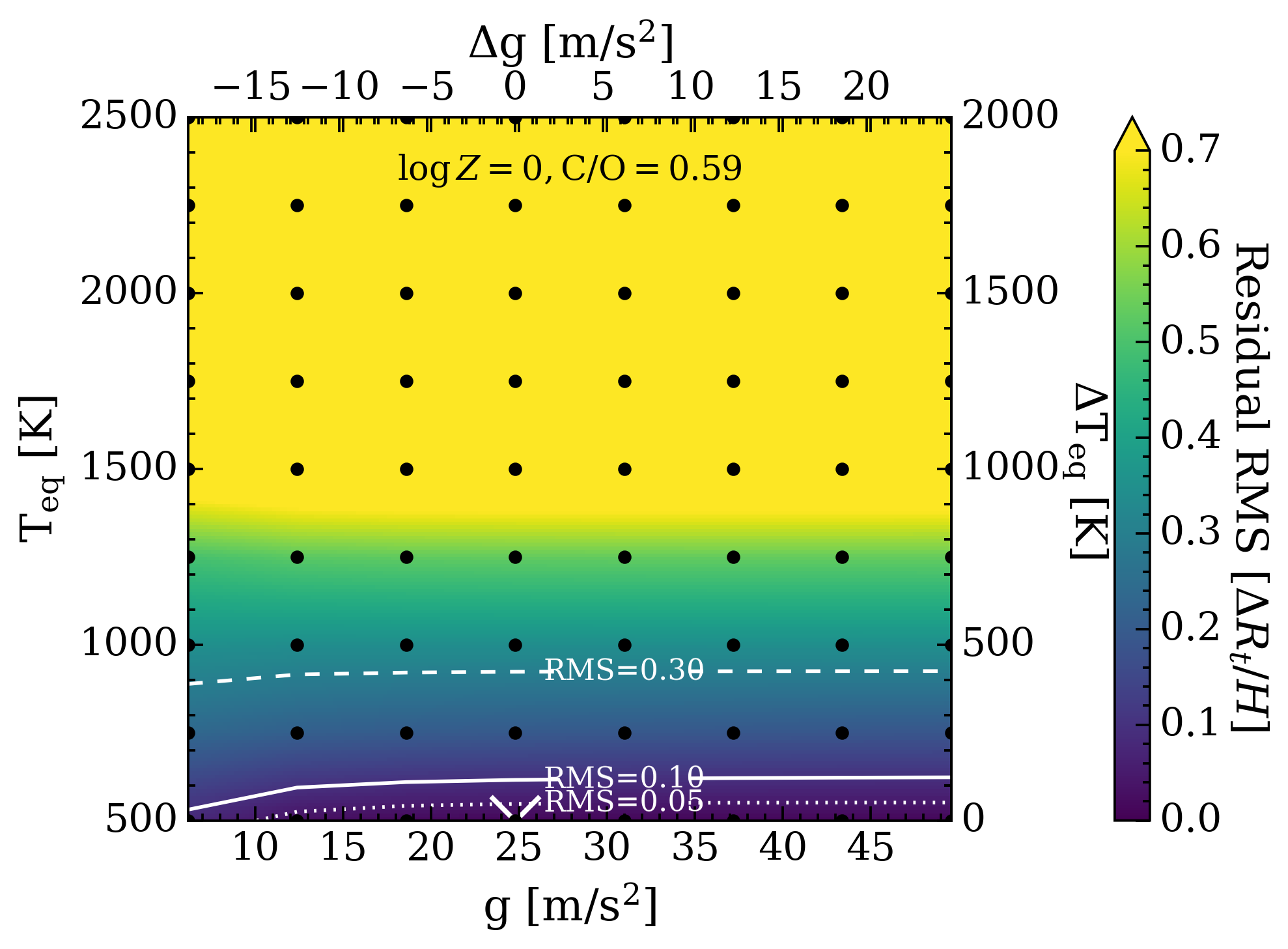}
    \caption{These figures are the results from similar tests to those presented in Figures \protect\ref{fig:two_planet_grid_identical} and \protect\ref{fig:two_planet_grid_variable} but with a cooler reference planet (shown by the white cross) that always forms one of the two planets in the stack. See the caption of Figure \protect\ref{fig:two_planet_grid_identical} for details.}
    \label{fig:two_planet_grids_cool_reference_planet}
\end{figure}

In this appendix, we repeat the test of Section \ref{sec:stacking_different_planets_identical_compositions} and Section \ref{sec:stacking_different_planets_different_compositions} but with a cooler reference planet (T$_{\mathrm{eq}} = 500$\,K). When the two stacked planets' compositions are identical, changes in temperature can cause $>1 \sigma$ differences between the geometric mean abundance spectrum and stacked spectrum once the temperature difference exceeds $\sim 1250$\,K (Figure \ref{fig:two_planet_grids_cool_reference_planet}, top panel), which builds upon our findings in Figure \ref{fig:stacking_test_2}. Similar to what we showed in Figure \ref{fig:two_planet_grid_variable}, this temperature dependence is amplified when allowing the compositions to vary according to equilibrium chemistry, with it being inappropriate to stack two planets that span the CH$_4$/CO boundary (Figure \ref{fig:two_planet_grids_cool_reference_planet}, bottom panel). 

\bsp	
\label{lastpage}
\end{document}